\documentclass[a4paper,12pt]{article}
\usepackage{amsmath,amsfonts,epsfig,color,latexsym}
\usepackage{jheppub,esint,shuffle,psfrag}
\usepackage[utf8]{inputenc}
\usepackage{bm}
\usepackage{amssymb}

\newcommand\eps{\epsilon}
 
 \usepackage{color}
\usepackage{esint} 
\usepackage{breqn}

\usepackage{booktabs}

\usepackage{hyperref}

\numberwithin{equation}{section}

\def\la{\langle}
\def\ra{\rangle}

\def \tr {\mathop{\rm tr}\nolimits}

\def \Re {\mathop{\rm Re}\nolimits}

\def \e  {\mathop{\rm e}\nolimits}
\def\cG{  {\cal G}  }
\def\cO{  {\cal O}  }
\newcommand\lr[1]{{\left({#1}\right)}}
\newcommand \widebar [1] {\overline{#1}}

\newcommand \vev [1] {\langle{#1}\rangle}

\newcommand \ket [1] {|{#1}\rangle}

\newcommand\re[1]{(\ref{#1})}
\def \qqquad {\qquad\quad}
\def \qqqquad {\qquad\qquad}

\newcommand{\q}{\theta}
\newcommand{\bn}{\bar n} 
\newcommand{\pa}{\partial}
\newcommand{\om}{\omega}

\newcommand{\cF}{{\cal F}}
\newcommand{\cN}{{\cal N}}

\newcommand{\cE}{{\cal E}}
\newcommand{\cK}{{\cal K}}

\newcommand{\mD}{{\mathbb D}}
\newcommand{\ep}{\epsilon} 
\newcommand{\nt}{\notag\\} 
 
\newcommand{\p}[1]{(\ref{#1})}

\newcommand{\redq}{}

\def\be#1\ee{\begin{align}#1\end{align}}
\def\nn{{\nonumber}}

\def\numberbysection{\@addtoreset{equation}{section}
                     \def\theequation{\thesection.\arabic{equation}}}
 
\makeatletter
\def\timenow{\@tempcnta\time
  \@tempcntb\@tempcnta
  \divide\@tempcntb60
  \ifnum10>\@tempcntb0\fi\number\@tempcntb
  \multiply\@tempcntb60
  \advance\@tempcnta-\@tempcntb
  :\ifnum10>\@tempcnta0\fi\number\@tempcnta}
\makeatother

\makeatletter
\DeclareFontFamily{OMX}{MnSymbolE}{}
\DeclareSymbolFont{MnLargeSymbols}{OMX}{MnSymbolE}{m}{n}
\SetSymbolFont{MnLargeSymbols}{bold}{OMX}{MnSymbolE}{b}{n}
\DeclareFontShape{OMX}{MnSymbolE}{m}{n}{
    <-6>  MnSymbolE5
   <6-7>  MnSymbolE6
   <7-8>  MnSymbolE7
   <8-9>  MnSymbolE8
   <9-10> MnSymbolE9
  <10-12> MnSymbolE10
  <12->   MnSymbolE12
}{}
\DeclareFontShape{OMX}{MnSymbolE}{b}{n}{
    <-6>  MnSymbolE-Bold5
   <6-7>  MnSymbolE-Bold6
   <7-8>  MnSymbolE-Bold7
   <8-9>  MnSymbolE-Bold8
   <9-10> MnSymbolE-Bold9
  <10-12> MnSymbolE-Bold10
  <12->   MnSymbolE-Bold12
}{}

\let\llangle\@undefined
\let\rrangle\@undefined
\DeclareMathDelimiter{\llangle}{\mathopen}%
                     {MnLargeSymbols}{'164}{MnLargeSymbols}{'164}
\DeclareMathDelimiter{\rrangle}{\mathclose}%
                     {MnLargeSymbols}{'171}{MnLargeSymbols}{'171}
\makeatother

\def\lla{\llangle}
\def\rra{\rrangle}
\begin{document}

\title{Energy correlations in heavy states}

\author{Dmitry Chicherin$^{a}$, Gregory P. Korchemsky$^{b,c}$, Emery Sokatchev$^{a}$, and  Alexander Zhiboedov$^{d}$}

\affiliation{$^a$LAPTh-CNRS-USMB, 9 chemin de Bellevue, 74940, Annecy, France }
\affiliation{$^b$Institut de Physique Th\'eorique\footnote{Unit\'e Mixte de Recherche 3681 du CNRS}, Universit\'e Paris Saclay, CNRS, F-91191 Gif-sur-Yvette, France}
\affiliation{$^c$Institut des Hautes \'Etudes Scientifiques, 91440 Bures-sur-Yvette, France}
\affiliation{$^d$CERN, Theoretical Physics Department, CH-1211 Geneva 23, Switzerland}

\abstract{
We study energy correlations in states created by a heavy operator acting on the vacuum in a conformal field theory. We argue that the energy correlations in such states exhibit two characteristic regimes as functions of the angular separations between the calorimeters: power-like growth at small angles described by the light-ray OPE and slowly varying, or ``flat'', function at larger angles. The transition between the two regimes is controlled by the scaling dimension of the heavy operator and the dynamics of the theory. We analyze this phenomenon in detail in the planar ${\cal N}=4$ SYM theory both at weak and strong coupling. An analogous transition was previously observed in QCD in the measurement of the angular energy distribution of particles belonging to the same energetic jet. In that case it corresponds to the transition from the light-ray OPE, perturbative regime described in terms of correlations between quarks and gluons to the flat, non-perturbative regime described in terms of correlations between hadrons.  
}

\begin{flushleft}
 \hfill \parbox[c]{40mm}{CERN-TH-2023-109 \\ IPhT--T23/051 \\ LAPTH-034/23}
\end{flushleft}

\maketitle
 
\setcounter{page}{1}\setcounter{footnote}{0}
\pagestyle{plain}
\renewcommand{\thefootnote}{\arabic{footnote}}

\thispagestyle{empty}
 
\setcounter{page}{1}\setcounter{footnote}{0}

\section{Introduction and summary}

The energy correlations are among the best studied observables both experimentally and theoretically \cite{Neill:2022lqx}. They measure the flux of energy deposited in calorimeters located at different points on the celestial sphere and carry information about the dynamics of the underlying theory. A lot of activity has recently been devoted to studying the energy correlations in QCD and in the maximally supersymmetric $\mathcal N=4$ SYM theory. The latter serves as a very useful toy model that one can use to develop new techniques for computing these observables in QCD. Moreover, it allows us to understand certain properties of the energy correlations that cannot be explained within the conventional perturbative QCD approach. 

As an example, consider the energy-energy correlation (EEC) measuring the angular distribution of the energy of the particles that enter into two calorimeters separated by the relative angle  $0\le \theta\le \pi$ \cite{Basham:1978bw,Basham:1978zq}. At large total energy $Q$, the hadronic final states consist of collimated beams of energetic particles, or jets. In the energy correlations, jets manifest themselves as peaks located at small angles $\theta$, as well as at finite $\theta$ corresponding to the angular separation between the jets.

For a small angle $\theta$,  the EEC describes the correlation between  particles belonging to the same jet. The analysis of the experimental data shows \cite{Komiske:2022enw} that it behaves differently at small angles, depending on how $\theta$ compares with the non-perturbative parameter $\theta_0 \simeq \Lambda_{\rm QCD}/Q$ given by the ratio of the QCD hadronization scale and the total energy,\footnote{The {\bf flat} region corresponds to energy correlations which are slowly-varying functions of the angle. The {\bf OPE} region corresponds to energy correlations which exhibit a simple power-like behavior controlled by the operator product expansion (OPE) between the energy calorimeters.}  
\begin{align}
\label{eq:tworegimes}
{\bf Flat}:~~~\text{EEC} \sim \text{const}, \qqqquad {\bf OPE}:~~~\text{EEC}\sim 1/\theta^{2-\gamma}\,.
\end{align}

In QCD, the energy correlations are flat for $\theta \lesssim \theta_0 \ll 1$, and they exhibit a power-like growth for $\theta_0\lesssim  \theta \lesssim1$. The scaling behavior $\text{EEC}\sim 1/\theta^{2-\gamma}$ can be derived from the light-ray OPE  applied to the energy calorimeters \cite{Hofman:2008ar,Dixon:2019uzg,Korchemsky:2019nzm,Kologlu:2019mfz,Chang:2020qpj},  and the exponent $\gamma$ can be computed at weak coupling as a power series in the QCD coupling constant.\footnote{ In CFT $\gamma$ is the anomalous dimension of the spin-3,  signature plus (continued from even spins) operator on the stress-energy tensor Regge trajectory \cite{Hofman:2008ar,Caron-Huot:2017vep,Kravchuk:2018htv}.}

Such a change of behavior, from OPE to flat as $\theta$ decreases, corresponds to the transition from the perturbative regime described in terms of correlations between quarks and gluons to the non-perturbative regime described in terms of correlations between hadrons. 
To see it, notice that the energy scale that characterizes the branching of quarks and gluons at small angles is determined by their relative transverse momenta $Q\theta$. For small transverse momenta $Q\theta =O(\Lambda_{\rm QCD})$
the theory becomes strongly coupled. Conversely, the perturbative approach is justified for $\theta\gg {\Lambda_{\rm QCD}/ Q}$.
The behavior $\text{EEC}\sim  \theta^{0}$ can be reproduced if one thinks about the final state as consisting of a dense cloud of hadrons weakly interacting with one another. Describing the transition to this regime requires control over the non-perturbative QCD.

 A similar change of  behavior of the energy-energy correlation at small angles also takes place in the $\mathcal N=4$ SYM theory as the 't Hooft coupling varies, but the underlying mechanism is slightly different. This theory is conformal and it looks alike at short and large distances. At weak coupling, the EEC in $\mathcal N=4$ SYM has the same power-like behavior at small angles as in QCD. This behavior changes as one goes to the limit of strong 't Hooft coupling constant $\lambda =g_{\rm YM}^2N_c  \gg 1$. Increasing the value of the coupling constant, one enhances the production of particles (gauge, gaugino and scalars) in the final state. At strong coupling, the final state in $\mathcal N=4$ SYM consists of an infinite number of soft particles whose energy is distributed homogeneously on the celestial sphere. As a consequence, for $\lambda\to\infty$  the energy-energy correlation does not depend on the angle, $\text{EEC}(\theta)=1+O(1/\lambda)$.  
As in the case of QCD, the change of behavior of the EEC as a function of the angle is associated with the presence of a large number of particles in the final state. To describe the transition in detail, one needs to know the energy correlation in $\mathcal N=4$ SYM for an arbitrary  't Hooft coupling,  which is out of reach at present. 
  
{In this paper, we point out that there exists another physical mechanism of transitioning  between the two characteristic regimes \eqref{eq:tworegimes}. For simplicity we restrict our attention to conformal field theory (CFT), but the basic mechanism should be applicable to theories with dimensionful scales, such  as generic four-dimensional gauge theories including QCD. Because the transition is driven by the large number of particles produced in the final state, we can create such a final state by exciting the vacuum with a ``heavy'' operator $O_H(x)$  carrying a large scaling dimension $\Delta_H \gg 1$. Physical intuition suggests that for $\Delta_H\to\infty $ the state $\ket{H(q)}= \int d^4 x \,e^{i q x} O_H(x) | 0 \rangle$ should contain  arbitrarily many soft particles which are largely uncorrelated. As a consequence, for $\Delta_H\to\infty $ the energy correlations are expected to be angle-independent, up to corrections suppressed by a power of $1/\Delta_H$.

In order to formulate this property, it is convenient to introduce the energy flow operator $\mathcal E(n)$ which measures the energy flux  in the direction specified by a null vector $n^\mu=(1,\vec n)$ with $\vec n^2=1$ \cite{Sveshnikov:1995vi,Korchemsky:1997sy,Korchemsky:1999kt}. In the rest frame of the source, for $q^\mu=(Q,\vec 0)$,  its expectation value in the state $\ket{H(q)}$  is determined by the total momentum, $\vev{\mathcal E(n)}_H=Q/(4\pi)$. It does not depend on the choice of the source operator and yields the total energy after integration over the celestial sphere, $\int d^2 \vec n\, \vev{\mathcal E(n)}_H=Q$.

The multi-point energy correlations
are given by the expectation value of the product of several flow operators, in the state $\ket{H(q)}$ created by a local operator with scaling dimension $\Delta_H$,  
\begin{align}
\label{eq:1.2}
   \underbrace{\text{EE$\dots$E}}_k \text{C} = \vev{ { \mathcal E}(n_1)\dots { \mathcal E}(n_k) }_{ {H}} =\lr{Q\over 4\pi}^k \vev{ \widehat{ \mathcal E}(n_1)\dots \widehat{ \mathcal E}(n_k) }_{ {H}}\,,
\end{align}
where $\widehat {\mathcal E}(n) = \mathcal E(n)/\vev{\mathcal E(n)}_H$ is the normalized energy flow operator with $\vev{\widehat {\mathcal E}(n)}_H=1$.
To study their behavior in the limit $\Delta_H \to\infty$, we find it convenient 
to apply the cumulant expansion 
\be
\label{eq:1.3}
\vev{ \widehat{ \mathcal E}(n_1)\dots \widehat{ \mathcal E}(n_k) }_{ {H}} = 1 &+ \sum_{i<j} \vev{  \widehat{ \mathcal E}(n_i)  \widehat{ \mathcal E}(n_j)}_{c} + \sum_{i<j<m} \vev{  \widehat{ \mathcal E}(n_i)  \widehat{ \mathcal E}(n_j) \widehat{ \mathcal E}(n_m)}_c + \dots  \ ,
\ee
where $\vev{ \dots }_c$ denotes the connected correlation.
Here the terms $\vev{\widehat{ \mathcal E}(n_i)  \widehat{ \mathcal E}(n_j)}_{c}$ depend on the single relative angle between the $i$th and $j$th detectors; $\vev{  \widehat{ \mathcal E}(n_i)  \widehat{ \mathcal E}(n_j) \widehat{ \mathcal E}(n_m)}_c$ depend on the three relative angles between the $i$th, $j$th, and $k$th detectors, etc.  Starting from four detectors, for $k\geq 4$,  the expansion \eqref{eq:1.3} involves non-linear terms,  the simplest being $\vev{  \widehat{ \mathcal E}(n_i)  \widehat{ \mathcal E}(n_j)}_c \vev{\widehat{ \mathcal E}(n_m)\widehat{ \mathcal E}(n_q)}_c$, see e.g. \cite{LeBellac:1991cq} and Appendix \ref{sA}. 
 
We conjecture that when the source becomes heavy, $\Delta_H \gg 1$, each subsequent term in \eqref{eq:1.3} is further suppressed compared to the previous one, namely
\be
\vev{  \widehat{ \mathcal E}(n_1) \dots \widehat{ \mathcal E}(n_k)}_c &=O\bigg({1\over\Delta_H^{h_k}}\bigg)\,,\qquad
0< h_2< h_3< \dots \ ,
\label{eq:1.4}
\ee
The values $h_k$ and the precise form of $\vev{  \widehat{ \mathcal E}(n_1) \dots \widehat{ \mathcal E}(n_k)}_c$ depend on the theory and the physical state in question. Notice that in \eqref{eq:1.4},  when taking $\Delta_H \to \infty$, we keep the angles between the detectors, as well as the number of detectors, fixed.

The same property was observed before in strongly coupled gauge theories with gravity duals \cite{Hofman:2008ar}. As mentioned above, in this case the multi-particle final state is not generated by considering a heavy source, but by the strongly coupled dynamics of the theory. The role of $\Delta_H$ is played by the 't Hooft coupling $\lambda$, and $h_k = {k \over 2}$. In contrast, we would like to emphasize that the relation  
\re{eq:1.4} holds in the limit of large $\Delta_H$ for an arbitrary coupling (including the free theory!).
 
In this paper we study the energy correlations  \re{eq:1.2}  in planar ${\cal N}=4$ SYM. To define the heavy state $\ket{H(q)}$, we choose the operator $ O_H(x)$ to be a half-BPS scalar operator of the form $ O_H(x)=\tr[\phi^K(x)]$.  Its scaling dimension is protected from quantum corrections, $\Delta_H=K$, and the heavy state limit
corresponds to $K\to\infty$. Another advantage of this choice is that the two-point correlation $\vev{  \widehat{ \mathcal E}(n_1)  \widehat{ \mathcal E}(n_2)}_{c}$, defining the leading angular-dependent contribution in \re{eq:1.3}, can be computed explicitly both at weak and strong coupling for arbitrary $K \geq 2$. This computation is done most efficiently by the method based on correlation functions and Mellin transforms, developed in \cite{Hofman:2008ar,Belitsky:2013xxa,Belitsky:2013bja,Belitsky:2013ofa,Belitsky:2014zha,Henn:2019gkr}.

Using the explicit expression for $\vev{  \widehat{ \mathcal E}(n_1)  \widehat{ \mathcal E}(n_2)}_{c}$, we can verify that its dependence on the relative angle $\cos \theta = (\vec n_1 \,\vec n_2)$ is a function of  the scaling dimension of the source $K$. We find that for $K=2$, the energy-energy correlation  at weak coupling in ${\cal N}=4$ SYM has a shape  very similar to that in perturbative QCD. Namely, it is peaked around the end points, $\theta=0$ and $\theta=\pi$, and is flat in between. This shape corresponds to a final state containing two jets, one per each scalar field in  the definition of the source operator $\tr[\phi^2(x)]$. As $K$ increases, we observe that the peak at $\theta=\pi$ disappears and the function flattens out for $0<\theta<\pi$. Moreover, $\vev{  \widehat{ \mathcal E}(n_1)\widehat{ \mathcal E}(n_2)}_{c}$ vanishes  as $1/K$ in the limit $K\to\infty$. To the lowest order in the coupling we find, up to corrections suppressed by powers of $\lambda$ and $1/K$,
\begin{align}\label{EE-w}
\vev{  \widehat{ \mathcal E}(n_1)\widehat{ \mathcal E}(n_2)}_{c}= {\lambda\over 8\pi^2 K}\left[ \frac{3}{z_+}+ 2 \text{Li}_2(1-z)-6 \log z-2\zeta_{2}-\frac{13}{2} +\frac52\delta(z)\right] + \dots\,,
\end{align}
where $z=(1-(\vec n_1\vec n_2))/2=\sin^2(\theta/2)$ in the rest frame of the source and $1/z_+$ is the plus-distribution. 
The result \re{EE-w} is in agreement with our expectation \re{eq:1.4} for $\Delta_H=K$ and $h_k=k-1$.

The relation \re{EE-w} holds at weak coupling for $K\gg 1$. The situation changes at strong coupling for 
$\lambda \gg K\gg 1$. In this limit, we find that the energy-energy correlation does not depend on $K$ and it vanishes as $1/\lambda$,
\begin{align}\label{EE-s}
    \vev{  \widehat{ \mathcal E}(n_1)\widehat{ \mathcal E}(n_2)}_{c}= {4 \pi^2 \over \lambda} \left(1 - 6 z(1-z) \right) \,.
\end{align}

The relations \re{EE-w} and \re{EE-s} illustrate two different mechanisms of producing  {slowly-varying}  energy distributions. At weak coupling, 
the leading correction is controlled by the scaling dimension of the source. At strong coupling, it comes from the evolution of the state and is controlled by the coupling constant. 

 Below we show  that at large $K$ and finite coupling constant the energy-energy correlation in ${\cal N}=4$ SYM has the following behaviour at small angles $z=\sin^2(\theta/2) \ll 1$
\begin{align}\label{pred}
 \vev{  \widehat{ \mathcal E}(n_1)\widehat{ \mathcal E}(n_2)}_{c} \sim {C\over K} \theta^{-2+\gamma}\,,
\end{align}
where $C$ and $\gamma$ depend on the 't Hooft coupling.  
In the case of QCD we expect that the EEC should have the same behavior in the OPE region \re{eq:tworegimes}, for $\theta>\theta_0$. An important difference compared to $\mathcal N=4$ SYM is that the ``kinematical'' large parameter $K$ in \re{pred} is replaced in QCD by a dynamical one defined by the multiplicity of the produced hadrons within a jet.

Above we focused our discussion on energy correlations in QCD and its conformal cousin ${\cal N}=4$ SYM. It is interesting to ask what happens for event shapes, or matrix elements of light-ray operators, in a general interacting CFT when the source becomes heavy.\footnote{In an upcoming work \cite{RattazziToAppear}, event shapes in the large charge limit of CFTs with global $U(1)$ symmetry that admit a superfluid description are studied. The results of \cite{RattazziToAppear} are compatible with the proposal of the present paper.} We present some evidence that the clustering structure \re{eq:1.4} is valid for energy correlations in any CFT. Let us briefly explain the reason for that. According to the state-operator correspondence, a local operator $O_H(x)$ with scaling dimension $\Delta_H$ can be associated with an energy eigenstate
$E_{\text{cyl}}={\Delta_H / R}$ in a CFT defined on a cylinder $\mathbb{R} \times S^{d-1}$, where $R$ is the radius of the sphere. A heavy operator with large scaling dimension $\Delta_H \gg 1$ corresponds to a highly excited energy eigenstate. In an interacting theory it is expected to look thermal when simple enough observables are considered \cite{Lashkari:2016vgj} (the statement known as the Eigenstate Thermalization Hypothesis). The clustering structure of energy correlations \re{eq:1.4} in a heavy state is then related to the clustering of correlation functions of stress-energy tensors in the thermal state as the separation between the operators becomes large \cite{El-Showk:2011yvt}.

The paper is organized as follows.
In Section~\ref{sect2} we compute the multi-point energy correlations $\vev{\mathcal E(n_1) \dots \mathcal E(n_k)}_H$ in a free theory and show that they satisfy 
\re{eq:1.4} in the limit of large scaling dimension of the source.  
In Section~\ref{sect3} we consider the energy-energy correlation $\vev{\mathcal E(n_1)\mathcal E(n_2)}_H$ in planar $\mathcal N=4$ SYM at weak coupling. We choose the source operator to be the simplest half-BPS operator built out of $K$ scalar fields and examine the dependence of the energy correlation on $K$ at the leading order  in the coupling constant (Born level). In Section~\ref{sect:weak} we compute the one- and two-loop perturbative corrections to the energy-energy correlation for arbitrary $K$ and show that they verify the relation \re{eq:1.4} at large $K$. In Section~\ref{s5} we present an approach that allows us to compute the $O(1/K)$ correction to \re{eq:1.4} directly, without going through the details at finite $K$.  In Section~\ref{s6} we obtain the contact terms  necessary to describe the singular behavior of the energy correlations at the end points, for finite $K$ and for $K \to \infty$. In Section~\ref{sect6} we compute the same observable at strong coupling including the first stringy correction. In Section~\ref{sect8} we present arguments in favor of
\re{eq:1.4} in a generic  interacting CFT.  The paper contains several appendices and ancillary files. 
 
\section{Energy correlations in the free {scalar} theory}\label{sect2}

The energy correlations for a heavy source are expected to have the general form \re{eq:1.4}.  
In this section, we show that the relations \re{eq:1.4} are indeed satisfied in the free scalar theory. 

We consider a free massless scalar field $\phi(x)$ and choose the source operator to be $O_K(x) = \phi^K(x)$. It creates the state
$\ket{H(q)}= \int d^4 x \,e^{i q x} O_K(x) | 0 \rangle$ containing $K$ massless scalar particles with the total momentum $q^\mu$ (with $q^0>0$ and $q^2>0$). 

\begin{figure}[h!]
 \centerline{
{\parbox[c]{85mm}{ \includegraphics[width = 85mm]{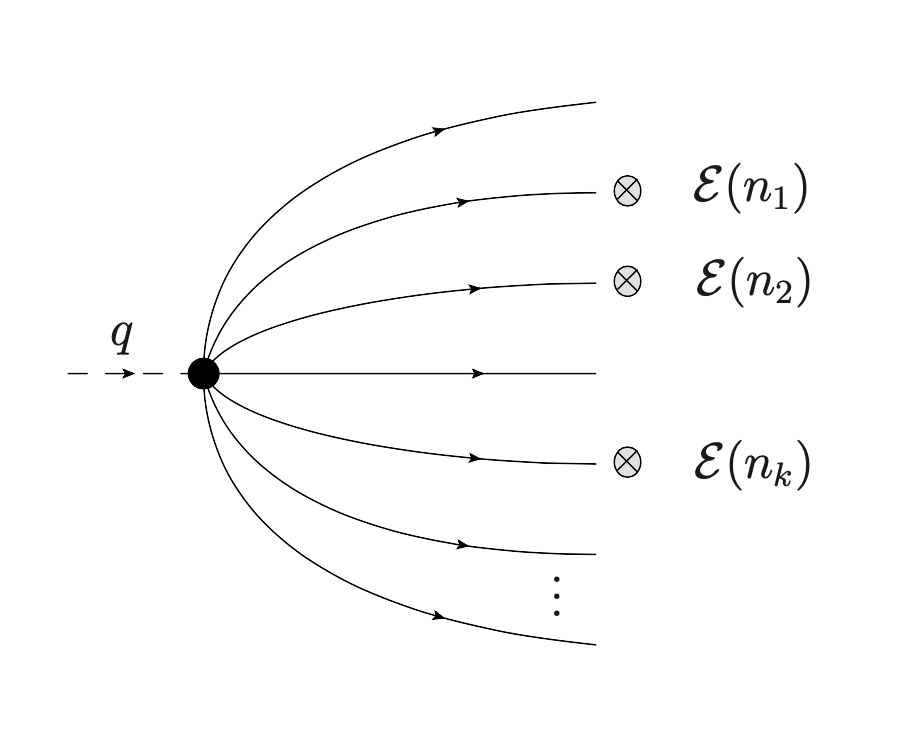}}}
}
\caption{Multi-point energy correlation $\vev{\mathcal E(n_1) \dots \mathcal E(n_k)} $ in the final state created by a heavy source.}\label{fig:FS}
\end{figure}

The differential cross-section describing the probability to find $K$ particles in a final state with on-shell momenta $p_i$ (with $p_i^2=0$) is given by
\begin{align}\label{dsigma}
d\sigma_K =    (2\pi)^4 \delta^{(4)} (q-\sum_{i=1}^K p_i)\prod_{n=1}^K d {\rm LIPS}(p_n)\,,
\end{align}
where %$q^\mu$ is the total momentum (with $q^0>0$ and $q^2>0$) and
 $d {\rm LIPS}(p)$ is the Lorentz invariant phase space measure of a particle with  momentum $p$,
\begin{align}
d {\rm LIPS}(p) =  {d^4 p\over (2\pi)^4} 2\pi \delta_+(p^2) \,,
\end{align}
with $\delta_+(p^2)=\delta(p^2)\theta(p^0)$.
The total cross-section is given by
\begin{align}\label{sigma}
\sigma_K(q) = \int d\sigma_K = \int d^4 x\, e^{iqx} \prod_{n=1}^K \int d {\rm LIPS}(p_n)\e^{-ip_n x}\,.
\end{align}
In this representation, the phase space integral factorizes into a product  of two-point Wightman functions
\begin{align}\label{prop}
D(x) =\vev{\phi(x)\bar \phi(0)}=  \int d {\rm LIPS}(p)\e^{-ip x} = {1\over 4\pi^2 (-x^2+ i0 x^0)}\,,  %\qquad c_D={1 \over 4\pi^2}\,. 
\end{align}
 where the `$+i0x^0$' prescription ensures that the Fourier transform is different from zero for $p^0>0$.    The calculation shows that
\begin{align}\label{sigmaK}
\sigma_K(q)=  \int  d^4 x\, \e^{iqx} [D(x)]^K = \theta(q^0)\theta(q^2){2\pi^3 (q^2/4)^{K-2}\over (2\pi)^{2K}\Gamma(K) \Gamma(K-1)} \,.
\end{align}
Notice that $\sigma_K(q)$ grows as a power of $q^2$ with the exponent being a linear function of the  weight (or scaling dimension) $K$ of the operator $O_K$.

Let us now examine the $k-$point energy correlation in the $K-$particle final state described by the differential distribution \re{dsigma}. It is given by 
\begin{align}\label{manyE}
\vev{\mathcal E(n_1) \dots \mathcal E(n_k)} = {K!/(K-k)!\over \sigma_K(q)}\int d\sigma_K\, w_{n_1}(p_1) \dots w_{n_k}(p_k)\,,
\end{align}
where the weight factor $w_n(p)$ selects the particle in the final state  moving along the null direction $n^\mu=(1,\vec n)$ (with $\vec n^2=1$), 
\begin{align}\label{w-E}
w_n(p) = p^0\delta^{(2)}\lr{{\vec p\over p^0}-\vec n}\,.
\end{align}
The combinatorial factor  $K!/(K-k)!$ in the numerator of \re{manyE} is due to the Bose symmetry of the scalar particles. It counts the total number of events in which $k$ out of $K$  particles enter the calorimeters. The diagrammatic representation of the relation \re{manyE} is shown in Figure~\ref{fig:FS}.

Substituting \re{dsigma} into \re{manyE} we can express $\vev{\mathcal E(n_1) \dots \mathcal E(n_k)}$ as an integral over  the phase space of the $K$ particles. The integral over the undetected $(K-k)$ particles gives rise to the total cross-section $\sigma_{K-k} (\hat q  )$ where $\hat q^\mu=q^\mu-\sum_{i=1}^k p_i^\mu$ is the total momentum of these particles. Then, $\vev{\mathcal E(n_1) \dots \mathcal E(n_k)}$ is given by 
the integral over the phase space of the detected $k$ particles, 
\begin{align}
\vev{\mathcal E(n_1) \dots \mathcal E(n_k)} ={K!/(K-k)!\over \sigma_K(q)} \int d\sigma_k\, w_{n_1}(p_1) \dots w_{n_k}(p_k)\, \sigma_{K-k} (\hat q  )\,.
\end{align} 
%where $\hat q^\mu=q^\mu-\sum_{i=1}^k p_i^\mu$ is the total momentum of $(K-k)$ undetected particles. 
Using \re{w-E}, this expression   can be written as a $k-$fold integral over the energies $w_i=p_i^0$ of the  detected particles,  
\begin{align}\label{int1}
\vev{\mathcal E(n_1) \dots \mathcal E(n_k)} ={c_K \over c_{K-k} }\int_0^\infty \prod_{i=1}^k {d\omega_i\,\omega_i^2\over 2q^2 (2\pi)^3}\theta(\hat q^0)\theta(\hat q^2) 
(\hat q^2/q^2)^{K-k-2} \,,
\end{align}
where $\hat q^\mu=q^\mu-\sum_{i=1}^k \omega_i n_i^\mu$ and 
$c_K =  (4\pi)^{2K} K! (K-1)! (K-2)!\, $. 

The integral \re{int1} involves the ratio of the invariant mass of the undetected particles and the total energy,
\begin{align}
{\hat q^2\over q^2} = 1- \sum_{i=1}^k {2(qn_i)\over q^2} \omega_i + 2\sum_{i<j} {(n_i n_j)\over q^2}\omega_i\omega_j\,.
\end{align}
Notice that $\hat q^2/q^2<1$ and, therefore, for large $K$ the dominant contribution to the integral \re{int1} comes from the integration over the region $ \hat q^2/q^2=1+O(1/K)$, or equivalently $\omega_i= O(1/K)$. Changing  variables as
\begin{align}
\omega_i =  {q^2 \over 2(K-k-2) (qn_i)} \varepsilon_i \,,
\end{align}
we find in the limit $K\to\infty$ with $k$ held fixed
\begin{align}\label{qq}
(\hat q^2/q^2)^{K-k-2}  {}&=e^{-\sum_i \varepsilon_i}\left[1-{1\over K}\lr{\sum_{i<j} \varepsilon_i\varepsilon_j (1-z_{ij})+\frac12 \sum_i \varepsilon_i^2}  +O(1/K^2) \right] ,
\end{align}
where $z_{ij}$  are dimensionless variables
\begin{align}\label{z-var}
z_{ij} = {q^2(n_in_j)\over 2(qn_i)(qn_j)} \,.
\end{align} 
In the rest frame of the source,  for $q^\mu=(Q,\vec 0)$ and $n_i=(1, \vec n_i)$, these variables are related to the relative angles between the detectors on the sphere, $z_{ij} = (1-\cos\theta_{ij})/2$.

Combining together \re{int1} and \re{qq} we find that the energy correlations in the large $K$ limit take a remarkably simple form, 
\begin{align}\notag\label{hom}
{}&  \vev{\mathcal E(n_1) \dots \mathcal E(n_k)}  = \lr{\prod_{m=1}^k {(q^2)^2\over 2(qn_m)^3}} 
\int_0^\infty \prod_{i=1}^k  {d\varepsilon_i\,\varepsilon_i^2 \over 4\pi}\,e^{- \varepsilon_i}
\\
{}&\qqquad \times  \left[1-{1\over K}\lr{\sum_{i<j} \varepsilon_i\varepsilon_j (1-z_{ij})+\frac12 \sum_i \varepsilon_i^2-\frac32{k(k+3)} }  +O(1/K^2) \right] .
\end{align}
The integral in this relation describes an ensemble of $k$ noninteracting particles 
%on the celestial sphere 
whose energies   $\omega=\varepsilon Q/(2K)$  are distributed according to the law $d P(\varepsilon)=d\varepsilon \, \varepsilon^2 e^{-\varepsilon}$. The first term in the brackets in \re{hom} is independent of the angular variables $z_{ij}$ defined in \re{z-var}. It describes a homogeneous distribution of the energy on the celestial sphere. The angular dependence comes from the second term, which is suppressed by a factor of $1/K$ and involves a quadratic polynomial in the energy variables $\varepsilon_i$. To higher orders in $1/K$, the coefficient of $1/K^p$ is given by a polynomial in $\varepsilon_i$ of degree $2p$. As we show below, it contributes to the connected part of the $p-$point correlations in \re{eq:1.3}.

The calculation of the integral in \re{hom} yields 
\begin{align}\label{E-cor} 
\vev{\mathcal E(n_1) \dots \mathcal E(n_k)} {}& = \lr{\prod_{i=1}^k {(q^2)^2\over 4\pi(qn_i)^3}}
\left[ 1+{1\over K}\lr{ \sum_{1\le i<j\le k} 9z_{ij} -3k(k-1)} +O(1/K^2) \right].
\end{align}
%In the rest frame of the source 
 Switching to the normalized operators $\widehat{\cE}(n_i)$ on the left-hand side of \p{E-cor}, the term of order $1/K$ on the right-hand side is identified with the sum of the two-point connected correlations in \p{eq:1.3}, 
\begin{align}\label{eq:2.16} 
\vev{\widehat\cE(n_i)  \widehat\cE(n_j)}_c  &= {1 \over K} \left( 9 z_{ij} - 6 \right) + O(1/K^2)\, . 
\end{align}
{It is straightforward to compute the subleading terms in \re{E-cor}. Bringing them to the form \re{eq:1.3}, we can determine the higher-point connected correlations, e.g. }
\begin{align}\label{eq:2.16three}
\vev{\widehat\cE(n_i)  \widehat\cE(n_j) \widehat\cE(n_m)}_c  = {1\over K^2} \Big[{}& 108(z_{i j} z_{i m} + z_{i j} z_{j m} + z_{i m} z_{j m}) \nn \\ {}& - 126(z_{i j} + z_{i m} + z_{j m}) + 114 \Big]  
 + O(1/K^3)\,.
\end{align}
The relations \re{eq:2.16} and \re{eq:2.16three} take the expected form  \re{eq:1.4} with $\Delta_H=K$ and $h_k=k-1$.
 
The relations \re{E-cor}, \re{eq:2.16} and \re{eq:2.16three} are valid for $n_i\neq n_j$ and they do not take into account the contribution of contact terms localized at  $n_i=n_j$. The presence of such terms can be detected as follows.
By definition, $\vev{\mathcal E(n_1) \dots \mathcal E(n_k)}$ measures the energy flux  on the celestial sphere in the directions specified by the unit vectors $\vec n_1,\dots,\vec n_k$. 
The integral of $\mathcal E(n) n^\mu$ over the unit sphere $\vec n^2=1$ yields the total momentum of the final state. As a consequence, the energy correlation has to satisfy the sum rule 
\begin{align}\label{sr0}
\int d^2 n_k \, n_k^\mu\, \vev{\mathcal E(n_1) \dots \mathcal E(n_k)} = q^\mu \vev{\mathcal E(n_1) \dots \mathcal E(n_{k-1})}\,.
\end{align}
One can check that the leading $O(K^0)$ term in \re{E-cor} verifies this relation
but this is not the case for the $O(1/K)$ term. The reason for this is that the integral in \re{sr0} receives contributions from  contact terms localized at coincident directions  $n_i$. 

A contact term proportional to the product of $L$
delta functions on the sphere,  
$\delta(n_1,n_2)\dots \delta(n_1,n_{L+1})$, comes with the combinatorial factor $\lr{K\atop k} /\lr{K\atop k-L}\sim 1/K^L$.
To the leading order in $1/K$,  only the contact terms with $L=1$ contribute.  They are proportional to $ \sum_{1\le i<j\le k}\delta^{(2)}(\vec n_i,\vec n_j)$. The sum rule \re{sr0} allows us to fix their contribution to the energy correlation \re{E-cor}. 

For instance, the contact term for the two-point connected correlation \re{eq:2.16} is given by ${3 \over 2K} \delta(z_{ij})$.
Adding it to \eqref{eq:2.16}, one can verify that the two-point correlation $\vev{\mathcal E(n_1)\mathcal E(n_2)}$ satisfies the sum rule \re{sr0}.
 A simple corollary of the sum rule \re{sr0} for the three-point energy correlation is that
\begin{align}\label{sreps}
\int d^2 n_m \, \epsilon_{\mu \nu \rho \sigma} q^{\mu} n_i^{\nu} n_j^{\rho} n_m^\sigma \, \vev{\mathcal E(n_i) \mathcal E(n_j) \mathcal E(n_m)} = 0  \ .
\end{align}
In distinction to \re{sr0}, it is not sensitive to  contact terms. We have checked that $\vev{\widehat\cE(n_i)  \widehat\cE(n_j) \widehat\cE(n_m)}_c$  in \eqref{eq:2.16three} satisfies \p{sreps}.

We conclude this section by reiterating our main conjecture \p{eq:1.4}: The connected correlations in the free theory have the expected heavy weight behavior with $\Delta_H=K$ and $h_k=k-1$ (with $k=2,3,\dots$).
 
\section{Energy correlations in $\mathcal N=4$ SYM}\label{sect3}

In the previous section, we demonstrated that the energy correlation in a final state consisting of a large number $K$ of free scalar particles is given by simple expressions like \eqref{eq:2.16} and \eqref{eq:2.16three}. The question arises, how does the interaction between the particles in the final state affect this result?

\subsection{Energy-energy correlation}

In order to address this question, we consider the energy correlations in a particular four-dimensional gauge theory -- the maximally supersymmetric $\cN=4$ Yang-Mills theory with  gauge group  $SU(N_c)$. For the sake of simplicity we shall concentrate on the two-point correlation $\vev{\mathcal E(n_1)\mathcal E(n_2)}$. To create a  final state with a large number of particles produced, we excite the vacuum by acting on it with a half-BPS operator of the form
\begin{align}\label{e11}
O_K(x) = \tr[\phi^K(x)] \,, \qqquad \phi(x)=\sum_{I=1}^6 Y^I X^I\,.
\end{align}
It is built of six real scalar fields $X^I$  with $Y^I$ being an auxiliary null complex vector of $SO(6)$ defining the orientation of $O_K(x)$ in the isotopic $R-$space. This operator has the $R-$charge of the $[0,K,0]$ representation of $SO(6) \sim SU(4)$, as well as conformal weight (or scaling dimension) $K$, the latter being protected from quantum corrections by  $\cN=4$ superconformal symmetry.   It creates $K$ scalar particles  out of the vacuum. Going to the limit $K\to \infty$, we encounter the final state discussed in the previous section.

The two-point    correlation function of the operators \re{e11} is protected from quantum corrections and is given by the product of $K$ free scalar propagators \re{prop} multiplied by an $SU(N_c)$ color factor. In the planar limit we have
\begin{align}\label{e13}
\vev{O_K(x)\bar O_K(0)} = {(Y\bar Y)^K\over (4\pi^2)^K} { K N_c^{K} \over (-x^2+i0x^0)^K } \,,
\end{align}
where $(Y\bar Y)=\sum_{I=1}^6 Y^I  (Y^I)^*$. Notice that the operators are not time ordered. 
To simplify the formulae, we put $(Y\bar Y)=1$ in what follows.
Then, the total cross-section for producing an arbitrary number of  particles in the final state is given by   (see (A.4) in \cite{Belitsky:2013xxa}) 
\begin{align}\label{e14}
& \sigma_{\rm tot}(q) = \int d^4 x\, \e^{iqx} \vev{O_K(x)\bar O_K(0)} = K N_c^{K} \sigma_K(q) \,,  
\end{align}
where $\sigma_K(q)$ is defined in \re{sigmaK}. Like \re{e13}, it is protected from quantum corrections. 

The relation \re{e14} expresses the total cross-section as the Fourier integral of the two-point Wightman correlation function \re{e13}. It is equivalent to \re{sigma} after  inserting the sum over the final states between the operators.
In a similar manner, the energy correlations \re{manyE} also admit a representation in terms of correlation functions involving two scalar operators \re{e13}, as well as the energy flow operators $\cE(n_i)$ (for the definition see Appendix \ref{sA}). The latter play the role of calorimeters detecting the particles in the final state. 

For instance, the two-point energy correlation is given by
\begin{align} \label{EE-corr}
\vev{\cE (n_1)\cE (n_2) } _K  = \sigma^{-1}_{\rm tot}(q)  \int d^4 x\, \e^{iqx} \vev{O_K(x)\cE (n_1)\cE (n_2)\bar O_K(0)} \,.
\end{align}
Lorentz symmetry allows us to fix the form of the correlation,   
\begin{align}\label{e1.6}
\vev{\cE (n_1)\cE (n_2) } _K  
= { (q^2)^4\over  2(qn_1)^3 (qn_2)^3}\, { \cF_K(z) \over (4\pi)^2}  \,, 
\end{align}
 up to an arbitrary function $\cF_K(z)$ of the Lorentz covariant angular separation $z\equiv z_{12}$ between the detectors (recall \p{z-var}),
\begin{align}\label{e18}
z={q^2 (n_1n_2)\over 2(qn_1)(qn_2)}=\frac12(1-(\vec n_1 \vec n_2))=\frac12(1-\cos\theta_{12})\,.
\end{align}
Here $\theta_{12}$ is the angle between the unit vectors $\vec n_1$ and $\vec n_2$ in the rest frame of the source, $q^\mu=(Q,\vec 0)$.
The kinematical  factor on the right-hand side of \re{e1.6} has Lorentz weight $(-3)$ under the independent rescaling of the detector vectors $n_i^\mu$, matching the corresponding weight of 
the energy operators $\cE(n_i)$ on the left-hand side (see \p{E-new}--\p{x}). The factor $(q^2)^4$ in the numerator corresponds to the scaling dimension $(+1)$ of $\cE(n_i)$.

For $K=2$ the energy correlation \re{e1.6} has been studied both at weak and strong coupling   \cite{Hofman:2008ar,Belitsky:2013bja,Belitsky:2013ofa,Belitsky:2013xxa,Belitsky:2014zha,Goncalves:2014ffa,Korchemsky:2015ssa,Henn:2019gkr,Korchemsky:2021okt,Korchemsky:2021htm}. Our goal in this paper is to extend these results to $K\ge 3$ and to understand the properties of the energy correlation in the limit of large $K$.
We show below that at large $K$ the function $\cF_K(z)$ satisfies
\begin{align}
\cF_K(z) =  2 + O(1/K)\,.
\end{align}
Substituting this relation in \re{e1.6} we reproduce  \re{eq:1.4} for $\Delta_H=K$ and $h_2=1$.

In distinction with the total cross-section, the function $\cF_K(z)$ is not protected in $\mathcal N=4$ SYM and depends on the coupling constant $g^2_{\rm YM}$, as well as on the rank of the gauge group $N_c$. In the planar limit, for $N_c\to\infty$ with the 't Hooft coupling constant $\lambda=g^2_{\rm YM} N_c$ held fixed, it admits a weak coupling expansion, 
\begin{align}\label{e1.8}
\cF_K(z)  = \cF^{(0)}_K(z)  + \sum_{\ell=1}^\infty \lr{\lambda\over 4\pi^2}^\ell \,  \cF^{(\ell)}_K(z) \,,
\end{align}
where  $\cF^{(0)}_K(z)$ refers to the Born approximation and $\cF^{(\ell)}_K(z)$ with $\ell\geq1$ denotes the $\ell-$loop perturbative correction. 

In the Born approximation, the energy correlation \re{e1.6} is given by \re{int1} evaluated for $k=2$. 
Matching \re{e1.6} with \re{E-cor} we find that in the limit of large $K$ the scaling function $\cF^{(0)}_K(z)$ is given by 
\begin{align}\label{F0}
\cF^{(0)}_K(z) = 2 +{1\over K} \lr{18 z - 12 +3\delta(z) }+O(1/K^2)\,.
\end{align}
As was explained in the previous section, this relation describes an ensemble of $K$ non-interacting particles whose energies 
are distributed according to the law \re{hom}.  Turning on the interaction between these particles, one generates additional correlations of their energies. 
They are described by the functions $\cF^{(\ell)}_K(z)$ in \re{e1.8} which we compute below for $\ell=1,2$. We will show that at large $K$ the loop corrections scale as $\cF^{(\ell)}_K(z)\sim 1/K$ for $\ell \ge 1$.

\subsection{Sum rules} \label{s1.4}

The function $\mathcal F(z)$ satisfies nontrivial conditions that follow from the sum rules \re{sr0}. 

We recall that 
the integral of the energy flow operator over the celestial sphere yields the total energy-momentum of the final state,
$\int d^2 n \,n^\mu \vev{\mathcal E(n)}_K=q^\mu$. Combined with \re{sr0},  this yields the following relations in the rest frame of the source,
\begin{align} \label{e1.16}\notag
& \int d^2 n_1 \int d^2 n_2 \vev{\cE (n_1)\cE (n_2) } _K=Q^2\,,
\\
& \int d^2 n_1 \int d^2 n_2 \, (1-(\vec n_1 \vec n_2)) \vev{\cE (n_1)\cE (n_2) } _K=Q^2\,.
\end{align}
Substituting the general expression \re{e1.6} of $\vev{\cE (n_1)\cE (n_2) } _K$  and using \p{e18}, we find that $\cF_K(z)$ has to satisfy the sum rules \cite{Kologlu:2019mfz,Korchemsky:2019nzm,Dixon:2019uzg}
\begin{align}\label{sr}
\int_0^1dz\, z\, \cF_K(z) =\int_0^1dz\, (1-z)\, \cF_K(z) =1 \,.
\end{align}
These relations  hold for an arbitrary coupling constant $\lambda$. 

Replacing $\cF_K(z)$ with its perturbative expansion \p{e1.8} and matching the coefficients of the powers of $\lambda$ on both sides of \re{sr} we find that the Born approximation  $\cF^{(0)}_K(z)$ alone produces the required right-hand side of \re{sr}, 
\begin{align}\label{e1.11}
\int_0^1dz\, z \cF^{(0)}_K(z) =\int_0^1dz\, (1-z)\, \cF^{(0)}_K(z) = 1 \,.
\end{align}
The sum rules for the loop corrections $ \cF^{(\ell)}_K(z) $ are  
\begin{align}\label{e1.12}
\int_0^1dz\, z\cF^{(\ell)}_K(z) =  \int_0^1dz\,(1- z)\, \cF^{(\ell)}_K(z) =0\,, \qquad   \ell\geq1 \,.   
\end{align}
It is straightforward to verify that \re{F0} satisfies the sum rules \re{e1.11}. Notice that the function accompanying $1/K$ in \re{F0} gives zero contribution to the sum rules \re{e1.11}.

\subsection{Born contribution}\label{s2}

In this subsection, we compute the Born contribution to \re{e1.8} at finite $K$ and study the transition from $K=2$ to the large $K$ behavior \re{F0}. 

In the previous section, we used the conventional amplitude approach to obtain the integral representation \re{int1} for 
this contribution. Let us show how the same result follows from the representation \re{EE-corr} of the energy correlations in terms of correlation functions.

The energy flow operator $\cE (n)$ is given by an integral involving the stress-energy tensor of $\mathcal N=4 $ SYM, see \eqref{E-new} in Appendix \ref{sA}. The latter is bilinear in the scalar fields $X^I$ constituting the source operators \re{e11}. As a consequence, 
the calculation of the four-point correlation function on the right-hand side of \re{EE-corr} in the Born approximation reduces to performing Wick contractions between the $K$ scalar fields in the operators $O_K(x)$ (source) and $\bar O_K(0)$ (sink) and the two pairs of scalar fields in the energy flow operators $\cE (n_1)$ and $\cE (n_2)$.\footnote{A simple example of  such a calculation is shown in \p{eA10} in Appendix~\ref{sA}. }

In this way, in the planar limit the correlation function $\vev{O_K(x)\cE (n_1)\cE (n_2)\bar O_K(0)} ^{(0)}$ can be factorized  into a product of $(K-2)$ propagators connecting the source and sink and the same correlation function evaluated for $K=2$, 
\begin{align}\label{e21}
\vev{O_K(x)\cE (n_1)\cE (n_2)\bar O_K(0)} ^{(0)} &=\frac14 K^2(K-1) \big[ N_cD(x)\big]^{K-2} \vev{O_2(x)\cE (n_1)\cE (n_2)\bar O_2(0)} ^{(0)} \,,
\end{align}
where the scalar propagator $D(x)$ is given by \re{prop} and the superscript `$(0)$' indicates the Born approximation. 
The relation \re{e21} is  represented diagrammatically in Figure~\ref{fig1}.
\begin{figure}[t!]
 \centerline{
{\parbox[c]{125mm}{ \includegraphics[width = 125mm]{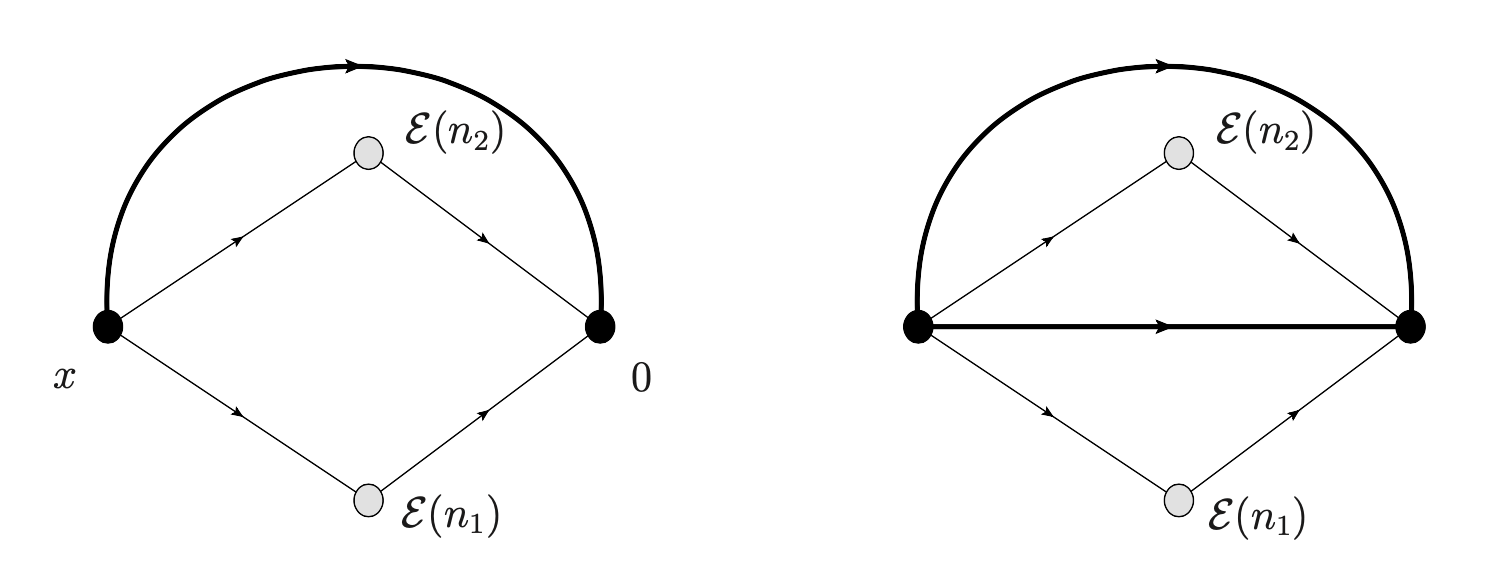}}}
}
\caption{The correlation function \re{e21} in the Born approximation and in the planar limit. 
Black and grey blobs depict scalar operators and energy detectors, respectively. A thin line represents a scalar propagator, a thick line represents a collection of scalar propagators. The total number of scalar propagators attached to the black blobs equals $K$. 
In the planar limit, the contributing diagrams have the topology of a sphere with four marked points. The two types of diagrams shown on the left and right panels differ by the number of lines separating the two energy detectors (no line on the left, at least one line on the right).} \label{fig1}
\end{figure}
The combinatorial $K-$dependent factor in \re{e21} counts the number of contributing diagrams.

Substituting \re{e21} into \re{EE-corr} we find that $\vev{\cE (n_1)\cE (n_2)}^{(0)}_K$ is  the convolution of the Fourier transform of $D^{K-2}(x)$ and $\vev{\cE (n_1)\cE (n_2)}^{(0)}_{K=2}$. The former is given by the function $\sigma_{K-2}$ defined in \re{sigmaK} and the latter is given in \re{e27} below. This leads to 
\begin{align}\label{e22}
\vev{\cE (n_1)\cE (n_2) } _K^{(0)}
&={K^2(K-1)\over (4\pi)^2 \sigma_{\rm tot}(q)} N_c^{K} 
\int_0^\infty d\tau_1 d\tau_2 \, \tau_1^2 \tau_2^2\, \sigma_{K-2}(q-n_1 \tau_1 -n_2 \tau_2) \,,
\end{align}
where $n_1\tau_1$ and $n_2\tau_2$ are the momenta of the particles entering the calorimeters. 
Replacing $\sigma_{K-2}$ with its expression \re{sigmaK} and changing the integration variables as $\tau_i=\omega_i q^2/(2(qn_i))$,  
we find that \re{e22} takes the expected form \re{e1.6} with
\begin{align}\notag
\cF^{(0)}_K(z)=& \frac1{2}
(K-3) (K-2)^2 (K-1)^2 K
\\
{}&\times
 \int_0^1 d\omega_1d\omega_2\, \omega_1^2 \omega_2^2\,(1-\omega_1-\omega_2+z \omega_1\omega_2)^{K-4} \,,
\end{align}
where the integration is restricted to the region $1-\omega_1-\omega_2+z \omega_1\omega_2\ge 0$.

The result can be expressed in terms of a hypergeometric function,
\begin{align}  \label{F-hyper}
\cF^{(0)}_K(z) 
&= \frac{2(K-2) (K-1)}{(K+1)(K+2)}  \,  {}_2F_1\left({3,3 ; K+3}\vert z\right) .
\end{align}
This relation is valid for $0<z<1$. The values of $z=0$ and $z=1$ correspond to the situation where the two detectors are located, respectively, atop of each other or antipodal to each other. In general, $\cF_K(z)$ contains   contact terms proportional to $\delta(z)$ and $\delta(1-z)$. We discuss them below in Section~\ref{s2.1}. 

For specific values of $K$ the relation \re{F-hyper} simplifies, e.g. \footnote{For $K=2$ the function $\cF^{(0)}_{K=2}(z)$ is given by a sum of contact terms, see \re{e27}.}
\begin{align}  \label{e24}\notag
\cF^{(0)}_{K=2}(z){}&=0 \,,  
\\[1.5mm]\notag
 \cF^{(0)}_{K=3}(z){}&= \frac{18(z-2)}{z^4} -6(z^2-6z +6)\frac{\log(1-z)}{z^5} \,,
 \\
 \cF^{(0)}_{K=4}(z){}&= \frac{8 \left(10 z^2-39 z+30\right)}{z^5}+\frac{24 (1-z) \left(z^2-8 z+10\right) \log
   (1-z)}{z^6}\,,
\end{align}
where $0<z<1$.  

The function $\cF^{(0)}_{K}(z)$ is plotted in Figure~\ref{fig:EECborn}  for several values of $K$. 
For $z\to 0$ it approaches a finite value,
\begin{align}
 \cF^{(0)}_{K}(z) =\frac{2 (K-2) (K-1)}{(K+1) (K+2)}+O(z)\,.
\end{align}
For $z\to 1$ it grows logarithmically for $K=3$ and stays finite for $K\ge 4$,
\begin{align}\notag\label{limits}
{}& \cF^{(0)}_{K=3}(z) = -6 \log(1-z) - 18 +O(1-z)\,,
\\
{}& \cF^{(0)}_{K\ge 4}(z) = {2K\over K-3} + O(1-z)\,.
\end{align}
In the large $K$ limit we get instead
\begin{align}\label{2.6}
\cF^{(0)}_K(z) \stackrel{K\gg 1}{=}2 +{6\over K} \lr{3 z - 2}+O(1/K^2)\,,
\end{align}
in agreement with \re{F0}. We repeat that the $z-$dependence only appears at the level of  the $O(1/K)$ corrections. 

We observe that the function $\cF^{(0)}_{K}(z)$ flattens out as $K$ increases and becomes constant for $K\to\infty$. This property is in agreement with our expectation \re{eq:1.4} that the correlations between the particles in the final state are suppressed at large $K$. 

\begin{figure}[t!]
\begin{center}
\includegraphics[width=8cm]{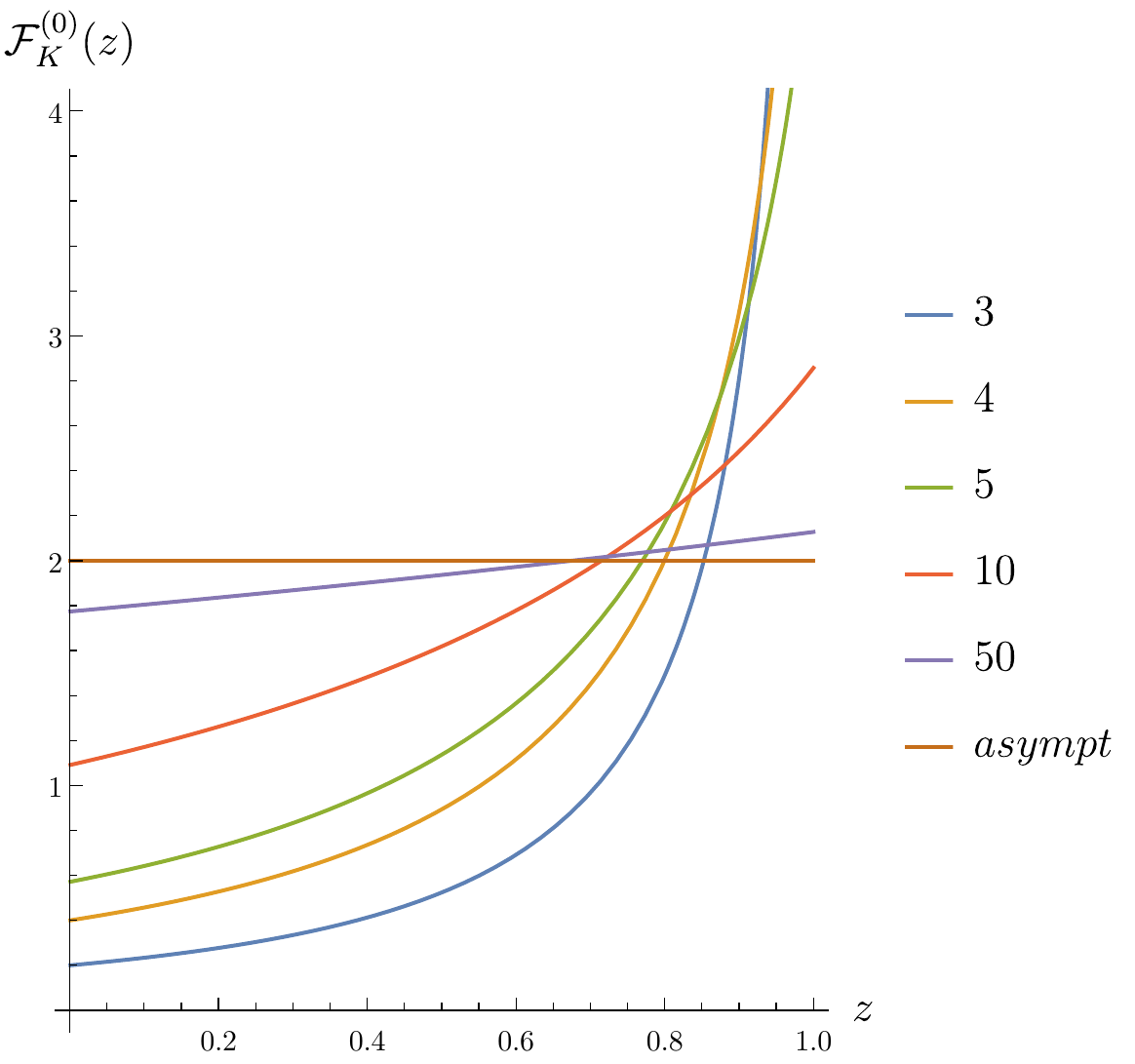}
\end{center}
\caption{Born approximation of the energy-energy correlation ${\cal F}^{(0)}_{K}(z)$ for several values of the source weight $K$ as a function of the angular separation $z=(1-\cos\theta_{12})/2$, see \p{F-hyper}.
For finite $K$, it is peaked around the end-point $z=1$. It flattens out as $K$ increases and approaches the constant value $2$ for $K \to \infty$ (the curve labeled  `asympt').}
\label{fig:EECborn}
\end{figure}
 
\subsection{Contact terms for $K=2$}\label{s2.1}

As was mentioned above, the relation \re{F-hyper} is valid for $0<z<1$. One of the reasons for this is that $\cF^{(0)}_K(z)$ is expected to contain contact terms localized at $z=0$ and $z=1$.  The easiest way  to reveal their presence is to apply the sum rules \re{e1.11}.

For $K=2$   the naive answer $\cF^{(0)}_{K=2}(z) =0$ in \p{e24}   does not satisfy the sum rules \re{e1.11}. In reality,  in this case the final state consists of two particles that move back-to-back in the rest frame of the source. As a consequence, the energy-energy correlation is different from zero only if the two calorimeters are atop of each other on the celestial sphere $(z=0)$ or antipodal $(z=1)$. 
This means that it is given by the sum of two contact terms, $\cF^{(0)}_{K=2}(z) = C_1\delta(z)+C_2 \delta(1-z)$, with coefficients that can be determined from  the sum rules \re{e1.11},
\begin{align}\label{e27}
\cF^{(0)}_{K=2}(z)=  \delta(z)+ \delta(1-z) \,.
\end{align}
Substituting this relation into \re{e1.6}, one reproduces the known expression for the energy correlation $\vev{\cE (n_1)\cE (n_2) } _{K=2}$ in the Born approximation.

For $K\ge 3$ we use \re{F-hyper} to get 
\begin{align}\label{e2.7}\notag
{}& \int_{\epsilon}^{1-\epsilon}dz\, z\, \cF^{(0)}_K(z) =1+ O(\epsilon)\,,
\\
{}& \int_{\epsilon}^{1-\epsilon} dz\,(1-z) \cF^{(0)}_K(z) = 1-{3\over  K+1} + O(\epsilon)\,,
\end{align}
where a small parameter $\epsilon>0$ was introduced to exclude the contribution of the end-points $z=0$ and $z=1$. 
In  close analogy with the case $K=2$, the latter is expected to be given by a sum of contact terms $\delta(z)$ and $\delta(1-z)$.
Comparing \re{e2.7} with \re{e1.11}, we observe that the contact terms should contribute zero to the first sum rule in \re{e1.11}
and  $3/(K+1)$ to the second sum rule. Put together, these conditions fix the coefficients of the contact terms. 

The resulting expression for the function $\cF^{(0)}_{K}(z)$ is
\begin{align}\label{F-Born}
\cF^{(0)}_{K\ge 3}(z) &=  {2(K-1)(K-2) \over  (K+1)(K+2)}  \   {}_2F_1\left({3,3 ; 3+K}\vert z\right) 
+ {3\delta(z)\over  K+1 } \,.
\end{align}
As compared with \re{F-hyper}, it is valid for $0\le z \le 1$.  At large $K$ the contact term in \re{F-Born} agrees with \re{F0}.
Combined  with \re{e1.6}, the relation \re{F-Born} yields the expression for the energy-energy correlation in planar $\mathcal N=4$ SYM, to the lowest order in the coupling constant. 

For finite $K$, the function \re{F-Born} is peaked around the end-point $z=1$ indicating that, like in the two-jet final state, the most of the energy in the final state is deposited at 
the calorimeters located at two antipodal points on the celestial sphere.  Going to the limit $K\to\infty$ we find that, in agreement with \re{eq:1.4}, the energy-energy correlation ceases to depend on the angular separation of the calorimeters. This means that the jets disappear and the energy is homogeneously distributed on the celestial sphere.

\section{Energy correlations at weak coupling}\label{sect:weak}

In this section, we study the corrections to the energy correlation \re{e1.6} in the planar $\mathcal N=4$ SYM theory
due to the interaction between the particles in the final state. At weak coupling, the Feynman diagrams contributing to \re{e1.6} can be obtained from those shown in  Figure~\ref{fig1} by adding   interaction vertices. In the planar limit, the interaction can only occur between  adjacent scalar particles. Depending on the choice of these particles we can distinguish three cases of interaction between:
(i)  undetected particles, (ii)  one detected particle and undetected ones, and (iii)  two detected and undetected particles. The first and the second class of diagrams constitute the total cross-section \re{sigmaK} and the single energy correlation $\vev{\mathcal E(n)}$. Because both quantities are protected from quantum corrections in $\mathcal N=4$ SYM, the above mentioned diagrams do not contribute to the unprotected energy correlations \re{e1.6}. We are therefore left with the Feynman diagrams in which two detected particles interact among themselves and with other undetected particles. To the first few orders in the 't Hooft coupling constant, examples of such diagrams are shown in Figure~\ref{fig2}. 

\begin{figure}[t!]
 \centerline{
{\parbox[c]{75mm}{ \includegraphics[width = 75mm]{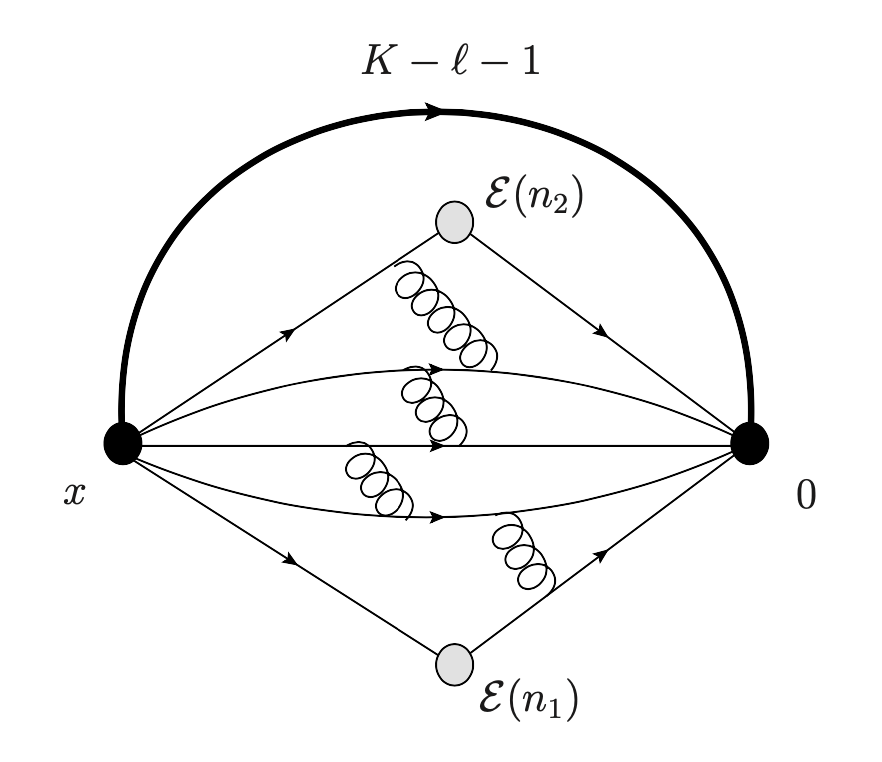}}}
}
\caption{Diagrammatic representation of the relation \re{fact-L}. The heavy operator creates $K$ particles. 
Two detected particles interact with $(\ell-1)$ particles, the remaining $(K-\ell-1)$ undetected particles propagate freely from the source to the sink. }\label{fig2}
\end{figure}

As follows from the form of these diagrams, at order $O(\lambda^\ell)$ the interaction can affect $(\ell+1)$ particles at most. At order $O(\lambda)$ the two detected particles can only interact with each other. At order $O(\lambda^2)$ there is an additional possibility for them to interact with one undetected particle (spectator). At order $O(\lambda^{\ell})$ the number of spectators cannot exceed $(\ell-1)$. 

For $\ell<K$ the number of interacting particles $\ell+1$ is smaller or equal to the number $K$ of particles produced (this is equivalent to the absence of wrapping corrections to the energy correlation \re{EE-corr}). The remaining $(K-\ell-1)$ particles 
propagate freely from the source to the sink. In close analogy with \re{e21}, their contribution to the correlation function $\vev{O_K(x)\cE (n_1)\cE (n_2)\bar O_K(0)}$ is just a power of the free scalar propagator,
\begin{align}\label{fact-L} 
\vev{O_K(x)\cE (n_1)\cE (n_2)\bar O_K(0)} ^{(\ell)} &=
{ K^2\over (\ell+1)^2}[N_c D(x)]^{K-\ell-1}\vev{O_{\ell+1}(x)\cE (n_1)\cE (n_2)\bar O_{\ell+1}(0)} ^{(\ell)} .
\end{align}
The combinatorial factor on the right-hand side gives the ratio of the number of Wick contractions of scalars in the two correlators in the planar limit. We would like to emphasize that the relation \re{fact-L} is only valid for $\ell\le K-1$. In particular, in the limit $K\to\infty$ it holds to any given order of the weak coupling expansion.

At one and two loops  the relation \re{fact-L} reads
\begin{align}\label{e35}
\vev{O_K(x)\cE (n_1)\cE (n_2)\bar O_K(0)} ^{(1)} &=
{ K^2\over 4}[N_c D(x)]^{K-2}\vev{O_{2}(x)\cE (n_1)\cE (n_2)\bar O_{2}(0)} ^{(1)} \,,
\\ \label{e351}
\vev{O_K(x)\cE (n_1)\cE (n_2)\bar O_K(0)} ^{(2)} &=
{ K^2\over 9}[N_c D(x)]^{K-3}\vev{O_{3}(x)\cE (n_1)\cE (n_2)\bar O_{3}(0)} ^{(2)} 
\,.
\end{align}
Notice that the first relation holds for $K\ge 2$ and the second one for $K\ge 3$.

Comparing the last two relations with \re{e21} we observe an important difference. At large $K$ the expression on the right-hand side of  \re{e35} and \re{e351}   is suppressed by a factor of $1/K$ as compared with \re{e21}. The reason for this is that the leading $O(K^3)$ contribution to the Born level result \re{e21} comes from the diagrams shown in Figure~\ref{fig1} on the right panel. These diagrams do not contribute to \re{fact-L} in the planar limit as explained above. In application to the energy correlation \re{EE-corr}, \re{e1.6} this implies that the loop corrections to the scaling function $\cF_K(z)$ are suppressed by a factor of $1/K$ as compared to the Born contribution \re{F0}.   

\subsection{Recurrence relation for the  perturbative  corrections} 

The relations \re{e35} and \re{e351} can be used to compute the correction to the energy correlation \re{e1.8} at one and two loops. 
Let us introduce the auxiliary functions
\begin{align}\label{eq4.4}\notag
{}& G_K^{(0}(q) =\int d^4 x\, e^{iqx}  [D(x)]^{K-2} N_c^{-2} \vev{O_{2}(x)\cE (n_1)\cE (n_2)\bar O_{2}(0)}^{(0)} \,,
\\
{}& G_K^{(\ell)}(q) =\int  d^4 x\, e^{iqx}  [D(x)]^{K-\ell-1}  N_c^{-\ell-1} \vev{O_{\ell+1}(x)\cE (n_1)\cE (n_2)\bar O_{\ell+1}(0)}^{(\ell)} \,,
\end{align}
where $\ell=1, 2$ and the scalar propagator $D(x)$ is given by \re{prop}. 
They satisfy the differential equation
\begin{align}\label{e37}
\Box_q \, G_K^{(\ell)}(q)  =G_{K-1}^{(\ell)}(q) \,.
\end{align}
Below we show that it yields a recurrence relation for the functions $\mathcal F_K^{(\ell)}(z)$ (for $\ell$ fixed) that allows us to efficiently determine the loop corrections to the energy correlation \re{e1.8}.

We combine equations \re{e21} and \re{fact-L} together with \re{EE-corr} and \re{e1.6} to get 
\begin{align}\notag
{}& \mathcal F_K^{(0)}(z) =  \pi^2(K-1)K\frac{ (n_1n_2)^3}{ z^3}
{G_K^{(0)}(q)\over q^2\sigma
   _K(q^2)}\,,
\\
{}& \mathcal F_K^{(\ell)}(z) =4\pi^2 K\frac{(n_1n_2)^3}{(\ell+1)^2 z^3} {G_K^{(\ell)}(q)\over q^2\sigma
   _K(q^2)}\,,
\end{align}
where $\sigma_K(q^2)$ is given by \re{sigmaK}. Next, the relation \re{e37} leads to
differential equations  for the functions $\mathcal F_K^{(0)}(z)$ and $\mathcal F_K^{(\ell)}(z)$, 
\begin{align}\label{EQ0}
&\left[\mD -  K (K-1) \right] \cF^{(0)}_{K}(z) + K(K-1) \cF^{(0)}_{K-1}(z)=0\,, 
\\[2mm] \label{EQ}
&\left[\mD -  K (K-1) \right] \cF^{(\ell)}_{K}(z) + K(K-2) \cF^{(\ell)}_{K-1}(z)=0\,, 
\end{align}
where $\ell\ge 1$ and $ \mD $ is the second-order differential operator 
\begin{align}\label{e527}
 \mD = \frac1{z} {d\over dz}  (1-z) {d\over dz} z^3 \,.
\end{align}
The relations \re{EQ0} and \re{EQ} are valid for $K\ge 3$ and $K\ge 2+\ell$, respectively.

One can check that the Born approximation result \re{F-hyper} verifies \re{EQ0}. In fact,  we could have obtained \re{F-hyper}
(up to an overall normalization) by solving the recurrence relation \re{EQ0} starting from $\cF_{K=2}^{(0)}(z) =0$ (up to contact terms) and requiring  $\cF_{K}^{(0)}(z)$ to be regular as $z\to 0$.

\subsection{One-loop correction}

In this subsection we show that the recurrence relation \re{EQ} can be effectively used to compute the energy correlation function $\mathcal F_K^{(1)}(z)$
at one loop. The starting point of the recurrence is the one-loop expression for this function at $K=2$ \cite{Belitsky:2013bja} \footnote{This result has been obtained by the Mellin method which we review in Section~\ref{s5}.}   
\begin{align}\label{1loop}
\cF^{(1)}_{K=2}(z) = - { \log(1-z)\over z^2(1-z)} \,,
\end{align}
that is valid for $0<z<1$.

\subsubsection*{Boundary conditions} %\label{s4.2.1}

The solutions to the second-order differential equations  \re{EQ} are defined up to the zero modes of the operator $(\mD -  K (K-1))$,
\begin{align}   \label{316}
\cF_K ^{(1)}\to \cF_K^{(1)}  &+  c_1\, z^{K-3} \, _2F_1\left({K,K \atop 2K}\Big\vert z\right)
  +c_2 \, z^{-2-K} \,    _2F_1\left({1-K,1-K\atop 2-2 K}\Big\vert z\right) \,.  
\end{align}
This freedom can be fixed    by imposing boundary conditions. Notice that the last term on the right-hand side of \re{316} grows as $z^{-K-3}$ for $z\to 0$, thus producing a divergent contribution to the sum rules \re{e1.12}. To avoid it, we put $c_2=0$. 

According to \re{1loop}, the function $\cF_{K=2}^{(1)}(z)$ behaves as $O(1/z)$  at small $z$. Examining the differential equation \p{EQ} for $K=3,4,\dots$, one finds that its solution has to have the following asymptotic behavior  at small $z$,
\begin{align}\label{e4.2}
 \cF^{(1)}_K(z) =  {a_K\over z} +  {b_K}\log z +O(z^0)\,.
\end{align}
For $K=2$ we find from \re{1loop} that $a_2=1$ and $b_2=0$. 

To fix the coefficient $c_1$ in \re{316} we examine the behavior of the function $\mathcal F_K^{(1)}(z)$ around $z=1$. In the Born approximation, we deduce from \re{limits} that $\mathcal F_K^{(0)}(z)$ grows logarithmically for $K=3$ and it approaches a finite value for $K\ge 4$. Requiring $\mathcal F_K^{(1)}(z)$ to stay finite for $K\ge 4$ as $z\to 1$ implies  $c_1=0$. This condition can be derived from the sum rules \re{e1.12} as follows. Multiplying \p{EQ} by $z$ and integrating by parts,  we obtain  
\begin{align}\label{E47}
&\int_0^{1-\epsilon} dz\,   \partial_z [(1-z)  \partial_z (z^3\, \cF^{(1)}_K(z))]  = [(1-z) \partial_z (z^3\, \cF^{(1)}_K(z)) ]\Big|_{z=1-\epsilon}=O(\epsilon) \,.
\end{align}
The boundary term  at $z=0$ vanishes due to \p{e4.2}. This derivation relies on the sum rules \p{e1.12}, i.e. we assume that for any $K \geq 3$ the relation
$\int_0^{1-\epsilon} dz\, z \cF^{(1)}_{K}(z)=O(\epsilon)$ holds. As we show below, it is satisfied for $K\ge 4$ only. We therefore deduce from \re{E47} that $\cF^{(1)}_K(z)$ should stay finite at $z\to 1$ and $K\ge 4$. This leads to $c_1=0$ because the term proportional to $c_1$ in \re{316} grows logarithmically as $z\to 1$.

For $K=3$ one can check using \re{1loop} that $\int_0^{1-\epsilon} dz\, z \cF^{(1)}_{2}(z)=\frac12\log^2\epsilon$ as $\epsilon\to 0$. In this case, in order to satisfy the sum rule \p{e1.12}, one has to add to \re{1loop} a contact term localized at $z=1$. As a result, for $K=3$ the right-hand side of \re{E47} scales as $\frac32\log^2\epsilon$ leading to $\cF^{(1)}_{K=3}(z)=\frac12\log^3(1-z)$ at $z\to 1$. The term proportional to $c_1$ in \re{316} yields a subleading contribution in this limit. To fix the coefficient $c_1$ we can use the sum rule \re{e1.12}.~\footnote{For $K>3$ the sum rule \re{e1.12} leads to $c_1=0$. }
 
We can now turn to solving the differential equation \re{EQ} supplemented with the boundary conditions specified above.
Our strategy is as follows: we first solve explicitly the differential equation for $\cF^{(1)}_{K=3}(z)$. Based on the form of $\cF^{(1)}_{K=3}(z)$ we then work out the general form of the solution for $K\geq4$.
Finally, we combine all $\cF^{(1)}_K(z)$ into a two-variable generating function and provide a closed-form expression for it. In this way we determine the one-loop energy correlation $\cF^{(1)}_{K}(z)$ for any weight $K$.

\subsubsection*{Solution for $K=3$} 

Solving the differential equation  \p{EQ} for $K=3$, we replace $\cF^{(1)}_{K=2}(z)$ with its expression \re{1loop} and pick the solution  that satisfies \p{e4.2}. 
The resulting expression for $\cF^{(1)}_{K=3}(z)$ is given by a linear combination of special functions of different transcendental weights 
accompanied by seven polynomials of $z$,
 \begin{align}\label{328}\notag
 \cF^{(1)}_{3}(z)   = {}& {1\over z^5} \Big[ c^{[3]}_{1}(z) L(z) + c^{[3]}_{2}(z)  \text{Li}_2(z) + c^{[3]}_{3}(z)  \log(z)  \log (1-z) 
 \\[1.5mm]
{}& +  c^{[3]}_{4}(z) \log ^2(1-z) + c^{[3]}_{5}(z) \log (1-z) 
  + c^{[3]}_{6}(z) \log(z) + c^{[3]}_{7}(z)\Big] \,.
\end{align}
Here  the function $L(z)$   contains only  terms of the maximal logarithmic  weight 3: \footnote{The terms in the second line are not included in the polynomials  $ c^{[3]}_{5}$ and $c^{[3]}_{7}$ in order to keep their coefficients  rational.}
 \begin{align}\label{}
L(z) & := \text{Li}_3(1-z)+\frac{1}{2} \text{Li}_2(z) \log (1-z)-\frac{1}{12}\log^3(1-z) + \frac{1}{2} \log^2(1-z)\log (z) \nt
&\ \ \ -  \zeta_2 \log (1-z)-\zeta _3 \,.
\end{align}
The coefficients $c^{[3]}_m(z)$ (with $m=1,\ldots,7$) are polynomials of degree $2$:
\begin{align} 
c^{[3]}_1 & =-6 \left(z^2-6 z+6\right),\quad & c^{[3]}_{2} & = 18 (3-2 z)\,, \nt
c^{[3]}_{3} & =9 \left(z^2-6 z+6\right),\quad  & c^{[3]}_{4} & = -9 (z-3) (z-1)\,, \nt
c^{[3]}_5 & =9 (z-1) (4 z-9)\,,  \quad & c^{[3]}_{6} & = 27 (2-z) z, \nt
c^{[3]}_{7} & =9 (3 - 2 z) z \, . & & 
\end{align}
The coefficients $c^{[3]}_5$ and $c^{[3]}_{7}$ depend linearly on the zero-mode coefficient $c_1$ in \re{316}. As explained above, its value is fixed by the sum rule $\int_0^1 dz\,z \cF^{(1)}_{3}(z) =0$. 

The  function $\cF^{(1)}_{3}(z)$  is plotted in Figure~\ref{fig:EEC1loop}.  For $z\to 0$ it has 
the asymptotic behavior 
\begin{align}\label{e411}
\cF^{(1)}_{K=3}(z) &=\frac{3}{4 z}-\frac{3}{10} \log (z)+ \frac{24}{25} +O\left(z\right)\,,
\end{align}
 in agreement with \re{e4.2}.  For $z\to 1$ we find 
\begin{align}
\cF^{(1)}_{K=3}(z) & = \frac{1}{2} \log ^3(1-z)+\frac{1}{2} \pi ^2 \log (1-z)+6 \zeta (3)+3 \pi ^2+9+O(1-z) \,. \label{FK1Lz1}
\end{align}
As expected, this function grows as a power of $\log(1-z)$.

It is straightforward to verify that the function \re{328} does not satisfy the second sum rule in \re{e1.12}, $\int_0^1 dz\, (1-z) \cF^{(1)}_{3}(z) \neq 0$. Like what happens in the Born approximation \re{e2.7}, this indicates that $\cF^{(1)}_{3}(z)$ should contain a contact term $\sim \delta(z)$. It is needed to regularize the contribution of the pole in \p{e411} to $\int_0^1 dz\, (1-z) \cF^{(1)}_{3}(z)$. More details about the contact terms can be found in Section~\ref{s6}.

\subsubsection*{General solution for $K\geq4$} %\label{s412}

The solution of the recurrence relation \re{EQ} for any $K\geq4$ takes the same form as \p{328},
\begin{align}\label{e412'}\notag
  \cF^{(1)}_{K}(z) = {}&{1\over z^{K+2}} \Big[  
 c^{[K]}_{1}(z) L(z) + c^{[K]}_{2}(z)  \text{Li}_2(z) + c^{[K]}_{3}(z)  \log(z)  \log (1-z) 
 \\[1.5mm]
& +  c^{[K]}_{4}(z) \log ^2(1-z) + c^{[K]}_{5}(z) \log (1-z) 
  + c^{[K]}_{6}(z) \log(z) + c^{[K]}_{7}(z)\Big] \,,
\end{align}
where   $c^{[K]}_m(z)$ (with $m=1,\ldots,7$) are polynomials of degree $K-1$. They  are determined recursively by solving \p{EQ} and
requiring $\cF^{(1)}_{K}(z)$ to be finite for $z\to 1$ and to satisfy \p{e4.2} for $z\to 0$. In particular, the polynomial $ c^{[K]}_{1}(z) $ has a simple closed form,  
 \begin{align}\label{e414}
 c^{[K]}_{1}(z) &= -   K(K-2) \left(z-1\right)^{K-3}   \left[ K(K+1)-4 K z+2 z^2 \right]  \,.
\end{align}
Instead of presenting explicit formulas for the remaining polynomials at each weight $K$,  in the next subsection we provide  a generating function for all  $\cF^{(1)}_{K}(z)$.

The functions ${\cal F}_{K}^{(1)}(z)$ are plotted in Figure~\ref{fig:EEC1loop}  for several values of the weight $K$. For small $z$ they behave as
\begin{align}\label{EEC1loopz0}
{\cal F}_{K}^{(1)}(z) = \frac{3}{K+1}z^{-1} -\frac{6(K-2)}{(K+1) (K+2)} \log z+O(z^0)\,,
\end{align}
in agreement with \re{e4.2}. For $K=2$ and $K=3$, they also agree with \re{1loop} and \re{e411}, respectively. 
For $z\to 1$, we find 
\begin{align}\label{EEC1loopz1}
{\cal F}_{K}^{(1)}(z=1) = & -\frac{9}{(K-3)^2} + \frac{6}{K-2} + \frac{24\zeta_2 -15 -12 H^{(2)}_{K-4}}{12(K-1)}  - \frac{24\zeta_2 + 33 - 12 H^{(2)}_{K-4}}{4(K-3)} \,, 
\end{align}
where $H^{(2)}_{K-4}=\sum_{p=1}^{K-4} p^{-2}$ is the generalized harmonic number of order two. 
The relation \re{EEC1loopz1} is valid for $K\ge 4$.

\begin{figure}
\begin{center}
\includegraphics[width=10cm]{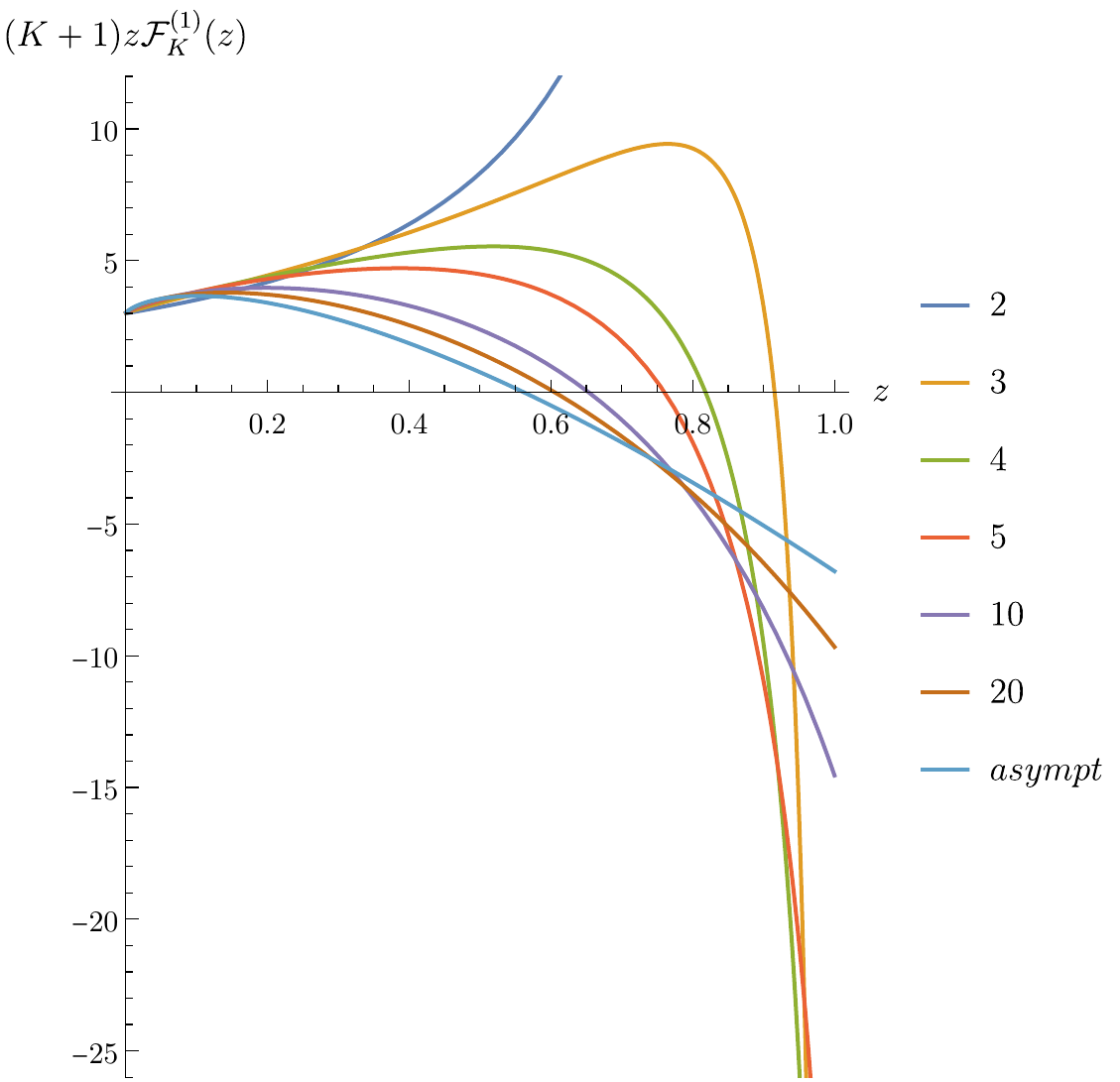}
\end{center}
\caption{The one-loop correction $(K+1)\,z\,{\cal F}^{(1)}_{K}(z)$ for several values of the weight $K$. The limiting curve for $K\to\infty$ is labeled  `asympt', see \p{phi1}.
% The leading term of the large $K$ asymptotics $\varphi^{(1)}(z)$ labeled `asympt' on the plot, eq.~\p{phi1}. 
 Owing to the normalization prefactor, all plots originate from $3$ at $z=0$, see  \p{EEC1loopz0}. For $z\to 1$, the one-loop correction has a pole at $K=2$, a logarithmic singularity at $K=3$ and it is finite for $K \geq 4$, see \p{1loop}, \p{FK1Lz1} and \p{EEC1loopz1}. Its value at $z=1$ decreases with $K$ and approaches $-\frac{7}{2} - \frac{\pi^2}{3} \approx -6.7899$ in the limit $K \to \infty$, see \p{asymptPhi1}.}
\label{fig:EEC1loop}
\end{figure}

\subsubsection*{Generating function} %\label{s413}

It is convenient to introduce an auxiliary variable $t$  and combine the one-loop functions ${\cal F}_{K}^{(1)}(z)$ of all weights $K$ in the generating function
\begin{align} \label{GenFun}
G(z,t) := \sum_{K \geq 2} t^K {\cal F}^{(1)}_K(z) \,.
\end{align}
Replacing ${\cal F}^{(1)}_K(z)$ with its expression \re{e412'}, we find that each of the seven polynomials $c^{[K]}_m(z)$ (with $m=1,\dots,7$) gives rise
to a function
\begin{align}
G_m(z,t) := \frac{1}{z^{2}} \sum_{K \geq 3} \left(\frac{t}{z}\right)^K c^{[K]}_m(z) \;, \quad m=1,\ldots,7\,.
\end{align}
 Then the generating function \re{GenFun} takes the form
\begin{align}\label{phisum}
G(z,t) = & G_1(z,t) \, L(z) + G_2(z,t) \, {\rm Li}_2(z) + G_3(z,t) \log(z) \log(1-z) + G_4(z,t) \log^2(1-z) \nt 
& + G_5(z,t) \log(1-z) + G_6(z,t) \log(z) + G_7(z,t) + t^2 {\cal F}_{K=2}^{(1)} \,. 
\end{align}
We have found  closed-form expressions for the functions  $G_m(z,t)$. 
They are given by linear combinations of classical polylogarithms of weights up to three, decorated with rational terms. More precisely, $G_1$, which resulted from the summation of the polynomials \p{e414}, is a rational function; $G_{2},\,G_{3},\,G_{4}$ contain logarithms and rational terms; $G_5,\,G_6$ contain dilogarithms, logarithms and rational terms; and $G_7$ is the most complicated function containing tri-logarithms, dilogarithms, logarithms and rational terms.
The arguments of the polylogarithms depend  both on $z$ and $t$ in such a way that the generating function $G(z,t)$ is given by 2dHPL functions \cite{Gehrmann:2000zt}. We provide an explicit expression for $G(z,t)$ in the ancillary file.

We can apply the generating function \re{phisum} to show that ${\cal F}^{(1)}_K(z)$ behaves as $O(1/K)$ at large $K$,
\begin{align}\label{F-large-K}
{\cal F}^{(1)}_{K}(z) = \frac{1}{K} \varphi^{(1)}(z)  + O(1/K^2) \,.
\end{align}
Combining this relation with \re{GenFun}, we expect that $G(z,t)$ should scale as $-\log(1-t) \varphi^{(1)}(z)$ for $t\to 1$. Indeed,  
examining  the generating function $G(z,t)$ for $t\to 1$ we reproduce the expected behavior and identify the function
\begin{align}\label{phi1}
\varphi^{(1)}(z) =  \frac{3}{z}+ 2 \text{Li}_2(1-z)-6 \log (z)-2\zeta_{2}-\frac{13}{2}\,.  
\end{align}
We therefore conclude that, as announced earlier, the one-loop correction to the energy correlation \re{e1.8} is suppressed at large $K$ by a  factor of $1/K$, as compared with the Born contribution \re{F0}.

For $z\to 0$ the relation \re{F-large-K} takes the expected form \re{e4.2} with $a_K=3/K+ O(1/K^2)$ and $b_K=-6/K+O(1/K^2)$,
\begin{align}
{\cal F}^{(1)}_{K}(z) = \frac{1}{K}\lr{ \frac{3}{z}-6 \log z + O(z^0) }\,.
\end{align}
It also agrees with \p{EEC1loopz0}. For $z\to 1$ we have, in agreement with \p{EEC1loopz1},
\begin{align}
{\cal F}^{(1)}_{K}(z) = \frac{1}{K}\lr{ -\frac{7}{2} -2\zeta_2 +O\left(z-1\right)}. \label{asymptPhi1}
\end{align}
In the previous relations we keep only the leading term at large $K$.

\subsection{Two-loop correction}\label{s42}

The two-loop corrections ${\cal F}^{(2)}_K(z)$ satisfy the recurrence differential equations \p{EQ}. These equations are valid for $K\ge 4$ and they allow us to express ${\cal F}^{(2)}_K(z)$ for arbitrary $K\ge 4$ in terms of ${\cal F}^{(2)}_{K=3}(z)$. The latter can be determined by extending the relation \p{EQ} to the case $K=3$ with a properly defined inhomogeneous term ${\cal F}_{\rm aux}^{(2)}(z)$. 

\subsubsection*{Two-loop solution of the recurrence relation}

Below we show  that  ${\cal F}^{(2)}_K(z)$ can be expanded over a basis of polylogarithm functions ${\cal L}_i^{(w)}(\sqrt{z})$, namely 
\begin{align}
z^{K+2} {\cal F}^{(2)}_K(z) = {}& \sum_{w=0}^{5} \sum_{i=1}^{l_w} a^{(w)}_{i,K}(z)\, {\cal L}_i^{(w)}(\sqrt{z}) + \sqrt{z}\, b_{K}(z)\, {\cal L}_1^{(2)}(\sqrt{z}) \notag\\
& + \delta_{K,3}\; z^2\left[ -6 z {\cal L}^{(3)}_3(\sqrt{z}) + 6 z {\cal L}^{(2)}_3(\sqrt{z}) + 6 \sqrt{z} \, {\cal L}^{(2)}_1(\sqrt{z})\right] \,. \label{EEC2loop}
\end{align}
Here $a^{(w)}_{i,K}(z)$ and $b_K(z)$ are polynomials in $z$ of degree $(K-1)$ and $(K-2)$, respectively, with rational $K$-dependent coefficients. The   second line in \p{EEC2loop} is relevant only for $K=3$.

The functions ${\cal L}_i^{(w)}(\sqrt{z})$ are pure polylogarithms of  transcendental weight $w$. They are given by multi-linear combinations  (with rational coefficients) of harmonic polylogarithms (HPL) \cite{Remiddi:1999ew} with argument $\sqrt{z}$ and zeta-values $\zeta_2$, $\zeta_3$, $\zeta_5$, which are graded by the transcendental weight. The functions ${\cal L}_i^{(w)}(\sqrt{z})$ do not depend on $K$. At weight $w$ we employ $l_w$ linearly independent polylogarithmic combinations, ${\cal L}^{(w)}_i$ with $i=1,\ldots,l_w$, in the expression \p{EEC2loop}. The counting is as follows: 
\begin{align}
l_0=1\,, \qquad l_1 = 2 \,, \qquad l_2 = 5 \,, \qquad l_3= 8 \,, \qquad l_4 = 5 \,, \qquad l_5 = 3\,.
\end{align}
In total, the ansatz \p{EEC2loop} contains $\sum_{w=0}^{5} l_w = 24$ polylogarithmic functions  ${\cal L}_i^{(w)}(\sqrt{z})$. 
They take the following form
\begin{itemize}
\item
Weight-zero function ${\cal L}^{(0)}_1 \equiv 1$;
\item 
Weight-one functions
\begin{align}
\{ {\cal L}^{(1)}_{i=1,2} \}= \{ H_{-1} - H_{1}, \,2 H_0 \} = \{ \log(1-z) , \log(z) \}\,;
\end{align} 
\item 
Weight-two functions
\begin{align}
\{ {\cal L}^{(2)}_{i=1,\dots,5} \}= \{ & 2H_{-1,0}+2 H_{1,0} , \, - H_{-1,-1} + H_{-1,1} + H_{1,-1} -4 H_{1,0} - H_{1,1} , \notag \\ 
& H_{1,1} - H_{1,-1} +2 H_{0,1} - 2 H_{0,-1} - H_{-1,1} + H_{-1,-1} + \zeta_2 ,\notag\\
& H_{0,0} + H_{1,0} , \, \zeta_2 \} \,.
\end{align} 
\end{itemize}
Here we omit the arguments of the HPLs for the sake of brevity, $H_{a,b,\ldots} \equiv H_{a,b,\ldots}(\sqrt{z})$ \cite{Remiddi:1999ew}. The remaining higher-weight combinations ${\cal L}^{(w)}_i$ have a similar form. They are defined in the ancillary file.

The polylogarithmic combinations of HPLs of   transcendental weights up to four, ${\cal L}_i^{(w)}$ with $0 \leq w \leq 4$, can be expressed as classical polylogarithms. However, this does not apply to the weight-five combinations ${\cal L}_i^{(5)}$, so we prefer to present all weights using the same HPL notations.

\subsubsection*{Solution procedure}
 
In this subsection we present some details of the derivation of \re{EEC2loop}. Namely, we first compute ${\cal F}^{(2)}_{K=3}(z)$ and then apply \re{EQ} for $\ell=2$ and $K\ge 4$ to obtain ${\cal F}^{(2)}_{K}(z)$.

Before we turn to the computation of  ${\cal F}^{(2)}_{K=3}(z)$, let us examine the function ${\cal F}^{(2)}_{K=2}(z)$ which defines the two-loop correction to the energy  correlation \re{EE-corr} for $K=2$. It was computed in \cite{Belitsky:2013ofa} using the Mellin approach  summarized in Section~\ref{s5.1} and Appendix~\ref{appA}.   As   shown there, the function ${\cal F}^{(2)}_{K=2}(z)$ admits a Mellin integral representation  \re{MellinF2phys}.
The result of  the Mellin integration in \re{MellinF2phys}  can be expanded
over the basis of special functions ${\cal L}^{(w)}_i\equiv {\cal L}^{(w)}_i(\sqrt{z})$ defined above,
\begin{align}
& z^3 {\cal F}_{K=2}^{(2)}(z) = 2z(1+\sqrt{z})\, {\cal L}^{(2)}_1 + 2 z  {\cal L}^{(2)}_2 + 2 z^2  {\cal L}^{(2)}_3 \notag\\
& + \frac{z}{(1-z)} \left[ \left(4+\frac{13z}{2}\right){\cal L}^{(3)}_1 - z {\cal L}^{(3)}_2 + z(1+2z) {\cal L}^{(3)}_3 + {\cal L}^{(3)}_4 + z {\cal L}^{(3)}_5 \right]. \label{EEC2loopK2phys}
\end{align}
 
To find the function ${\cal F}_{K=3}^{(2)}$, we have to solve the differential equation \p{EQ} for $K=3$. This may seem contradictory, since earlier we declared that  at two loops equation \p{EQ} is valid only for $K\geq4$.  In reality, we can extend its validity to $K=3$ by making the following observation. As explained in Appendix~\ref{AB},  the correlator \p{211'}  takes a factorized form for all $K\geq3$ with the common function $F^{(2)}_{K\geq 3}$ given in \p{Fs}. 
As compared with the function $F^{(2)}_{K=2}$ that defines the correlator for $K=2$, it is given by a different linear combination of the same two-loop conformal integrals.  We recall that the factorized form \p{211'} is at the origin of the differential recursion  \p{EQ} (see \p{eq4.4} and \p{e37}). However, in the case at hand the starting point of the recursion is not the `physical' ${\cal F}_{K=2}^{(2)}(z) $  from \p{EEC2loopK2phys} but a different `auxiliary' function ${\cal F}_{\rm aux}^{(2)}(z)$ that is defined below in \re{F-aux}.

In this way we arrive at  the differential equation
\begin{align}\label{eq-aux}
\left(\mD -  6 \right) \cF^{(2)}_{K=3}(z) + 3\cF^{(2)}_{\rm aux}(z)=0\,.
\end{align}
This is a particular case of the general recursion \p{EQ}, specified to $\ell=2$ and $K=3$, but with a different inhomogeneous term $\cF^{(2)}_{\rm aux}(z)$, not to be confused with ${\cal F}_{K=2}^{(2)}(z)$.   In other words, we have extended the relation \re{EQ}, initially derived for $K\ge 4$, to the case $K=3$.

 To find the explicit expression for $\cF^{(2)}_{\rm aux}(z)$ we need to repeat the Mellin integration in \p{MellinF2phys} with a different linear combination of the same Mellin amplitudes, see \p{MellinF2nonphys}.   
The result has an expansion in the basis of polylogarithmic functions ${\cal L}^{(w)}_i(\sqrt{z})$, similar to \re{EEC2loopK2phys}:
\begin{align}
z^3 {\cal F}_{\rm aux}^{(2)}(z) = & (1+\sqrt{z})\, {\cal L}^{(2)}_1 + {\cal L}^{(2)}_2 + z  {\cal L}^{(2)}_3 + \frac{1}{(1-z)} \Bigl[ {\cal L}^{(3)}_1 + z {\cal L}^{(3)}_2 + z^2 {\cal L}^{(3)}_3 \Bigr]  . \label{EEC2loopK2nonphys}
\end{align}

At two loops, the differential equations \re{eq-aux} and \p{EQ} can be solved recursively for $K=3,4,\dots$ in the same manner as in the one-loop case. The general solution takes the form \re{EEC2loop}. The polynomials  $a^{(w)}_{i,K}(z)$ and $b_{K}(z)$ in \re{EEC2loop} can be found by substituting the ansatz \re{EEC2loop} into \re{eq-aux} and \p{EQ} and by taking into account that the polylogarithmic functions ${\cal L}^{(w)}_i({\sqrt{z}})$ satisfy the differential equations
\begin{align}\label{dr}
\frac{d}{dz} {\cal L}^{(w)}_i({\sqrt{z}}) = \sum_{j=1}^{l_{w-1}} \left[ \frac{c^{(w)0^2}_{i,j}}{z} + \frac{c^{(w)0}_{i,j}}{\sqrt{z}} + \frac{c^{(w)-}_{i,j}}{\sqrt{z}-1} + \frac{c^{(w)+}_{i,j}}{\sqrt{z}+1} \right]{\cal L}^{(w-1)}_j({\sqrt{z}}),
\end{align}
where $c^{(w)}_{i,j}$ are rational coefficients. Since the functions ${\cal L}^{(w)}_i({\sqrt{z}})$ are linearly independent, the relations \re{dr}  lead to recurrence relations for the coefficients of the polynomials $a^{(w)}_{i,K}(z)$ and $b_{K}(z)$. 

The two  boundary conditions  needed to  fix the solution of the second-order differential equations are the absence of $z^{-2-K}$ singularities at $z \to 0$, and the finiteness of ${\cal F}^{(2)}_{K\geq4}(z)$ at $z \to 1$.~\footnote{Let us note that in order to impose the boundary conditions it is sufficient to work out only the leading terms in the expansions $z\sim 0$ and $z\sim 1$ of the HPL combinations ${\cal L}^{(w)}_i$, and they do not depend on $K$. Then fixing the boundary conditions is completely algebraic.} For $K=3$, the function 
${\cal F}^{(2)}_{K=3}(z)$ has a logarithmic singularity at $z=1$ (see \re{F2-K=3} below). As before, we use the first sum rule in \p{e1.12} to fix the remaining freedom in the solution for ${\cal F}^{(2)}_{K=3}(z)$. The sum rule requires the evaluation of integrals involving HPLs. In Section~\ref{s64}, we describe   a semi-numerical implementation of the sum rules for the two-loop energy correlation \p{EEC2loop} that yields exact values of the remaining unknown.

We have solved explicitly the recurrence relations for the polynomials $a^{(w)}_{i,K}(z)$ and $b_{K}(z)$, up to $K=100$. 
In the anciliary file we provide a \texttt{Mathematica} code  that constructs the solutions recursively.
The procedure is completely algebraic and it only requires solving a system of linear equations with rational coefficients at the $K$-th step. Its solution is then fed in the linear system of the $(K+1)$-th step, etc.  

Unlike the one-loop case, we have not attempted to find a closed-form expression for ${\cal F}_{K}^{(2)}(z)$ for arbitrary $K$. Instead, we used  the obtained solutions for several small values of $K$ to deduce the asymptotic behavior of ${\cal F}_{K}^{(2)}(z)$ as $z\to 0$ for any $K$,   
\begin{align}
{\cal F}_{K}^{(2)}(z) = \frac{3}{K+1}\frac{\log(z)}{z} + \frac{c^{(2)}_{1/z,K}}{K+1}\frac{1}{z} + O(\log z) \,,\label{FK2loopz0}
\end{align}
where
\begin{align}
c^{(2)}_{1/z,K} =  -\frac{3}{2} \zeta_3 + 3\zeta_2- \frac{7}{2} - \frac{3}{K-1} - \frac{3}{K} - \frac{3}{K+1} \,. \label{c21zK}
\end{align}
The expression on the right-hand side of \re{FK2loopz0} contains a pole at $z=0$. To make ${\cal F}_{K}^{(2)}(z)$ integrable at the origin, one has to add to \re{FK2loopz0} the contact terms localized at $z=0$. These terms are derived in  Section~\ref{s64}  for any $K$. {The two-loop EEC  ${\cal F}_{K=2}^{(2)}$, calculated in \cite{Belitsky:2013ofa}, has a similar asymptotic behavior at $z \to 0$, see \p{eqF2L2}.}

As compared with the one-loop result \re{EEC1loopz0}, the two-loop correction \re{FK2loopz0} is enhanced at small $z$ by a factor of $\log z$. Adding together \re{EEC1loopz0} and \re{FK2loopz0}, we keep the most singular terms at each loop order to  get
\begin{align}\label{1+2-loops}
a {\cal F}_{K}^{(1)}(z) + a^2 {\cal F}_{K}^{(2)}(z) + \dots = {3\over K+1} \left[ {a\over z} + {a^2\over z}\log z +\dots \right],
\end{align}
where $a=\lambda/(4\pi^2)$.
This relation is in agreement with the expected small $z$ behavior of the energy correlation 
\begin{align}\label{low-z}
 {\cal F}_{K}(z) \sim z^{-1+\gamma(\lambda)/2} \,,
\end{align}
which follows from the operator product expansion on the celestial sphere (see \re{small-z} below). Here $\gamma(\lambda)=\lambda/(2\pi^2) +O(\lambda^2)$ 
does not depend on $K$ and coincides with the anomalous dimension of the twist-two operators of spin $3$. 

For $z\to 1$, the two-loop correction to the energy correlation ${\cal F}_{K}^{(2)}(z)$ approaches a finite value for $K\ge 4$. At large $K$ it decreases as $O(1/K)$, see \re{cF-LO} and \re{asymptPhi2} below.
For $K=3$, the function ${\cal F}_{K=3}^{(2)}(z)$ grows for $z\to 1$ as a power of $\log(1-z)$,
\begin{align}\label{F2-K=3}
{\cal F}_{K=3}^{(2)}(z) = & -\frac{3}{160} \log^5(1-z) - \frac{\zeta_2}{2}\log^3(1-z) + \frac{3 \zeta_3}{2} \log^2(1-z)- \frac{33}{4} \zeta_4\log(1-z) \notag\\
& + 57\zeta_5 -72\zeta_2\zeta_3 -\frac{513}{4}\zeta_4 -\frac{216}{2}\zeta_3 + 270\zeta_2 \log(2)- 45\zeta_2 + 3+O(1-z)\,.
\end{align}
{For $K=2$, the function ${\cal F}_{K=2}^{(2)}(z)$ has a stronger singularity at $z\to 1$, namely it grows as $O(\log^3(1-z)/(1-z))$ at the end point, see \p{eqF2L2}.}
  
\begin{figure}
\begin{center}
\includegraphics[width=10cm]{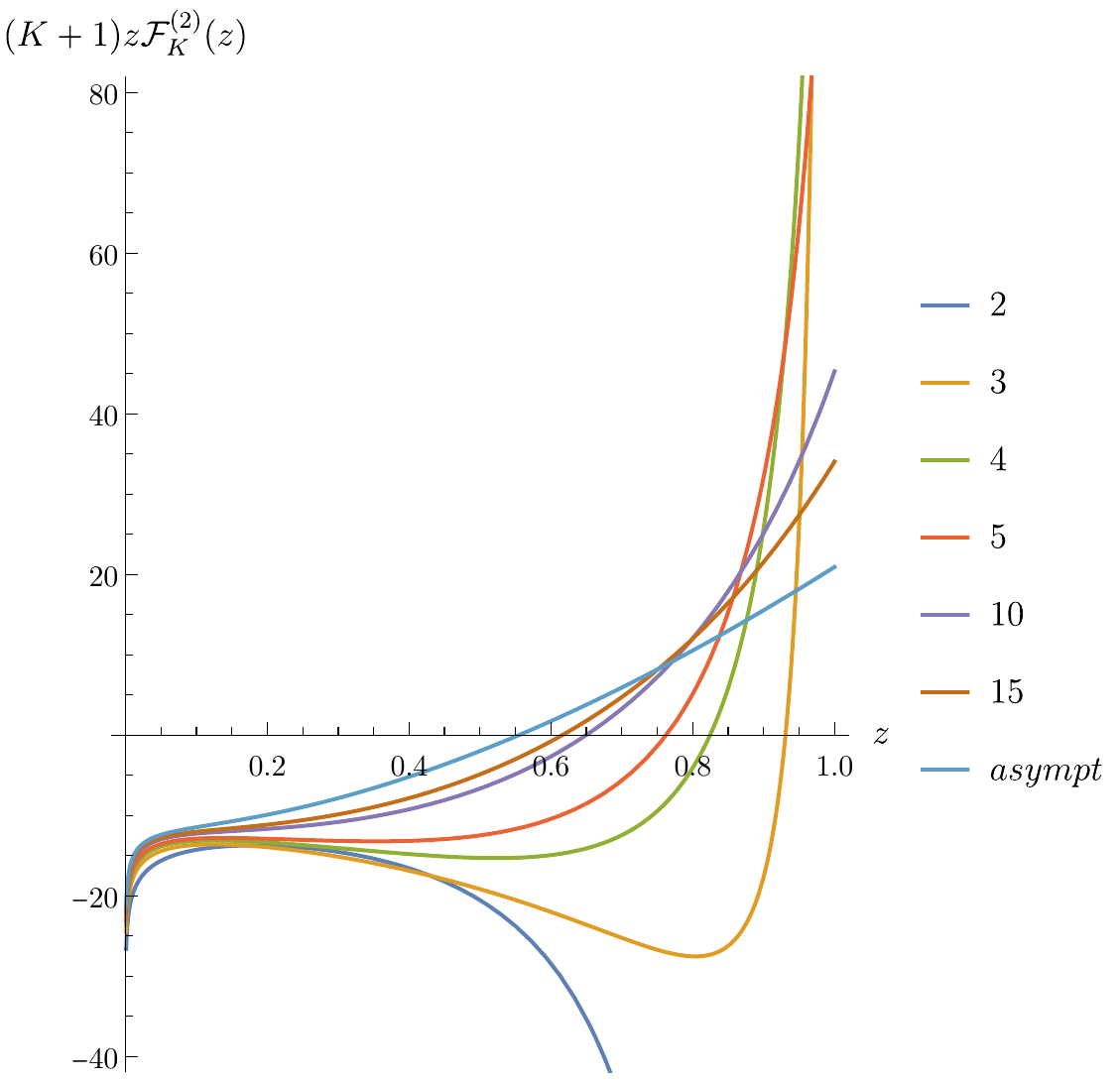}
\end{center}
\caption{The two-loop correction to the energy correlation  $(K+1)\,z\,{\cal F}^{(2)}_{K}(z)$ for several values of the weight $K$ of the source. 
The limiting curve for $K\to\infty$ is labeled  `asympt'. Owing to the normalization prefactor, all plots scale as $3\log z +O(z^0)$ at $z\to 0$, see \re{FK2loopz0}.  For $z\to 1$ the two-loop correction develops a pole at $K=2$, scales as a power of $\log(1-z)$ at $K=3$ (see \re{F2-K=3}) and approaches a finite value for $K\ge 4$. The latter decreases with $K$ and is given by $\frac{49}{144} + \frac{181}{180}\zeta_2 + \frac{85}{6}\zeta_3 + \frac{7}{4} \zeta_4  \approx 20.9176$  in the limit $K \to \infty$, see \p{asymptPhi2} below.}
\label{fig:EEC2loop}
\end{figure}
 
The functions $(K+1)\,z\,{\cal F}^{(2)}_{K}(z)$ are plotted in Figure~\ref{fig:EEC2loop} for several values of $K$. The additional factor of $(K+1)z$ is inserted to soften the singularity at $z=0$, see \p{FK2loopz0}, and to ensure that the function stays finite as $K\to\infty$. As follows from \re{e1.12}, the integral of  $(K+1)z {\cal F}^{(2)}_K(z)$ over the interval $0<z<1$ has to vanish. This explains why the functions shown  in Figure~\ref{fig:EEC2loop} change sign at some value of $z$ for $K\ge 3$. For $K=2$ the function takes negative values for $0<z<1$ and its integral vanishes after one takes into account the contact term proportional to $\delta(1-z)$. We recall that such contact terms are absent for $K\ge 3$.

We  observe that, in agreement with \re{low-z}, the curves in Figure~\ref{fig:EEC2loop} look alike for small $z$. The situation is different for finite $z$, the functions $(K+1){\cal F}^{(2)}_K(z)$ become more and more flat as $K$ increases. We have  observed  the same pattern at one loop. We recall that the $z-$dependence describes the angular distribution of the energy on the celestial sphere. The flattening of the $z-$dependence of the energy correlation at large $K$ implies that the energy distribution becomes more homogeneous. 

\section{Heavy sources at weak coupling}  \label{s5}

In the previous section, we computed the energy correlations at weak coupling for an arbitrary weight $K$ of the source operators and then examined their asymptotic behavior at large $K$. In this section, we present another approach that allows us to obtain this asymptotic behavior directly, without going through the details at finite $K$. It is based on the method developed in \cite{Belitsky:2013xxa,Belitsky:2013bja,Belitsky:2013ofa,Henn:2019gkr} for the calculation of event shapes with sources of weight $K=2$. The main advantage of this method is that it is applicable to the energy correlations \re{EE-corr} for an arbitrary coupling constant.

\subsection{Mellin approach} \label{s5.1}

The energy flow operator $\mathcal E(n)$ in \re{EE-corr} is given by the stress-energy tensor integrated along the light-ray defined by the null vector $n^\mu$ (see \re{E-new}).  As a consequence, the energy correlation \re{EE-corr} is related to the four-point function 
$\vev{O_K(1) T(2) T(3) O_K(4)}$
involving two heavy source operators and two stress-energy tensors. In the maximally supersymmetric $\mathcal N=4$ SYM theory, the stress-energy tensor belongs to the same supermultiplet as the simplest half-BPS operator \re{e11} of weight $K=2$. 

The $\mathcal N=4$ superconformal  Ward identities relate the above mentioned four-point correlation function to that of  four half-BPS operators $\vev{O_K(1) O_2(2) O_2(3) O_K(4)}$. The properties of these correlators are summarized in Appendix~\ref{AB}. In particular,
for an arbitrary coupling constant they are expressed in terms of a single function $\Phi_K(u,v)$ depending on the two conformal cross-ratios\footnote{This choice of cross-ratios is convenient for the Mellin procedure described in Appendix~\ref{appA}.} 
\begin{align}\label{cr}
u= {x_{12}^2 x_{34}^2\over x_{14}^2 x_{23}^2 }\,,\qqqquad
v= {x_{13}^2 x_{24}^2\over x_{14}^2 x_{23}^2 }\,.
\end{align}
It proves convenient to use the Mellin representation for this function,
\begin{align}\label{Phi}
\Phi_K(u,v) =  \int \frac{dj_1 dj_2}{(2\pi i)^2} M_K(j_1,j_2) \, u^{j_1} v^{j_2} \,,
\end{align}
where the integration contours run parallel to the imaginary axis to the left of the origin, $\Re j_i =-\delta$ with $0< \delta <1$. The Mellin amplitude $M_K(j_1,j_2)$ depends on the coupling constant. {The expansion coefficients $M_K^{(\ell)}$ of $M_K(j_1,j_2)$ at weak coupling in the planar limit,}
\begin{align}
M_K(j_1,j_2) = \sum_{\ell \geq 0} \left(\frac{\lambda}{4\pi^2}\right)^\ell M_K^{(\ell)}(j_1,j_2) \,,   \label{eq:Mexpandla}
\end{align}
{are not all independent. As we have already discussed in the beginning of Section~\ref{sect:weak}, and provide more details in Appendix~\ref{AB} (see  \p{M12}), the $\ell$-loop perturbative corrections with $K \geq \ell+1$ are all the same,}  
\begin{align} \label{eq:5.4}
M^{(\ell)}_{K=\ell+1} = M^{(\ell)}_{K=\ell+i} \ , ~~~ i \geq 2 . 
\end{align}

Applying the $\mathcal N=4$ superconformal Ward identities, one can express the correlation function in \re{EE-corr} in terms of the function \re{Phi} or equivalently the Mellin amplitude $M_K(j_1,j_2)$. The result is (see \cite{Belitsky:2014zha,Korchemsky:2015ssa} for details)
\begin{align}\label{e54}
\vev{O_K(x)\cE (n_1)\cE (n_2)\bar O_K(0)}=
{}& {8 \over  (n_1n_2)^{3}} {1\over (x^2)^{K+1}} {d^2 \over d\gamma^2} 
 (1-\gamma)^2 \gamma^2{d^2 \over d\gamma^2} \cG_K(\gamma) \,,
\end{align}
where the function $\cG_K(\gamma)$  is given by a Mellin integral involving the same amplitude $M_K(j_1,j_2)$ as in \re{Phi}, 
\begin{align}\label{e55}
\cG_K(\gamma) = - {1\over 16\pi^3}\int {dj_1 dj_2\over (2\pi i)^2} \left[{\Gamma(1-j_1-j_2)\over \Gamma(1-j_1)\Gamma(1-j_2)} \right]^2 M_K (j_1,j_2)\,  \gamma^{j_1+j_2-1} \,,
\end{align}
and $\gamma$ is a dimensionless Lorentz scalar variable invariant under the independent rescaling of the null vectors $n_i$, 
\begin{align}\label{gamma}
{\gamma={2(xn_1)(xn_2) \over  x^2 (n_1n_2) }}\,.
\end{align}
The differential operator acting on $\cG_K(\gamma)$  in \re{e54} stems from the relation between the correlators with different spins mentioned in the beginning of this subsection.

Substituting \re{e54} into \re{EE-corr} and going through the calculation we can obtain the expression for the energy correlation $\vev{\cE (n_1)\cE (n_2) }_K$.
It takes the expected form \re{e1.6} with the function $\cF_K(z)$ given by the double Mellin integral
\begin{align}
 \cF_K(z) &=    \int {dj_1 dj_2\over (2\pi i)^2} \left[{\Gamma(1-j_1-j_2)\over \Gamma(1-j_1)\Gamma(1-j_2)} \right]^2 M_K(j_1,j_2) \, \cK_K(j_1+j_2,z)  \,.\label{MK}
\end{align}
Here the Mellin amplitude $M_K(j_1,j_2)$ defines the four-point correlation function \re{Phi} and  depends on the coupling constant. The function $\cK_K(j_1+j_2,z)$ is  called  the detector kernel. It is independent of the coupling constant and is given by
\begin{align}\notag\label{K-2F1}
 & \cK_K(j,z) = \redq  \sum_{k=0,1,2}  z^{-j-k}\,  (-1)^k {{k}\choose{2}} \,  \frac{ \Gamma (j) \Gamma (j+k)}{  \Gamma (j-2) \Gamma (j+k-2) } 
 \\
&\times  { \Gamma (K-1) \Gamma (K+1) \over   \Gamma (K-2+j+k)  \Gamma (K+3-j-k)}   \,  _2F_1\left({3-j-k,3-j-k \atop  K+3-j-k}\Bigg| z\right)  .
\end{align}
The derivation of this relation can be found in Appendix~\ref{appA}, see \re{K-2F1'} for $s_1=s_2=2$.  
We see that the dependence on the angular variable $z$ in \p{MK} comes only through the detector kernel. We remark that the relation \re{MK} holds for any  value of the coupling constant.

One can check that the  function $\cF_K(z)$ defined in \p{MK}  satisfies the differential equation \re{EQ}, irrespectively of the choice of the Mellin amplitude $M_K(j_1,j_2)$, provided that the latter  does not depend on $K$. For example, this is the case at weak coupling in the planar limit for sufficiently large $K$, namely $K\geq \ell+1$, see \p{eq:5.4}.

In the special case  $K=2$  the kernel simplifies drastically and it is given by \re{4.17} (see \cite{Belitsky:2013xxa}--\cite{Belitsky:2014zha}). 
In the next subsection, we discuss the asymptotic behavior of the kernel \re{K-2F1} at large $K$. As we show below, the main advantage of 
the Mellin representation \re{MK} is that it allows us to demonstrate directly that the corrections to the energy correlation scale as $1/K$ for a finite coupling constant. In particular, at weak coupling
\begin{align}\label{cF-LO}
\cF^{(\ell)}_{K}(z) = \frac{1}{K} \varphi^{(\ell)}(z) + O(1/K^2)\,.
\end{align}
This relation generalizes \re{F-large-K} to any loop order. 

\subsection{Leading term of the large $K$ expansion}

For arbitrary $K$ the detector kernel \re{K-2F1} is given by a sum of three terms, each  containing a hypergeometric function. At large $K$ these functions 
can be replaced by $1$. In this way,  we obtain that, to the leading order in $1/K$, the detector kernel \re{K-2F1} takes the  form
\begin{align}\label{kappa}
\cK_{K}(j,z) = \frac{1}{K} \kappa(j,z) +O(K^{-2})\,,
\end{align}
where the leading coefficient function $\kappa(j,z)$   is given by
\begin{align}
\kappa(j,z) &= \redq (j-2) (j-1)\, z^{-2-j}\,  
   \left[ j(1+j) + 2j(1-j)z + (j-1)(j-2)z^2 \right]  \,.
\end{align}
This relation is conveniently rewritten in terms of a fourth-order differential operator,
\begin{align} 
\kappa(j,z) = {\cal D}  z^{-j+1} \,,  \qquad  {\cal D} := \frac{1}{z} \frac{d}{dz} z^2 \frac{d}{dz} \frac{(1-z)^2}{z^2} \frac{d}{dz} z^2 \frac{d}{dz} \,.  \label{e512}
\end{align}
It coincides with the analogous operator in \re{e54} after replacing $\gamma$ with $1/z$. 
Inserting \p{e512} into \p{MK} and changing the integration variable $j_2$ to  $j=j_1+j_2$, we arrive at the relation \re{cF-LO} with the function $\varphi^{(\ell)}(z)$ given by
\begin{align}\label{e5.13}
\varphi^{(\ell)}(z) &=  \redq  {\cal D}  \int {dj dj_1\over (2\pi i)^2} \left[{\Gamma(1-j)\over \Gamma(1-j_1)\Gamma(1-j+j_1)} \right]^2 M^{(\ell)}(j_1,j-j_1) \,   z^{-j+1} \,,
\end{align}
where $M^{(\ell)}(j_1,j-j_1)$ is the $O(\lambda^\ell)$   correction to the Mellin amplitude $ M_K(j_1,j-j_1)$.

To illustrate the relation \re{e5.13}, we repeat the calculation of the function $\varphi^{(\ell)}(z)$ at one loop, i.e. for $\ell=1$. The one-loop Mellin amplitude is given by \p{m1}.
After its substitution in \re{e5.13} the integration over $j_1$  can be done immediately leading to
\begin{align}
\varphi^{(1)}(z) &= {\cal D} \int {dj\over 2\pi i} {1\over 2j}\left[\pi\over \sin(\pi j) \right]^2  z^{-j+1} 
={\cal D}  \left[ {z\over 2}\left(\text{Li}_2(1-z)-\zeta_2\right) \right]
 .  \label{424}
\end{align}
Here in the second relation we closed the integration contour in the left half-plane and picked the residues at the poles $j=-1,-2,\dots$. Applying the differential operator 
\re{e512} we recover the expression \p{phi1}, obtained in the previous section from the generating function.

\subsection{Integral relation between the kernels at large $K$}

The  differential operator in \re{e512} allows us to establish a remarkably simple  relation between the detector kernels \re{K-2F1} for $K=2$ and for  $K \to \infty$.

We recall that  the kernel $\cK_{K=2}(j,z)$ is given by \p{4.17}. Applying the identity
\begin{align}\notag 
z^{-j+1} {}&= -   z^2 \frac{ \sin (\pi  j)}{\pi } \int_0^1 d\tau \,  \frac{(1-\tau)^j \tau^{-j-1}}{z+ \tau (1 - z)} 
\\
{}& = -\frac{z^2}{2  }  \int_0^1 d\tau\,    \frac{\tau(1-\tau)}{z+ \tau (1- z)}\  \cK_{K=2}(j,\tau)  \,,
\end{align}
we get from \re{e512}
\begin{align}
  &\kappa(j,z)  
  = - \frac1{2}\, {\cal D}
  \left[ z^2   \int_0^1 d\tau\,    \frac{\tau(1-\tau)}{ z + \tau (1- z)}\ \cK_{K=2}(j,\tau)     \right] .  \label{4.28}
\end{align} 

Notice that the Mellin parameter $j$ enters the right-hand side of \re{4.28} through the argument of the $K=2$ detector kernel. Then, substituting \re{kappa} and \re{4.28} into \re{MK} we can swap the Mellin integration with that in \p{4.28} to get
\begin{align}
 \varphi^{(\ell)}(z) &=   {\cal D} \left[ - \frac{z^2}{2} \int_0^1 d\tau\,    \frac{\tau(1-\tau)}{  z + \tau (1- z)} \cF^{(\ell)}_{\rm aux}(\tau) \right]  , \label{4.29}
\end{align}
where $\cF^{(\ell)}_{\rm aux}(\tau)$  is the convolution of the Mellin amplitude of the correlation function \re{Phi} and the $K=2$ detector kernel, 
\begin{align}\label{F-aux}
\cF^{(\ell)}_{{\rm aux}} (z) = \int {dj dj_1\over (2\pi i)^2} \left[{\Gamma(1-j)\over \Gamma(1-j_1)\Gamma(1-j+j_1)} \right]^2 M_K^{(\ell)}(j_1,j-j_1) \cK_{K=2}(j,z)  \,.
\end{align}
At one loop, the Mellin amplitude $M_K^{(1)}$ is independent of $K$ and is shown in  \re{m1}. As a consequence, the function $\cF^{(1)}_{\rm aux} (z)$ coincides with the one-loop correction to the energy correlation $\cF^{(1)}_{K=2} (z)$. Recalling the one-loop result \p{1loop}, we apply the relation \p{4.29} and reproduce  $\varphi^{(1)}(z)$ in \p{phi1}. 

{In summary, we have calculated the leading term  in  \p{cF-LO} in three different ways: (i) extracting it from the generating function \p{GenFun}; (ii) doing the Mellin integrations in \p{e5.13}; (iii)  doing the one-fold integration in \p{4.29}.}

At two loops, the Mellin amplitude $M_K^{(2)}$ takes different forms for $K=2$ and $K\ge 3$, see \p{M12}. In the former case, the function $\cF^{(2)}_{{\rm aux}} (z)$ coincides with the two-loop correction to the energy correlation $\cF^{(2)}_{K=2}(z)$ in \re{MellinF2phys}. In the latter case, 
$\cF^{(2)}_{{\rm aux}} (z)$ is independent of $K$ and it coincides with the auxiliary function  in \re{MellinF2nonphys}.

At higher loops, a similar transition happens at $K=\ell+1$. The underlying reason for this was explained in the beginning of Section~\ref{sect:weak}.  For $K\ge \ell+1$ the Mellin amplitude $M_K^{(\ell)}$ ceases to depend on $K$ and the same is true for the function $\cF^{(\ell)}_{{\rm aux}} (z)$. Then, it follows from \re{4.29} and \re{cF-LO} that the leading $O(1/K)$ correction to the energy correlation $ \varphi^{(\ell)}(z)$ is independent of $K$.

\subsection{Two-loop corrections to the large $K $ asymptotics} \label{sect:phi2}

In this subsection, we use the relation \p{4.29} at $\ell=2$ and compute the function $\varphi^{(2)}(z)$ defining the
leading term of the large $K$ asymptotics of the energy correlation at two loops. The evaluation of the integral in  \p{4.29} is 
much simpler than  the double  Mellin integral \p{e5.13}. 
As explained above, the function $\cF^{(2)}_{{\rm aux}} (z)$ in \p{EEC2loopK2nonphys}  should not be confused with the two-loop energy correlation ${\cal F}_{K=2}^{(2)}(z)$ defined in \p{EEC2loopK2phys}. 

According to \p{EEC2loopK2nonphys}, the function ${\cal F}_{\rm aux}^{(2)}(z)$ is a combination of   classical polylogarithms of    transcendental weight up to three. It can also be expressed in terms of harmonic polylogarithms \cite{Remiddi:1999ew}, whose arguments 
define an alphabet of three letters,  
\begin{align}
z \,, \quad 1-\sqrt{z} \,, \quad 
1+\sqrt{z}  \,. \label{HPLalph}
\end{align}  
Note that the two-loop correction  \p{EEC2loop} involves the same HPL alphabet.

The integration of ${\cal F}_{\rm aux}^{(2)}(z)$ in \p{4.29} is done with the help of \texttt{HyperInt} \cite{Panzer:2014caa}. The result %contains transcendental functions of weight up to four. They 
can be expressed in terms of classical polylogarithms of weight up to four but the HPL alphabet \p{HPLalph} is not sufficient anymore, the new letter $\sqrt{1-z} + i \sqrt{z}$ has to be added.\footnote{{This letter also appeared in the NLO calculation of EEC in QCD  \cite{Dixon:2018qgp}. There it contributes to individual Feynman diagrams representing real corrections but cancels out in their sum.}}
Schematically we can write
\begin{align} 
%f^{(2)}(z) = 
z^2 \int^1_0 d \tau \frac{\tau(1-\tau)}{z+\tau-z\tau} {\cal F}_{K=2}^{(2)}(\tau) = \frac{z}{(1-z)^2}\Bigl[ & g^{(3)}_1 + z g^{(3)}_2 + g^{(4)}_3 + z g^{(4)}_4  + z^2 g^{(4)}_5 \nt 
& + z h_1^{(4)} + i\sqrt{z(1-z)} h^{(3)}_2 \Bigr] , 
\label{intF2}
\end{align}
where $g^{(w)}_i$ and $h^{(w)}_i$ are multi-linear combinations (with rational coefficients) of  classical polylogarithms and zeta-values of homogeneous weight $w$. The $g-$functions are spanned over products of
\begin{align}
\log(z) \,, \quad {\rm Li}_2(z) \,, \quad {\rm Li}_3(z) \,, \quad {\rm Li}_3(1-z) \,, \quad {\rm Li}_4(z) \,, \quad {\rm Li}_4(1-z) \,, \quad {\rm Li}_4\left(-\frac{1-z}{z}\right) ,
\end{align}
whereas the $h-$functions are spanned over products of polylogarithms that depend on the new letter and their arguments are naturally expressed in terms of $x := i \sqrt{\frac{z}{1-z}}$,
%, which involve the new letter,
\begin{align}
\log(x) ,\, \; \log(1\pm x) ,\, \; {\rm Li}_2(x) ,\, \; {\rm Li}_3\left(\frac{x}{1+x}\right)  ,\, \; {\rm Li}_4\left(\frac{x}{1+x}\right).
\end{align}
%where $x := i \sqrt{\frac{z}{1-z}}$.  
Note that \p{intF2} is invariant upon the reflection $\sqrt{z} \to -\sqrt{z}$ that is equivalent to the complex conjugation $x\to x^*$.  

Substituting \p{intF2} into \re{4.29}, we apply the  differential operator \p{e512} and obtain a closed-form expression for the function $\varphi^{(2)}(z)$, to be found   in the ancillary file. We recall that this function defines the leading large $K$ behavior \re{cF-LO} of the energy correlation at two loops. It is interesting to note that, due to the appearance of the new letter in the HPL alphabet, its expression involves a larger set of functions than the two-loop energy correlation ${\cal F}_{K}^{(2)}(z)$ at finite $K$. Also, the maximal transcendental weight of these functions varies with $K$, as summarized in Table~\ref{tab:trans}.
\begin{table}
\begin{center}
\begin{tabular}{c|ccccc}
\toprule
loop order $\ell$ & $K=2$ & $K\geq 3$ & $K\to\infty$\\
\midrule
0 & 0 & 1 & 0 \\
1 & 1 & 3 & 2 \\
2 & 3 & 5 & 4\\
\bottomrule
\end{tabular}
\end{center}
\caption{Maximal transcendental weight of the polylogarithms in the expression for the energy correlation  ${\cal F}^{(\ell)}_K(z)$
at $\ell$ loops for finite $K$ and in the limit $K\to\infty$.
}
\label{tab:trans}
\end{table}

The dependence of $z \varphi^{(2)}(z)$ on the angular variable $0< z<1$ is shown in Figure~\ref{fig:EEC2loop} (see the curve labelled  `asympt').
%the leading term $\varphi^{(2)}(z)$ of the large $K$ two-loop asymptotics according to \p{4.29}. The explicit expression is provided in the ancillary file, and $\varphi^{(2)}(z)$ is plotted in Figure~\ref{fig:EEC2loop}.
For $z\to 1$ it approaches a finite value,
\begin{align}
\varphi^{(2)}(z) = \frac{7 \zeta_4}{4}+\frac{85 \zeta_3}{6}+\frac{181 \zeta_2}{180}+\frac{49}{144}+O\left(1-z\right) . \label{asymptPhi2}
\end{align}
For $z\to 0$ it grows as $\log z/z$,
%The expansions of the two-loop large $K$ asymptotics $\varphi^{(2)}(z)$ at $z\sim 0$ and $z \sim 1$ are as follows,
\begin{align} \label{asymptPhi21}
& \varphi^{(2)}(z) = \frac{3\log z}{z} + \left( -\frac{3}{2} \zeta_3 +3\zeta_2 -\frac{7}{2} \right) \frac{1}{z}+O\left(\log^2 z\right).
%& \varphi^{(2)}(z) = \frac{85 \zeta_3}{6}+\frac{7 \zeta_4}{4}+\frac{181 \zeta_2}{180}+\frac{49}{144}+O\left(1-z\right) . \label{asymptPhi2}
\end{align}
We verify that this relation is in agreement with the large $K$ limit of  \p{FK2loopz0}.
 
\subsection{Subleading corrections}\label{s5.2.4}

So far we have discussed only the leading term in the large $K$ expansion of the energy correlation \re{cF-LO}, 
\begin{align}\label{F-expansion}
\cF^{(\ell)}_K(z) = \frac{1}{K} \varphi^{(\ell)} (z) + \sum_{n=2}^\infty \frac{1 }{K^n}  \varphi^{(\ell)}_n (z) \,.
\end{align}
The differential equation \re{EQ} allows us to find the subleading coefficient functions $\varphi^{(\ell)}_n (z)$ in terms of the leading function $\varphi^{(\ell)}(z) \equiv \varphi^{(\ell)}_{n=1}(z)$.

Let us rewrite \re{EQ} as
\begin{align}\label{e525}
& \mD  \cF^{(\ell)}_{K}(z) =  K \left[ (K-1)\cF^{(\ell)}_{K}(z)  -  (K-2) \cF^{(\ell)}_{K-1}(z) \right]\,,
\end{align}
where the differential operator $\mathbb{D}$ is defined in \re{e527}. 
Substituting the expansion \p{F-expansion} into this equation and comparing the coefficients on both sides, we derive a recursion relation for $\varphi_n(z)$ with $n\geq 2$. The solution is
\begin{align} \label{e5.31}
& \varphi_{2}^{(\ell)}(z) = \left( 1 - \mathbb{D} \right) \varphi^{(\ell)}(z) \,, \notag\\[1.5mm]
& \varphi_{3}^{(\ell)}(z) = \left( 1 - \mathbb{D} +\frac{1}{2}\mathbb{D}^2 \right) \varphi^{(\ell)}(z)\,, \notag\\
&  \varphi_{4}^{(\ell)}(z) = \left( 1 - \mathbb{D} +\frac{1}{3}\mathbb{D}^2 -\frac{1}{6}\mathbb{D}^3 \right) \varphi^{(\ell)}(z) \,, \quad \ldots 
\end{align} 

For $z\to 1$, the functions  \re{F-expansion} approach a finite value. 
At small $z$, replacing $\varphi^{(\ell)}(z)\sim (\log z)^{\ell-1}/z$ (see \re{low-z}) in \re{e5.31}, it is straightforward to verify that the ratio $\varphi^{(\ell)}_n (z)/\varphi^{(\ell)}(z)$ goes to  $(-1)^{n-1}$ for $z\to 0$. Therefore, the subleading terms in \re{F-expansion} effectively modify the coefficient in front of the leading term   to $1/(K+1)$
\begin{align}\label{low-z1}
 {\cal F}_{K}(z) ={3\over K+1} z^{-1+\frac{\lambda}{4\pi^2}} +\dots\,,
\end{align}
where the dots denote the subleading corrections suppressed by powers of $\lambda$ and $z$.
At  two loops, the relation \re{low-z1} is in agreement with \re{1+2-loops}.  
 
\section{Contact terms}\label{s6}

The expressions for the energy correlations derived in the previous sections are valid only for $0<z<1$. To extend them to the end points $z=0$ and $z=1$,  
we have to add  contact terms. As was explained above, such terms are needed to ensure that the sum rules \re{sr} are satisfied. Furthemore, the obtained expressions for the energy correlations have non-integrable singularities at the end points $z=0$ (for $K\geq 2$ and $K\to\infty$) and $z=1$  (for  $K=2$) and require a careful treatment.

In this section we present  two complimentary approaches to computing the contact terms.   We start with the simplest example of the one-loop correction to the energy correlation $\cF^{(1)}_{K=2}(z)$ given by \p{1loop} and proceed to the case of arbitrary $K$ at one loop. In Section~\ref{s64} we compute the contact terms at two loops.

\subsection{Sum rule approach}\label{s512}

At $K=2$, the one-loop correction to the EEC \p{1loop} is singular at the end points $z= 0$ and $z= 1$. 
Following Section 4 in  Ref.~\cite{Korchemsky:2019nzm} (see also \cite{Dixon:2019uzg,Kologlu:2019mfz}), we   
 define the   regular  (i.e. integrable)  part   by subtracting  the non-integrable singularities from \p{1loop},
\begin{align}\label{7.6}
\cF^{(1)\, \text{reg}}_{K=2}(z) &= - \frac{\log (1-z)}{z^2(1-z)}  - \frac{1}{z} 
+   \frac{\log (1-z)}{1-z}   = -\frac{1}{z^2} \left( z+(z+1) \log (1-z)\right) .  
\end{align}
The resulting function $\cF^{(1)\, \text{reg}}_{K=2}(z)$ is integrable in the interval $0\le z\le 1$.
Next, we compensate the subtracted singular terms by adding them up but now interpreted as (integrable) plus-distributions (for the definitions see \p{D2} and \p{D3}).  In addition, we  allow for contact terms   at the end points with arbitrary coefficients, 
\begin{align}\label{6.9}
\cF^{(1)}_{K=2}(z)  =\cF^{(1)\, \text{reg}}_{K=2}(z)     +  \left[  \frac{1}{z} \right]_{+}   -  
 \left[  \frac{\log (1-z)}{1-z} \right]_{+}  +     C_1  \delta(z)+C_2  \delta(1- z)\,.
\end{align}
The above procedure is the most general regularization of the energy correlation at the end points consistent with the sum rules \re{sr}. Indeed, subtracting the poles makes the function integrable. The subsequent addition  of the subtracted term in the form of the plus-distributions 
does not modify the energy correlation  \p{1loop} for $0<z<1$, at the same time maintaining the integrability. At this stage the regularized expression \re{6.9} still contains the arbitrary coefficients $C_1$ and $C_2$,  which is the usual ambiguity in singular distributions. They can be determined from the sum rules \p{e1.12}.

Substituting  \re{6.9} into the sum rules \p{e1.12} we get for $K=2$
\begin{align}\label{6.12}
&   \left(1+\zeta_2\right) +C_1 + C_2 =  \zeta_2  + C_2 =0 \,,
%\nt & \Rightarrow \qquad %C_1=- 1\,, \quad C_2 =- \zeta_2 \,,
\end{align}
leading to $C_1=-1$ and $C_2 = -\zeta_2$.
%for the perturbative part of the EEC. 
%We obtain $C_1=-1$ and $C_2 = -\zeta_2$ from\footnote{When integrating the $[]_+$ distributions we choose the test functions $\phi(z)=1$ or $\phi(z)=z$, respectively.}
Finally, the complete one-loop expression for the energy correlation at weight $K=2$ is
\begin{align}
\cF^{(1)}_{K=2}(z)  =\cF^{(1)\, \text{reg}}_{K=2}(z)   +  \left[  \frac{1}{z} \right]_{+}   -  
 \left[  \frac{\log (1-z)}{1-z} \right]_{+}   - \delta(z)- \zeta_2 \delta(1- z) \,, \label{F1LK2contact}
\end{align}
where the regular part $\cF^{(1)\, \text{reg}}_{K=2}(z)$ is given by \re{7.6}.
  
For higher weights  $K \geq 3$, the contact terms for the energy correlation ${\cal F}^{(1)}_K(z)$ can be found in the same way. We recall that the function ${\cal F}^{(1)}_{K\geq3}(z)$ is singular for $z\to 0$ (see \p{EEC1loopz0}) but it is finite at $z\to1$ for $K\geq 4$ (see \p{EEC1loopz1}). 
Then we define the regular function ${\cal F}^{(1){\rm reg}}_K(z)$ by  subtracting the non-integrable singularity at $z=0$,
%we extract ${\cal F}^{(1){\rm naive}}_K(z)$ from the generating function \p{GenFun} and subtract its non-integrable singularity,
\begin{align}
{\cal F}^{(1){\rm reg}}_K(z) = \frac{1}{K!} (\partial_t)^K G(z,t)\Bigr|_{t=0} - \frac{3}{K+1}\frac{1}{z} \,,
\end{align}
where the first term on the right-hand side involves the generating function \p{GenFun}.
To get the complete one-loop result, we add to ${\cal F}^{(1){\rm reg}}_K(z)$ the plus-distribution term $[1/z]_+$ and 
calculate the coefficient of the contact term $\delta(z)$ with the help of the sum rules  \p{e1.12}. In this way, we find for $K\ge 3$
\begin{align}
{\cal F}^{(1)}_K(z) = {\cal F}^{(1){\rm reg}}_K(z) + \frac{3}{K+1}\left[\frac{1}{z}\right]_{+} + \frac{c^{(1)}_{\delta,K}}{K+1}\,\delta(z) \,, \label{F1dist}
\end{align}
where
\begin{align}
c^{(1)}_{\delta,K} = \frac{5}{2} - \frac{3}{K-1} - \frac{3}{K} - \frac{3}{K+1} \,.
\end{align}  
Let us emphasize that only the contact term at the end point $z=0$ is required, and its coefficient is fixed by one of the sum rules in \p{e1.12}. The other sum rule is not sensitive to the contact term in \p{F1dist}, and it should be automatically satisfied. This is a useful cross-check of our EEC calculation,
\begin{align}
\frac{3}{K+1} + \int^1_0 dz \, z\, {\cal F}^{(1){\rm reg}}_K(z) = 0\,.
\end{align}
At large $K$, we apply \p{phi1} to find the regular part of $\varphi^{(1)}(z)$,
\begin{align}\label{EEC-infty}
\varphi^{(1)}_{\text{reg}}(z)  & %= \varphi^{(1)}_{\text{naive}}(z) - \frac{3}{z} 
= 2 \text{Li}_2(1-z)-6 \log (z)-2\zeta_2-\frac{13}{2} \,.
%\nt
%&=   {1\over 4K}   {1 \over z^3}  \left[ (-12 \log (z)-13)+O\left(z^1\right) \right]
\end{align}
The sum rule \p{e1.12} enables us to calculate the contact terms,
\begin{align}\label{6.10}
\varphi^{(1)}(z) =\varphi^{(1)}_{\text{reg}}(z)    + \left[  \frac{3}{z} \right]_{+}  + \frac{5}{2} \delta(z)\, .
\end{align}
We see that the  distribution terms in \p{6.10} agree with those in \p{F1dist} at large $K$.
%We have already shown in Section~\ref{s413} that the expression for $\varphi^{(1)}_{\text{naive}}(z)$ agrees with EEC ${\cal F}^{(1)\text{naive}}_{K}(z)$  calculated at finite weight $K$.

\subsection{Mellin approach}\label{s513}

In this subsection, we follow Refs.~\cite{Korchemsky:2021okt,Korchemsky:2021htm} to compute the contact terms for the function $\varphi^{(1)}(z)$ which is given by the Mellin integral \p{424}. 

We computed the integral in \p{424} by closing the contour from the left and picking the poles at $j=-1,-2,-3,\ldots$. 
The integrand in \re{424} involves  ${\mathcal D}( z^{-j+1})$. It is given by the sum of three terms of the form $z^\alpha$ with $\alpha=-j-2,-j-1,-j$  
which
we now treat as distributions  $z^{\alpha}_+$, see \p{D1}. 
The key observation is that this distribution  has a contact term in its Laurent expansion, see \p{a16}. This creates an extra pole under the Mellin integral at
$\alpha=-1$, whose contribution accounts for the  contact term in the energy correlation. Of the three values of $\alpha$ listed above,
only $\alpha=-j-2$ is inside the integration contour. Replacing $z^{-j-2}_+= -\frac1{j+1}\delta(z) + \ldots$, we compute the residue at $j=-1$,
\begin{align}\label{615}
- \delta(z)  \underset{{j=-1}}{\text{Res}}\left(  \frac{1}{2}(j-2) (j-1)   
   \left[ \frac{\pi } { \sin (\pi  j)} \right]^2 \right) = \frac{5}{2}\delta(z) \,,
 \end{align}
in  agreement with \p{6.10}. The energy correlation $\cF^{(1)}_{K=2}$ with its two contact terms is treated similarly. 

%{\color{red}
%In the case of $\cF^{(1)}_{K=2}$ we expect two contact terms, see \p{6.9}. They originate from the Mellin integrand $\sim z^{-j-2} (1-z)^{j-1}$ of \p{6.3} treated as a singular distribution. This is done in two steps. Firstly, in the vicinity of $z=0$ we use $z^{-j-2}_+= -\frac1{j+1}\delta(z) + \ldots$, as before. The new residue is
%\begin{align}\label{}
% - \delta(z)  \underset{{j=-1}}{\text{Res}}\left[\frac1{(j+1)j} \frac{\pi } { \sin (\pi  j)}  \right] = - \delta(z) \,,
%\end{align}
%in agreement with \p{F1LK2contact}. Secondly, in the vicinity of $z=1$ we   close the contour from the right.  Then we use the expansion $(1-z)^{j-1}_+=  \frac1{j}\delta(1-z) + \ldots$ and pick a new residue at $j=0$:
%\begin{align}\label{}
%-\delta(1-z)  \underset{j=0}{\text{Res}}\left[ \frac1{j^2}   \frac{\pi } { \sin (\pi  j)}  \right] = -\zeta_2 \delta(1-z)  \,, 
%\end{align}
%in agreement with \p{F1LK2contact}.}

We see that the two approaches -- the sum rule method of fixing the contact terms and the approach based on the careful calculation of the Mellin integral -- give identical results for the one-loop energy correlation. However, technically the Mellin approach may be more difficult to implement beyond one loop. Indeed, there the function $M^{(\ell)}(j_1,j-j_1)$ in \p{e5.13} is given by a $2(\ell-1)-$fold Mellin integral. The sum rule approach is  more efficient, provided that the  function $\cF^{(\ell)}(z)$ is known explicitly for $0<z<1$. We apply it at the two-loop level in the next subsection. 

\subsection{Two-loop contact terms from the sum rules}\label{s64}

%We have derived the singular distribution terms of the two-loop EEC of any weight $K$ as well as at $K \to \infty$. 

Let us start with $K=2$. The two-loop function $\cF^{(2)}_{K=2}(z)$ \p{EEC2loopK2phys} was calculated in \cite{Belitsky:2013ofa} for $0<z<1$.
%assuming that $z\in (0,1)$, i.e. $\cF^{(2)\text{naive}}_{K=2}(z)$. 
 Like the one-loop case, this function has non-integrable singularities at $z=0$ and $z=1$. Promoting them to plus-distributions,  adding the contact terms $\delta(z)$ and $\delta(1-z)$ and using the sum rules  \p{e1.12} to fix their coefficients,  we  obtain  
\begin{align}\label{}
 \cF^{(2)}_{K=2}(z) &=\cF^{(2)\, \text{reg}}_{K=2}(z) + \left(-3+\zeta_2-{\zeta_3\over 2}  \right) \left[  \frac{1}{z} \right]_{+} +  \left[  \frac{\log (z)}{z} \right]_{+}\nt
 & +{\zeta_3\over 2} \left[  \frac{1}{1-z} \right]_{+} + {3\zeta_2\over 2}  \left[  \frac{\log (1-z)}{1-z} \right]_{+} + {1\over 2}  \left[  \frac{\log ^3(1-z)}{1-z} \right]_{+}\nt
 & +  \left(7 -3\zeta_2 +{11\over 4}  \zeta_4 \right)  \delta(z) + 5  \zeta_4 \delta(1-z) \,. \label{eqF2L2}
\end{align}
Here the regular function $\cF^{(\ell) \, \text{reg}}_{K} $ is obtained from $\cF^{(2)}_{K=2}(z)$ by subtracting the poles at $z=0$ and $z=1$. 
 
For $K\ge 3$, the two-loop function ${\cal F}_{K}^{(2)}(z)$ contains a non-integrable singularity at $z=0$ calculated in \p{FK2loopz0}. For $z\to 1$, it is regular  for $K \geq 4$ and has an integrable logarithmic singularity for $K=3$. As a result, only the contact term $\delta(z)$ is required. 
As before, we promote the poles at $z=0$ to  plus-distributions to get
%We promote it to distributions $[]_+$. At $z=1$, there is an integrable logarithmic singularity for $K=3$ and no singularity at all for $K \geq 4$. So only the contact term $\delta(z)$ is required,
\begin{align}
{\cal F}^{(2)}_K(z) ={}& {\cal F}^{(2){\rm reg}}_K(z) + \frac{3}{K+1}\left[\frac{\log(z)}{z}\right]_{+} + \frac{c^{(2)}_{1/z,K}}{K+1}\left[\frac{1}{z}\right]_{+}  + \frac{c^{(2)}_{\delta,K}}{K+1}\,\delta(z)   \,, \label{EEC2loopReg}
\end{align}
where $c^{(2)}_{1/z,K}$ is given in \p{c21zK}. 

To find the coefficient of the contact term $c^{(2)}_{\delta,K}$, we substitute \re{EEC2loopReg} into the sum rules \re{e1.12}.
%Evaluating the sum rule for the 
%two-loop EEC \p{EEC2loopReg}, we find the coefficient $c^{(2)}_{\delta}$ of the contact term. 
We implemented the integration in \re{e1.12} numerically and carried out the calculation for $K\leq 15$.
We interpolated the HPL expressions \p{EEC2loop}  by their generalized series expansions in the vicinity of $z=0$ and $z=1$, and integrated numerically the interpolation formulae achieving a precision of at least 40 digits. This level of precision was sufficient to reconstruct $c^{(2)}_{\delta,K}$ with $K \leq 15$ as rational linear combinations of the numbers $\{1, \zeta_2,\zeta_3,\zeta_4\}$ using the PSLQ algorithm. Exploiting these results, we arrived at  an expression for $c^{(2)}_{\delta,K}$ valid
for any weight $K$,
\begin{align}\label{c-numer}
c^{(2)}_{\delta,K} = {}& \frac{111}{16}\zeta_4 + \frac{1}{2}\left(-\frac{23}{2} + \frac{3}{K-1} + \frac{3}{K} + \frac{3}{K+1}\right) \zeta_3 + 3 H_{K+1}^{(2)} + \frac{15}{2} \notag\\
& - \left(\frac{7}{2} + \frac{3}{K-1} + \frac{3}{K} + \frac{3}{K+1}\right) \zeta_2  + \frac{8}{K-1} + \frac{7}{2K} - \frac{1}{K+1}\,,
\end{align}
where $H^{(2)}_n$ is the $n$-th generalized harmonic number of degree two. We have also checked numerically that the second sum rule in  \re{e1.12} is satisfied  for $K\leq 15$,
\begin{align}
-\frac{3}{K+1} + \frac{c^{(2)}_{1/z}(K)}{K+1} + \int^1_0 dz \, z\, {\cal F}^{(2){\rm reg}}_K(z) = 0\,.
\end{align}

Similarly, we use  \p{asymptPhi21} supplemented with the sum rule \re{e1.12} to calculate the contact terms for the two-loop function $ \varphi^{(2)}(z) $
defining the leading large $K$ asymptotics \re{cF-LO} of the energy correlation, 
\begin{align}\label{phi2contact}
 \varphi^{(2)}(z) &= \varphi^{(2)}_{\text{reg}}(z) + 3 \left[  \frac{\log (z)}{z} \right]_{+}+ \left(-{7\over 2} + 3\zeta_2-{3\over 2}\zeta_3  \right)  \left[  \frac{1}{z} \right]_{+} \nt
 & +  \left({15\over 2} - {\zeta_2\over 2} -{23\over 4}  \zeta_3 + \frac{111}{16} \zeta_4 \right)  \delta(z)  \,.
\end{align}
We would like to emphasize that, unlike \re{EEC2loopReg}, this relation  
was derived analytically.

 In the large $K$ limit the function \re{EEC2loopReg} takes the expected form \re{cF-LO}. Comparing the coefficients of the contact terms in  \re{EEC2loopReg} and \re{phi2contact} we obtain the consistency relation
\begin{align}
\lim_{K\to\infty} c^{(2)}_{\delta,K} = \frac{111}{16}\zeta_4 -\frac{23}{4} \zeta_3 -\frac{\zeta_2}{2} + \frac{15}{2}\,.
\end{align}
We verified using \re{c-numer} that it is indeed satisfied. Analogous consistency relations hold for $c^{(2)}_{1/z,K}$ in \p{c21zK} at large $K$.

\section{Event shapes at strong coupling}\label{sect6}

After exploring the dependence of the energy correlations on $K$ at weak coupling $\lambda \ll 1$, in this section we perform the same analysis at strong coupling $\lambda \gg 1$. 
We also consider event shapes in $\cN=4$ SYM with detectors other than energy calorimeters.

The energy correlations at strong coupling were analyzed in \cite{Hofman:2008ar}, where their computation  was mapped
to the propagation of a probe particle (dual to the source operator) on a shock wave background (dual to the energy calorimeters). In particular, it was found that in theories
which are dual to matter minimally coupled to gravity in AdS, the energy correlations do not depend on the angle,
\be
\label{eq:GR}
\vev{ \widehat{ \mathcal E}(n_1)\dots \widehat{ \mathcal E}(n_k) }_{\text{GR}} = 1 . 
\ee
This universality is the expression of the equivalence principle in the bulk since the energy correlations in this approximation effectively measure the Shapiro time delay, which in general
relativity does not depend on the type of particle in question. 

The leading stringy correction to the EEC was computed in \cite{Hofman:2008ar}, 
%and was found to be
\be
\label{eq:stringy}
\vev{ \widehat{ \mathcal E}(n_1) \widehat{ \mathcal E}(n_2) }_c ={4 \pi^2 \over \lambda} \left(1 - 6 z(1-z) \right)  . 
%_{\text{stringy}}
\ee
It was also found there that the connected energy correlations obey
\be
\vev{ \widehat{ \mathcal E}(n_1) \widehat{ \mathcal E}(n_2) ... \widehat{\mathcal E}(n_k) }_c \sim {1 \over \lambda^{k / 2}}  \,.
\ee
This relation is analogous to \eqref{eq:1.4} where the role of $\Delta_H$ is played by the coupling constant $\lambda$.

In this section we rederive and generalize these results starting from the Mellin space representation of the four-point functions $\vev{O_K O_2 O_2 \bar O_K }$ at strong coupling. We consider different event shapes measuring the correlations of various conserved charges (not just energy, see also footnote \ref{ftnt:spins0}). In close analogy with \eqref{e1.6} we introduce
\begin{align}\label{xyc}
\vev{\mathcal J_{s_1}(n_1)\mathcal J_{s_2}(n_2)}_K  
{}& = {(q^2)^{s_1+s_2}\over 2(4\pi)^2 (qn_1)^{s_1+1}(qn_2)^{s_2+1}} \cF_K^{s_1,s_2}(z)\,,
\end{align}
where the flow operator $\mathcal J_{s}(n)$ is built out of the conserved current of spin $s$.
The energy-energy correlation studied in the previous section corresponds to $\cF_K^{2,2}$. For $s=1$ we get the charge detector. The corresponding flow operator $\mathcal J_{s=1}(n)$ is given by the light-ray transform of the spin one conserved current (see \p{apB4}). For $s=0$ we get the scalar detector $\mathcal J_{s=0}(n)$, it is given by the light-ray transform of the  half-BPS operator of dimension  $\Delta=2$  (see \p{apB4}). More details about the correlations \p{xyc} can be found in  appendix \ref{appA}.

In this section we compute $\cF_K^{s_1,s_2}(z)$ in the supergravity approximation and derive the leading stringy correction to it, 
\be
\cF_K^{s_1,s_2}(z) = \cF_{K,\text{sugra}}^{s_1,s_2}(z) + {4 \pi^2 \over \lambda} \cF_{K,\text{stringy}}^{s_1,s_2}(z)+\dots \,,
\ee
{where the dots denote subleading corrections.}
We find that in the $K \to \infty$ limit the event shapes \re{xyc} do not depend on the angle and are given by the product of the one-point functions $\vev{\mathcal J_{s_1}(n_1)}_K\vev{\mathcal J_{s_2}(n_2)}_K$. We also analyze the event shapes away from the large $K$ limit, and we find that for event shapes other than the energy-energy correlation the dependence on $K$ is nontrivial.  For the EEC, in contrast to the weak coupling
results where the suppression factor is $\sim {1 / K}$, now the leading correction is suppressed by ${1 / \lambda}$. 

%To leading order the correlator is given by the supergravity approximation and event shapes do not depend on the angle as first observed in  \cite{Hofman:2008ar}. 
In our analysis we consider $K$'s that \textit{do not} scale parameterically with $\lambda$, see e.g. \cite{Aprile:2020luw}. Taking $K \sim \lambda^{1/4}$, which corresponds to a source dual to short massive string modes, we do not expect
to observe any change to \eqref{eq:GR} and \eqref{eq:stringy} above. For $K \gtrsim \sqrt{\lambda}$, in which case the source is described by a big classical string \cite{Buchbinder:2010ek}, we expect to get back the ${1 / K}$ suppression of the leading correction to the energy correlation \cite{Caetano:2011eb}. It would be interesting to check this explicitly. 

\subsection{Supergravity approximation}

The four-point function of the half-BPS operators $\vev{O_K O_2 O_2 \bar O_K}$ has been actively studied at strong coupling recently, starting from \cite{Rastelli:2016nze,Rastelli:2017udc}. We will be only interested in stringy corrections here, leaving quantum gravity corrections aside. 

The stringy corrections to the $\vev{O_K O_2 O_2 \bar O_K}$ Mellin amplitude were considered in \cite{Binder:2019jwn}. To utilize their results, let us notice that the Mellin variables $(s,t)$ used in that paper are related to $(j_1, j_2)$ used in the present paper as follows\footnote{The weight $K$ is denoted by $p$ in \cite{Binder:2019jwn}.}
\be
2 j_1 = K-s-t,~~~ 2 j_2 = t-K ~ .  
\ee
The Mellin amplitude ${\cal M}_K(s,t)$  in \cite{Binder:2019jwn} is related to our $M_K(j_1, j_2)$  as follows,
\be
{M_K(j_1, j_2) \over [\Gamma(1-j_1) \Gamma(1-j_2)]^2 \Gamma(2+j_1+j_2)}= {1 \over K} \Gamma(j_1+j_2+K) {\cal M}_K(-2(j_1+j_2) ,2 j_2+K)\, . 
\ee
Let us notice that while the normalization of the scalar  correlator $\vev{O_K O_2 O_2 \bar O_K}$ does not have an intrinsic meaning, it is important for us because we use superconformal Ward identities to relate it to the correlators with conserved currents which have canonical normalization, see appendix \ref{appA} for details. This explains the factor ${1 \over K}$ in the formula above.

The supergravity result for the Mellin amplitude ${\cal M}_K(s,t)$ takes the  form
\be
 {\cal M}_K^{\text{sugra}}(s,t) = {4 K \over \Gamma(K-1)} {1 \over (s-2)(t-K)(K-s-t)}\, .
\ee
%where $\tilde u = 2p - s- t$.
It is then easy to compute various event shapes by plugging this  Mellin amplitude into the master formula \eqref{e412}.
We get the following results
\be\notag
\cF_{K,\text{sugra}}^{0,0}(z) &=   2z^2 + {K(K-2)} z +\frac12 {K (K-1)^2 (K-2)} \, , 
\\ \notag
\cF_{K,\text{sugra}}^{1,1}(z) &= 2 z + \frac12{(K-2)(K+1)}\, , \\\notag
 \cF_{K,\text{sugra}}^{2,1}(z) &=-K\,, 
\\[1.5mm] 
 \cF_{K,\text{sugra}}^{2,2}(z) &= 2 \ . 
\ee
It is interesting to notice that while the energy correlations do not depend on $K$, in agreement with the arguments of  \cite{Hofman:2008ar}, the results for other event shapes require simply polynomial corrections.
We can also consider the large $K \to \infty$ limit. According to the general clustering arguments at the beginning of this paper all the event shapes effectively cluster, in other words they become angle-independent,
\be
\label{eq:scalingK}\notag
\lim_{K \to \infty}\cF_{K,\text{sugra}}^{0,0}(z) &\sim K^4 \ , \\
\lim_{K \to \infty} \cF_{K,\text{sugra}}^{1,1}(z) &\sim K^2 \ ,
\ee
with the correction to clustering being of order ${1 / K}$.

Let us comment on the power of $K$ that appears in \eqref{eq:scalingK}. As the two-point function clusters, the power is dictated by the square of the corresponding one-point event shape. This in turn is given by the light transform of the corresponding three-point function which in our case scales as $\sim K$. The light transform of a detector operator with quantum numbers $(\Delta,J)$ produces an extra factor $\Delta_H^{1-J}$. This is why starting from the three-point function which scales as $K$, we get $K^4$ for scalar-scalar correlations (detectors of spin $J=0$), $K^2$ for charge-charge correlations (spin $J=1$), and $K^0$ for energy-energy correlations (spin $J=2$).

\subsection{Stringy correction}

The computation of the stringy corrections to event shapes is subtle because the event shapes are sensitive to the high-energy
limit of the scattering amplitudes in AdS. In particular, computing the higher derivative stringy corrections to the Mellin amplitude
and plugging them into the formula \eqref{e412} produces divergent results.

To solve this problem we adopt the approach used in \cite{Goncalves:2014ffa} based on the Borel re-summation of the leading
high-energy corrections to the Mellin amplitude. The basic observation is that these are controlled by the scattering of strings
in flat space, see also \cite{Hofman:2008ar}. One can then use the formulas that relate the Mellin amplitude to its flat space limit
to predict the form of the relevant series \cite{Penedones:2010ue}.

%To compute the stringy correction we write down  the flat space of the amplitude
%\be
%-{s t u \ell_s^6\over 64} { \Gamma(-{\ell_s^2 s\over 4})  \Gamma(-{\ell_s^2 t\over 4})  \Gamma(-{\ell_s^2 u\over 4}) \over \Gamma(1+{\ell_s^2 s\over 4})  \Gamma(1+{ \ell_s^2 t\over 4})  \Gamma(1+{\ell_s^2 u \over 4})} = {1 \over {\cal N}} \lim_{L \to \infty} L^{14} \int_{- i \infty}^{i \infty} {d \alpha \over 2 \pi i} e^{\alpha} \alpha^{-4-p} {\cal M}_p \Big( {L^2 \over 2 \alpha} s,  {L^2 \over 2 \alpha} t \Big)  \ ,
%\ee
%where ${\cal N} = {2048 \pi^2 g_s^2 \ell_s^8 \over s t u } {p \over \Gamma(p-1)\Gamma(p+1)}$.

The leading stringy correction to the Mellin amplitude takes the form \cite{Binder:2019jwn}
\be
 {\cal M}_K^{\text{stringy}}(s,t) =  {4 K \over \Gamma(K-1)}   {(K+1) (K+2)(K+3) \over 4 \lambda^{3/2}} \zeta(3)  \ . 
\ee
If we try to plug this formula into the generating function for the event shapes, we find that the  $t$ integral takes the form $\int {d t \over 2 \pi i}  {\cal M}_K^{\text{stringy}}(s,t)$ and is divergent. Therefore we need to re-sum the stringy corrections. More precisely, rescaling $t \to \sqrt{\lambda} t$ we notice that infinitely many terms in the stringy expansion of the Mellin amplitude become of the same order $\int dt {t^{2n} \over \lambda^{3/2+n}} \sim {1 \over \sqrt{\lambda}}$, which is indeed the expected leading stringy correction to the event shape  \cite{Hofman:2008ar}. 

In \cite{Goncalves:2014ffa}, using the flat space limit of the amplitude, it was found that for $K=2$ the relevant series takes the form
\be
\sum_{n=0, n - \text{even}} c_n x^n, \qqqquad c_n ={1 \over 4} {\Gamma(6+n) \zeta(3+n) \over 2^{n+1}} \ .
\ee
For general $K$, using \cite{Binder:2019jwn}, we get instead
\be
\sum_{n=0, n - \text{even}} c_n x^n, \qqqquad c_n = {1 \over 4} {\Gamma(4+n+K) \zeta(3+n) \over 2^{n} \Gamma(1+K)} \, . 
\ee

To perform the Borel sum, we change the summand $c_n x^n \to {c_n (\sigma x)^n \over \Gamma(n+1)}$. We then substitute $\zeta(3+n) = \sum_{k=0}^{\infty} {1 \over (k+1)^{3+n}}$ and perform the sum over $n$.
As a result, we get the following integral when evaluating $\int {d t \over 2 \pi i}  {\cal M}_K^{\text{stringy}}(s,t)$
%\be
%{(p+1)(p+2)(p+3) \over 8 \lambda} \int {d t \over 2 \pi i} \int_0^{\infty} d z e^{-z} \sum_{k=0}^\infty {1 \over (k+1)^3} \left( {1 \over (1-{z t \over 2 (1+k)})^{4+p}} + {1 \over (1 + {z t \over 2 (1+k)})^{4+p}} \right) .
%\ee
%We can regularize this as follows
\be 
%&{(K+1)(K+2)(K+3) \over 8 \lambda} 
\int_0^{\infty} {d t \over \pi} \int_0^{\infty} d \sigma e^{-\sigma} &\sum_{k=0}^\infty {1 \over (k+1)^3} \left[ {1 \over (1-{i\sigma t \over 2 (1+k)})^{4+K}} + {1 \over (1 +{i\sigma t \over 2 (1+k)})^{4+K}} \right] =  {\pi^2 \over 3} {1 \over K+3}\,,
%\nn \\
%&=- {(K+1)(K+2) \pi^2 \over 24 \lambda}, 
\ee
where $\sigma$ is the integration of the Borel transform. To evaluate this integral we notice that if we rescale $x \to {x \over \sigma}$, which assumes $\sigma \neq 0$, we get zero. Therefore we can limit the integral over $\sigma$ to an infinitesimal interval around the origin and drop $e^{-\sigma}$ from the integrand.

In this way we get for the integral of the re-summed Mellin amplitude
\be
\int {d t \over 2 \pi i}  {\cal M}_K^{\text{stringy}}(s,t) =  {4 K \over \Gamma(K-1)}  {(K+1)(K+2) \pi^2 \over 24 \lambda}\, .
\ee
Using this result we can compute the stringy corrections to the event shapes:
\be \notag\label{FFs}
%{1 \over 2} \cF_{\text{stringy}}^{2,2}(z) &= 1- 6 z(1-z) , \\
 \cF_{\text{stringy}}^{0,0}(z) &=2 z^2 \left( 1- 6 z(1-z) + {K(K-2) \over 24} (18 z + K(K-2) -11)  \right) \ , 
 \\\notag
 \cF_{\text{stringy}}^{1,1}(z) &= 2 z \left( 1- 6 z(1-z) + {(K+1)(K-2) \over 8} (3 z - 2) \right) \ , 
 \\\notag
\cF_{\text{stringy}}^{2,1}(z) &= -{K + 2 \over 2} \left( 1- 6 z(1-z) \right) \ , \\[1.5mm]
\cF_{\text{stringy}}^{2,2}(z) &= 2 \Big( 1- 6 z(1-z) \Big) \ . 
\ee
As an extra consistency check, we verify that  the charge-charge correlation $\cF_{\text{stringy}}^{1,1}(z)$ satisfies the relation $\int_0^1 d z \,\cF_{\text{stringy}}^{1,1} = 0$, which follows from the 
conservation of the total charge. 
%current conservation Ward identity. 
Similarly, $\int_0^1 d z \cF_{\text{stringy}}^{2,2} = \int_0^1 d z z \cF_{\text{stringy}}^{2,2} = 0$ due to the energy-momentum sum rules \eqref{sr}. Putting $K=2$ in \re{FFs}, we recover the results of \cite{Belitsky:2013bja}  that, due to the superconformal Ward identities, all $\cF_{\text{stringy}}^{s_1,s_2}(z)$ are proportional to each other up to a power of $z$. This simple relationship between event shapes does not hold for $K>2$.

\section{Clustering in CFT}
\label{sect8}

In this section we present some arguments in favor of \eqref{eq:1.4} in a general CFT. We start by discussing the clustering of  correlation functions involving the stress-energy tensor in heavy states. It is analogous to familiar clustering of correlation functions in QFT as the spatial separation between operators becomes large. We then discuss the leading non-universal correction to the disconnected result which depends on the details of the theory and the heavy operator in question. We then proceed to event shapes which is the main topic of this paper, and we conjecture that the clustered structure of the stress-energy tensor correlators is not affected by the light-ray transform. Finally, we discuss the special case of heavy states in planar theories  $c_T \gg \Delta_H \gg 1 $, where $c_T$ is defined via the two-point function of stress tensors $\la T_{\mu \nu} T_{\rho \sigma} \ra \sim c_T$.\footnote{See, for example, \cite{Penedones:2016voo} for the precise definition.} The statements in this section are based on some basic physics intuition and we do not attempt to prove them rigorously starting from the CFT axioms \cite{Rychkov:2020rcd}.

\subsection{Clustering of local operators in a heavy state}

In QFT, the correlation functions of local operators factorize when the mutual separation between the local operators goes to infinity \cite{Streater:1989vi}.
In CFT a closely related property is expected to hold for the correlation functions of local operators in heavy states. We define
\be
\lla O_H(\infty) \prod_i O_L(x_i) O_H(0) \rra \equiv \lim_{x \to \infty} { \la O_H(x) \prod_i O_L(x_i) O_H(0) \ra \over \la  O_H(x) O_H(0) \ra} .
\ee
Let us also introduce the 
notation $T(x) \equiv T_{\mu \nu}(x) z^{\mu} z^{\nu}$ for the stress-energy tensor operator contracted with a null polarization vector $z^\mu$. We keep the dependence on the polarization implicit to avoid cluttering. 
%where 
We can now formulate the  clustering of the stress-energy tensor operators in a heavy state as follows
\be
\label{eq:clusteringCFT}
%\Delta_H^{- \alpha (\sum_{i} \Delta_{L_i})}
%\textbf{CFT clustering:~~~} \lim_{\Delta_H \to \infty}   { \la O_H(\infty) \prod_i O_L(x_i) O_H(0) \ra \over  \prod_i \la O_H(\infty) O_L(x_i) O_H(0) \ra} =1 .
\lim_{\Delta_H \to \infty}   { \lla O_H(\infty) \prod_i T(x_i) O_H(0) \rra \over  \prod_i \lla O_H(\infty) T(x_i) O_H(0) \rra} =1 .
\ee

Let us motivate why \eqref{eq:clusteringCFT} might be true. We consider a CFT on $\mathbb{R} \times S^{d-1}$ with the heavy operator defining a state of energy $E = {\Delta_H \over R}$ and
the physical distances between the operators being $L_{ij} = R | x_i - x_j |$. The key point is that we can naturally associate to the limit $\Delta_H \to \infty$ an infinite volume or thermodynamic limit by 
sending at the same time $R \to \infty$, see e.g. \cite{Jafferis:2017zna}. Such a limit is not uniquely defined because it requires specification of which quantities are kept fixed as we take the limit.

It is natural to keep certain local densities fixed as we take the limit. For our purposes we can keep the energy density $\epsilon = {E \over R^{d-1}} = {\Delta_H \over R^d}$ fixed. It is also
natural to keep the physical distances between the operators $L_{i j}$ fixed. Notice that it corresponds to the multi-OPE limit  $|x_{i j}| \to 0$ in the space of cross-ratios. On general grounds, we then expect to get nontrivial
correlation functions in the infinite volume $\mathbb{R}^{d-1}$ in an excited state characterized by a characteristic scale $L_{\epsilon}$. In  CFT, due to the absence of dimensionful parameters, such a correlator depends on the dimensionless ratios ${L_{i j} \over L_{\epsilon}}$. We expect that for  ${L_{i j} \over L_{\epsilon}} \ll 1$ we get nontrivial correlations in the excited state, whereas for ${L_{i j} \over L_{\epsilon}} \gg 1$ the correlator clusters as the relative spatial separation between the operators goes to infinity.\footnote{For the closely related finite temperature CFT correlators the clustering property was discussed in \cite{El-Showk:2011yvt}.}  Because the stress-energy tensor measures the energy density, the resulting one-point function $\la T \ra_\eps \neq 0$, and the disconnected piece indeed provides the leading answer to the correlator at large distances. 
 
There is an obvious generalization of the argument above for CFTs with global symmetries. Let us assume for simplicity a CFT with $U(1)$ global symmetry and denote the corresponding conserved current
$J(x) \equiv J_\mu(x) z^\mu$. We can then consider a source of large charge and apply the same argument to argue that
\be
\label{eq:clusteringCFTcharge}
%\Delta_H^{- \alpha (\sum_{i} \Delta_{L_i})}
%\textbf{CFT clustering:~~~} \lim_{\Delta_H \to \infty}   { \la O_H(\infty) \prod_i O_L(x_i) O_H(0) \ra \over  \prod_i \la O_H(\infty) O_L(x_i) O_H(0) \ra} =1 .
\lim_{Q \to \infty}   { \lla O_Q(\infty) \prod_i J(x_i) O_Q(0) \rra \over  \prod_i \lla O_Q(\infty) J(x_i) O_Q(0) \rra} =1 ,
\ee
where this time we keep the charge density  $q={Q \over R^{d-1}}$ fixed as we take the limit and we use the fact that $\la J \ra_q \neq 0$ because the conserved current measures charge density.

We expect similar arguments to hold for any local operators and not just conserved currents. However, in this case the non-vanishing property of the one-point function is not guaranteed.

Let us next discuss the leading correction to the disconnected result discussed above. For this purpose it is convenient to consider  the four-point function. At the level of the original correlation function on the plane the thermodynamic limit discussed above corresponds to the following limit
\be
\label{eq:macroscopic}
\lim_{\Delta_H \to \infty} \Delta_{H}^{-2}  \lla O_H(\infty)  T_{\mu \nu} \left(1 - {w \over \Delta_{H}^{1/d}}, 1 - {\bar w \over \Delta_{H}^{1/d}} \right) T_{\rho \sigma}(1) O_H(0)  \rra = \la T_{\mu \nu}(w, \bar w) T_{\rho \sigma}(0,0) \ra_{\eps} ,
\ee
where we have put all operators on a 2d plane with complex coordinates $(z, \bar z)$ and we set them to $\left(1 - {w / \Delta_{H}^{1/d}}, 1 - {\bar w / \Delta_{H}^{1/d}} \right)$ as we take $\Delta_H \to \infty$. The separation between the operators on the right-hand side after taking the limit is $x^2 = w \bar w$. The extra factor $\Delta_{H}^{-2}$ has a kinematic origin and comes from mapping the correlator on the plane to the one on the cylinder, see \cite{Jafferis:2017zna} for details. The two-point function $\la T_{\mu \nu}(w, \bar w) T_{\rho \sigma}(0,0) \ra_{\eps}$ stands for the two-point function in the microcanonical ensemble with energy density $\eps$ for the theory on $\mathbb{R}^{1, d-1}$. Locally, it is the same as the finite temperature correlator where the temperature is fixed to correctly reproduce the energy density $\eps$.

As we take the operator insertions apart, in an interacting theory the leading correction to the disconnected result is expected to decay exponentially, see e.g. \cite{Iliesiu:2018fao},
\be
\label{eq:clusteringthermal}
\la T_{\mu \nu}(w, \bar w) T_{\rho \sigma}(0,0) \ra_{\eps} = \la T_{\mu \nu} \ra_{\eps} \la T_{\rho \sigma} \ra_{\eps} + O(e^{- m_{\text{th}} |w|}) ,
\ee
where $m_{\text{th}}$ is the so-called thermal mass. 
%This behavior translates to the following behavior before taking the limit
%\be
%\label{eq:clusteringthermalcylinder}
% \la \phi  |  T_{\mu \nu}(x_1) T_{\rho \sigma}(x_2) | \phi  \ra =  \la \phi  |  T_{\mu \nu} | \phi  \ra \la \phi  |  T_{\rho \sigma} | \phi  \ra + O(e^{- \# \Delta_{\phi}^{1/d} | x_{12} |} ) . 
%\ee
It could also happen that the effective theory that emerges in the limit $\Delta_H \to \infty$ is gapless, in which case the leading correction to the disconnected piece is a power.
For the thermal case it happens for example in the case of the free scalar theory, and for the large charge case in the case of the three-dimensional $O(2)$ model \cite{Hellerman:2015nra,Monin:2016jmo}.  
In addition to the connected corrections in the macroscopic state we can imagine having corrections related to the finite curvature of the sphere, e.g. terms that go to zero as ${L_\eps \over R}$.

To summarize, while the leading result \eqref{eq:clusteringCFT} is universal, the corrections to it depend on the details of the theory and the nature of the heavy state. As such, they cannot be computed
 on general grounds and require a detailed case-by-case analysis. 

\subsection{Event shapes}

Going from the formulas for correlation functions of local operators to the calculation of event shapes includes the extra step of integrating over the detector times, or, equivalently, performing the light transform.
When we do that, the separation between the detector operators $|x_{12}|$  becomes arbitrarily small \cite{Kologlu:2019bco}. The small separation regime includes the light-cone limit and the Regge limit. It could then happen that
while the clustering holds for local operators at fixed separation, it fails for the light-ray operators. As a simple example, consider the charge-charge correlation related to \eqref{eq:clusteringCFTcharge}. 
As we perform the light-transform, the one-point charge correlation produces a finite result, whereas the two-point (or higher-point) function in general will diverge in the Regge limit.\footnote{It could be that due to an improved Regge behavior, the charge-charge correlation is finite. For example, this is expected to happen in the $O(2)$ Wilson-Fischer model \cite{Liu:2020tpf}, see also \cite{Caron-Huot:2022eqs}.} Therefore, the analog of \eqref{eq:clusteringCFTcharge}, in general will not hold for charge correlations. 

The situation is better for the energy correlations which are finite, or IR-safe, observables. In this case, the contribution of the Regge limit to the light transform produces a finite result. The question that remains then is whether this contribution
is suppressed, compared to the factorized result in the limit of the heavy source $\Delta_H \to \infty$. We do not know how to show it in general, but based on the explicit example analyzed in the paper and the intuition that the Regge limit is naturally suppressed for energy correlations, we conjecture that in a general CFT
%In this paper consider a closely related observable, namely energy correlations in states created by heavy sources. We argue that again 
\be
\label{eq:celcluster}
\textbf{Celestial clustering:~~~}\lim_{\Delta_H \to \infty} \vev{\mathcal E(n_1) \dots \mathcal E(n_k)} & = {\prod_{i=1}^k {(q^2)^2\over 4\pi(qn_i)^3}} .
\ee
Intuitively, \eqref{eq:celcluster} is consistent with the physical picture of a state created by a heavy operator on the celestial sphere having an effective angular correlation length that goes to zero as $\Delta_H \to \infty$.
% \footnote{Of course, there is a non-zero probability for the heavy operator to create a state with few particles. We expect this tunneling-like events to be exponentially suppressed in $\Delta_H$.} 
The same  picture suggests that the cumulant expansion of the energy correlations \eqref{eq:1.3} organizes itself in the hierarchical fashion \eqref{eq:1.4}. 

To understand this better in a non-perturbative setting it is interesting to consider the light-ray OPE, which captures the behavior of the energy correlations at small angles. For simplicity, let us start with the two-point energy correlation $\langle {\cal E}(n_1) {\cal E}(n_2) \rangle$. 
The leading OPE contribution at small angles
$z=\sin^2(\theta/2) \ll 1$ takes the form (including the $1$ from the disconnected contribution)

%\GK{I removed "c" and added $1$}
\be\label{small-z}
% {1 \over \Delta_H^2} 
\langle {\cal E}(n_1) {\cal E}(n_2) \rangle &\sim 1+ {\langle \mathbb{O}_3^+ \rangle_H \over z^{1 - {\gamma_3^+ / 2}} }\,,  \qquad z \ll 1 ,
\ee
where $\mathbb{O}_3^+$ is the spin three light-ray operator of positive signature belonging to the stress-energy tensor Regge trajectory and $\gamma_3^+$ is its anomalous dimension. 

As follows from \re{small-z}, the contribution of $\mathbb{O}_3^+$ to the energy-energy correlation diverges at small angles for $\gamma_3^+<2$. Assuming that this relation is satisfied, a necessary condition for \eqref{eq:1.4} to hold is  
\be
\langle \mathbb{O}_3^+ \rangle_H \lesssim {1 \over \Delta_H^{h_2}}\,, \qqqquad \Delta_H \to \infty \,. 
\ee
% \sout{Let us imagine that this inequality is saturated and that the leading small angle behavior of the energy-energy correlation is ${1 \over \Delta_H^{h_2}} {1 \over \theta^{2-\gamma_3^+}}$.} 
We then see from \re{small-z} that there is an emerging characteristic angle 
defined as ${1 \over \Delta_H^{h_2}} {1 \over \theta_*^{2-\gamma_3^+}} \simeq 1$. It plays the same role as $\theta_0$ in QCD. 
Namely, for $\theta \lesssim \theta_*$
the second term on the right-hand side of \re{small-z} dominates and the energy-energy correlation is in the OPE regime.
For $\theta \gtrsim \theta_*$ this term is subdominant and  the energy-energy correlation is in the flat regime. Notice that as the angle between the detectors decreases, the transition in QCD is OPE-to-flat, whereas for CFT energy correlations in heavy states it is flat-to-OPE.

Similarly, considering the multi-collinear limit of energy correlations, see e.g. \cite{Kologlu:2019mfz,Chen:2020vvp}, we get a spin $k+1$ light-ray operator
\be
\langle {\cal E}(n_1) {\cal E}(n_2) ...  {\cal E}(n_k) \rangle_c &\sim \langle \mathbb{O}_{k+1}^+ \rangle_H ,
\ee
and the conjecture  \eqref{eq:1.4} implies that
\be
\langle \mathbb{O}_{k+1}^+ \rangle_H \lesssim {1 \over \Delta_H^{h_k}}, \qqqquad \Delta_H \to \infty \ . 
\ee

Let us now see how this comes about in the simple multi-particle model of the final state. Imagine that the source carries total energy $E$ which is shared between $K$ particles.
In this case the operator $\mathbb{O}_{J}^+$ {schematically} measures the $(J-1)$-th moment of the energy distribution $\la E^{J-1} \ra = \sum_{i=1}^K E_i^{J-1}$. Taking $E_i = {E \over K}$, we get that
$\la \mathbb{O}_{J}^+ \ra \sim {E \over K^{J-2}}$. In the heavy limit we expect that $K(\Delta_H) \to \infty$, which creates the hierarchy between the different connected contributions. 
For example, in the free scalar theory the source $\phi^{K}$ creates a $K$-particle state, so that $h_k = k-1$.

\subsection{Planar theories}

In the discussion above we kept $c_T$ fixed as we took $\Delta_H \to \infty$. We expect that the suppression of connected energy correlations takes place in planar theories as well, where $c_T \gg \Delta_H \gg 1$. In this case, however, the argument about clustering in the thermodynamic limit does not apply and a different argument is needed. As we have explicitly shown in this paper using the example of half-BPS operators in ${\cal N} = 4$ SYM at large charge $K$, the suppression of connected correlations is different at weak \eqref{EE-w} and at strong \eqref{EE-s} coupling.

At weak coupling, perturbatively in $\lambda$, the connected energy correlations are suppressed by ${1 \over K}$. The mechanism is precisely the one described above, where the heavy operator creates a multi-particle state of weakly interacting particles with the number of particles $\sim K$. We expect this picture to be qualitatively correct for $\lambda \ll 1 \ll K$. 

At strong coupling, perturbatively in ${1 \over \lambda}$, the connected energy correlations are suppressed by ${1 \over \sqrt{\lambda}}$. In this case, the strongly coupled dynamics leads to a copious production of particles \cite{Hatta:2012kn}, with the characteristic energy controlled by $\lambda$ and not by $K$. We expect this picture to be correct for $1 \ll K \ll \lambda$. 

More generally, it would be interesting to exlpore different scalings between $K$ and $\lambda$ \cite{Aprile:2020luw} (in particular, see Figure 1 in that paper): $K \sim 1$ (SUGRA); $K \sim \lambda^{1/4}$ (short massive strings);  $K \sim \lambda^{1/2}$ (big classical strings). For example, the latter case is related to the so-called Frolov-Tseytlin limit \cite{Buchbinder:2010ek,Caetano:2011eb}, where the heavy state describes a classical ``fat'' string in AdS and to leading order the correlators are again simple, given by the one-point functions on this classical solution.  It would be also interesting to study energy correlations in the recently introduced large charge 't Hooft limit \cite{Caetano:2023zwe}. 

\section*{Acknowledgements}
We thank Benjamin Basso, Joao Caetano, Hao Chen, Vasco Goncalves, Jack Holguin, Shota Komatsu, Cyrille Marquet, Sasha Monin,  Ian Moult, Baur Mukhametzhanov, Kyriakos Papadodimas, Riccardo Rattazzi, Kai Yan and HuaXing Zhu  for useful discussions.
This project has received funding from the European Research Council (ERC) under the European Union's Horizon 2020 research and innovation programme (grant agreement number 949077).
DC is supported by the French National Research Agency in the framework of the ``Investissements d'avenir'' program (ANR-15-IDEX-02).

\appendix

\section{Event shapes from correlation functions}\label{sA}

In this appendix we recall the definition of an energy detector (and other flow operators) from \cite{Belitsky:2013xxa,Belitsky:2013bja}. We illustrate the procedure of calculating energy correlations from space-time correlation functions on the simple example of a single energy detector at Born level. 

The energy flow operator is defined by  the integral
\begin{align}\label{E-new}
\mathcal E(n) =  (n\bar n) \int_{-\infty}^\infty  dx_-  \lim_{x_+\to\infty} x_+^2 \,T_{++}(x_+ n  + x_- \bar n)\,,
\end{align}
in terms of the covariant light-cone component of the energy-momentum tensor
\begin{align}\label{eA2}
T_{++}\equiv \bar n^\mu\bar n^\nu T_{\mu\nu}(x)/(n\bar n)^2\,.
\end{align}
Here the lightlike vectors $n^\mu, \  \bar n^\mu$ (with $n^2=\bar n^2=0$) span a basis for the Lorentz covariant decomposition of a vector into light-cone components, e.g.
\begin{align}\label{x}
& x^\mu = x_+ n^\mu + x_- \bar n^\mu \,, \ \qqquad  x_+={(x\bar n)\over (n \bar n)}\,,\qquad x_-  ={(xn)\over (n \bar n)} \, .
\end{align} 
Lorentz covariance is maintained by requiring homogeneity under the rescaling of each light-like vector by an { arbitrary} positive scale, $n^\mu \to \rho\, n^\mu\,, \ \  \bar n^\mu \to \rho'\, \bar n^\mu $. For example, the operator \p{E-new} scales as $\rho^{-3} \cE(n)$. 
We can fix a Lorentz frame in which $ n^\mu=(1,\vec n)$ and $ \bar n^\mu=(1,-\vec n)$, where the unit vector $\vec n$ (with $\vec n^2=1$) defines the direction of the energy detector on the celestial sphere.  The vector $\bar n$ is auxiliary and it drops out from the expressions for the energy correlations.

We recall that in the maximally supersymmetric $\cN=4$ Yang-Mills theory the  energy-momentum tensor $T_{\mu\nu}$ together with the R-symmetry current $J_\mu$ and the half-BPS operator of weight two $O_2$ are members of the same supermultiplet. By analogy with the energy flow \p{E-new} one can define the R-charge and scalar flow operators (the R-symmetry indices are not displayed)
\begin{align}\label{apB4}
&\mathcal Q(n) =  (n\bar n) \int_{-\infty}^\infty  dx_-  \lim_{x_+\to\infty} x_+^2 \,J_{+}(x_+ n  + x_- \bar n)\,, \nt
&\mathcal O(n) =  (n\bar n) \int_{-\infty}^\infty  dx_-  \lim_{x_+\to\infty} x_+^2 \,O_2(x_+ n  + x_- \bar n)\,,
\end{align}
where $J_{+}\equiv \bar n^\mu J_{\mu}(x)/(n\bar n)$. We denote collectively these three flow operators as $\mathcal J_s(n)$ where the spin $s$ takes the values $s=0,1,2$.

The definition \p{E-new} of the flow operator involves two steps: (i) we send the detector, i.e. the projection $T_{++}$ of the energy-momentum tensor, weighted with the factor $x^2_+$, to spatial infinity; (ii) we integrate over the entire working time of the detector  $-\infty < x_- <  \infty$. These manipulations are  done with each insertion of the energy-momentum tensor  into the Wightman correlation function of the half-BPS scalar source and sink of weight $K$, 
\begin{align}\label{DD}
\vev{\mathcal E(n_1)\dots \mathcal E(n_k)}
&=\sigma_{\rm tot}^{-1} \int d^4 x\, \e^{iqx}  \vev{0|  O_{K}(x,Y) \, \mathcal E_1( n_1)\dots  \mathcal E_k(n_k)\,  O_{K}(0, \widebar Y)|0}_W\,.
\end{align}

{In  close analogy with the $k-$point correlation functions, the correlations \re{DD} can be decomposed into connected pieces (cumulants) $\vev{\mathcal E(n_1)\dots \mathcal E(n_k)}_c$ . {It is convenient to do it by introducing the energy correlations via a generating functional},
\begin{align}\label{Z}
{}&    \vev{\mathcal E(n_1)\dots \mathcal E(n_k)} = \partial_{J_1} \dots \partial_{J_k} Z(J)\Big|_{J_i=0}\,.
\end{align}
Here $J_i$ are sources coupled to the energy flow operators $\mathcal E(n_i)$ and the generating functional $Z(J)$ is defined by
\begin{align}
    Z(J) = e^{\sum_i J_i 
\vev{\mathcal E(n_i)}_c +   \sum_{i<j} J_i J_j
\vev{\mathcal E(n_i)\mathcal E(n_j)}_c+ \sum_{i<j<m} J_i J_j J_m
\vev{\mathcal E(n_i)\mathcal E(n_j)\mathcal E(n_m)}_c +\dots }
\end{align}
In particular,
\begin{align}
    \vev{\mathcal E(n_i)}_c=\vev{\mathcal E(n_i)}\,,\quad
    \vev{\mathcal E(n_i)\mathcal E(n_j)}_c = \vev{\mathcal E(n_i)\mathcal E(n_j)}- \vev{\mathcal E(n_i)}\vev{\mathcal E(n_j)}\,,\quad \dots
\end{align}
Note that the expressions for the connected correlations of four or more operators involve non-linear terms. 
Introducing the normalized energy flow operators $\widehat{\mathcal E}(n_i)= \mathcal E(n_i)/\vev{\mathcal E(n_i)}$, one finds that \re{Z} is equivalent to \re{eq:1.3}.}

In this paper we are mainly interested in the case of two energy flow operators, the so-called energy-energy correlation (EEC).\footnote{In Section~\ref{sect6} we also discuss the correlations of other flow operators.} For illustration purposes, in this appendix we show the details of the calculation of a single energy insertion at Born level \cite{Hofman:2008ar},\footnote{In $\cN=4$ SYM the three-point functions of the members of the stress-tensor multiplet are protected from quantum corrections. This is however not so for the four-point functions, the main subject of interest in this paper.}
\begin{align}\label{E-1}
\vev{{\cal E}(n)} = \sigma_{\rm tot}^{-1}  \int d^4 x\, \e^{iqx}  \vev{0|  O_{K}(x,Y) \, \mathcal E(n)\, O_{K}(0, \widebar Y)|0}_W^{(0)}\,.
\end{align}
The starting point   is  the  three-point function of two half-BPS operators of weight $K$ and one energy-momentum tensor. In Euclidean space it is given by 
\begin{align}\label{eA5}\notag
(G_E)^{\mu\nu}(1,2,3) &= \vev{0|O_{K}(x_1, Y) T^{\mu\nu}(x_2) O_{K}(x_3, \widebar Y)|0 }_E^{(0)} 
\\[2mm] &
 \sim  \frac{(Y \widebar Y)^K}{ x^2_{12} x^2_{2} (x^2_{1})^{K-1}}  \, \left(\frac{x_{12}^\mu}{ x^2_{12}}+\frac{x_{2}^\mu}{ x^2_{2}}\right) \left(\frac{x_{12}^\nu}{ x^2_{12}}+\frac{x_{2}^\nu}{ x^2_{2}}\right)  - \frac{\delta^{\mu\nu}}{4}  (\text{trace})\,,
\end{align}
where we have set $x_3=0$, as indicated in \p{E-1}. {The last term on the right-hand side of \re{eA5} ensures that $\delta_{\mu\nu} (G_E)^{\mu\nu}(1,2,3)=0$.}

We perform a Wick rotation to the Wightman function $(G_W)^{\mu\nu}$ in Minkowski space-time by inserting the prescription  $x^2_{ij} \to x^2_{ij}-i\ep  x^0_{ij}$ for $i<j$. Next we project the free vector indices  with $\bar n_\mu$, according to the definition \p{eA2} (this removes the trace part).  Using the decomposition \p{x} in the limit $x_{2+}\to\infty$ we obtain $x^2_{12} -i\ep   x_{12}^0 \to 2x_{2+} (  x_{2-} (n\bar n)-(x_{1}n) +i\ep )$ and $x^2_{2} -i\ep  x_{2}^0 \to 2x_{2+} (x_{2-} (n\bar n)- i\ep ) $. With this we take the limit
\begin{align}\label{eA6}
&\lim_{x_{2+} \to \infty}  x^2_{2+} (G_W)^{\mu \nu} \bar n_\mu \bar n_\nu \sim \frac{(x_{1}n)^2}{ ((x_{1}n)- x_{2-}(n\bar n) -i\ep ) ^3 (x_{2-} (n\bar n) - i\ep )^3 (x^2_1 -i\ep  x^0_1)^{K-1}} \,.
\end{align}
 We observe two poles in the variable $x_{2-}$  located on the opposite sides of the real axis. Closing the integration contour 
in, say, the upper half-plane, we get
\begin{align}\label{det3}
(n\bar n) \int_{-\infty}^\infty dx_{2-}   \lim_{x_{2+} \to \infty}  x^2_{2+} (G_W)^{\mu \nu} \bar n_\mu \bar n_\nu   \sim \frac{(Y\widebar Y)^K }{ ((x_1n) -i\ep )^3 (x_1^2 -i\ep  x_1^0)^{K-1}}\,.
\end{align}
Notice that  the auxiliary light-like vector $\bar n$ has dropped out of the right-hand side.  The last step in the procedure  is the Fourier transform in \p{E-1} which gives 
\begin{align}\label{HM}
\vev{{\cal E}(n)} =   {1\over 4\pi}\frac{(q^2)^2}{(qn)^3} \,.
\end{align}

In Section~\ref{s2} we deal with  the two-point correlation $\vev{\cE(n_1) \cE(n_2)}$  at Born level. The main contribution comes from the first diagram in Figure~\ref{fig1}. It factorizes into two one-point expressions of the type \p{det3}, with two different direction vectors $n_{1,2}$:
\begin{align}\label{eA10}
(n_1\bar n_1) (n_2\bar n_2) \int_{-\infty}^\infty dx_{2-}\, dx_{3-} {}& \lim_{x_{2+}, x_{3+} \to \infty}  x^2_{2+}  x^2_{3+}\  (G_W)^{\mu \nu; \lambda\rho }(1,2,3,4) \bar n_{1\mu} \bar n_{1\nu}\bar n_{2\lambda} \bar n_{2\rho} \nt
&   \sim \frac{(Y\widebar Y)^K }{ ((x_1n_1) -i0)^3  ((x_1n_2) -i0)^3 (x_1^2 -i0 x_1^0)^{K-1}}\,.
\end{align}
Introducing Schwinger parameters for the Fourier integral of the function in the second line, we arrive at Eq.~\p{e22} in the main text. There exist other diagrams where the two detectors exchange one propagator. They give rise to  a contact term, as explained below.

\subsection*{The origin of the contact term $\delta(z)$ at Born level}

Let us consider a simple example of an event shape, the Born level QSC (or charge-scalar correlation), defined in \p{xyc} with $s_1=1, s_2=0$. It can be obtained from the Born level correlator $\vev{O(1) J(2) O(3) O(4)}$, where $O=\bar\phi\phi$ is a scalar operator made of a charged scalar field $\phi(x)$,  and $J_\mu = i\bar\phi\overleftrightarrow{\pa_\mu}\phi$ is the $U(1)$ current. We want to turn the operators at points 2 and 3  into detectors according to \p{apB4}. The procedure becomes singular if the two detectors can exchange a particle (`cross talk'). At Born level this means to connect the detector points by a propagator. The result, as we show below, is the contact term $ \frac{1}{ (n_2 x_{14}) x^2_{14}}\, \delta( n_2 n_3)$. The subsequent Fourier transform $x_{14} \to q$ turns this into the expected contact contribution $\frac{q^2}{ (n_2 q)^2 (n_3 q)}\, \delta(z) $ (with $z$ as defined in \p{z-var}) to the event shape QSC.

\begin{figure}[t!]
 \centerline{
{\parbox[c]{120mm}{ \includegraphics[width = 55mm]{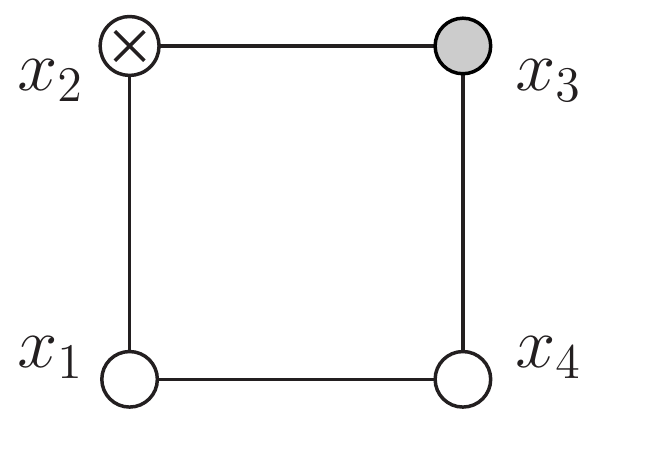} \qquad\qquad\includegraphics[width = 55mm]{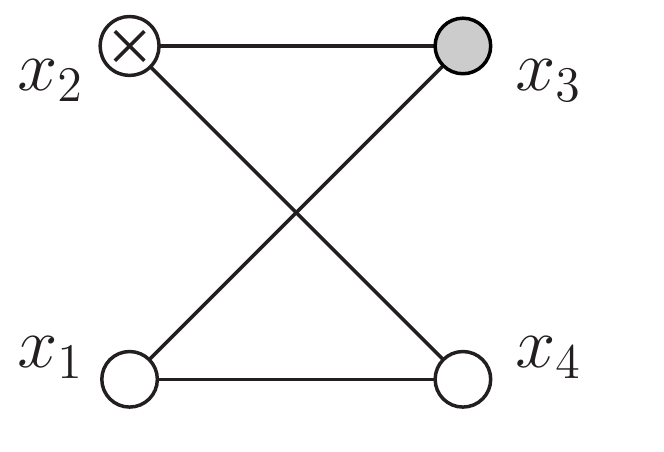}}}
}
\caption{Cross-talk diagrams for Born-level QSC. The blob with a cross and the grey blob denote the charge and scalar detectors, respectively.}\label{qsc}
\end{figure}

The  cross-talk Feynman diagrams  are shown in Figure \ref{qsc} and  have the expression\footnote{The remaining connected diagram, where both detectors are linked to the sources (see Figure~\ref{fig1}), gives rise to the contact term $\delta(1-z)$, as explained in \cite{Belitsky:2013xxa}.}
\begin{align}\label{b2.1}
&\vev{O(1) \bn_2^\mu J_\mu(2) O(3) O(4)}_{\rm Born} \ \Rightarrow \ \nt
 & \vev{\bar\phi(1) \phi(2)} i(\bn_2 \overleftrightarrow \pa_2)\vev{\bar\phi(2) \phi(3)} \vev{\bar\phi(3) \phi(4)} \vev{\bar\phi(4) \phi(1)}\nt 
&\qquad  +  \vev{\bar\phi(1) \phi(4)} \vev{\bar\phi(4) \phi(2)}  i(\bn_2 \overleftrightarrow \pa_2)\vev{\bar\phi(2) \phi(3)} \vev{\bar\phi(3) \phi(1)} \,.
\end{align}
Let us concentrate on the terms where the derivative $ (\bn_2 \pa_2)$ acts on the propagator connecting the two detector points $\vev{\bar\phi(2) \phi(3)}$ (up to normalization),
\begin{align}\label{b2.2}
 \frac{(\bar n_2 x_{23})} { (x^2_{23})^2 }\times \frac1{x^2_{14}}\left[   \frac{1}{ x^2_{12}x^2_{34}} + \frac{1}{ x^2_{42} x^2_{31}}\right]\,.
\end{align}
The remaining terms  differ by a total derivative  which vanishes under the detector  time integral. 
The analytic continuation from the Euclidean correlator \p{b2.2} to the Wightman one is done as explained after Eq.~\p{eA5}.  In the limit $x_+\to\infty$,  for a mixed pair of points source/detector we find   $ x^2_{ds} = 2x_{d+}\left[ x_{d-}-(nx_s))+i\ep\right]$. For the pair detector/detector the prescription is $x^2_{d1\, d2} - i\ep\, \varepsilon_{d1d2}$, where $\varepsilon_{d1d2} = {\rm sgn}(x_{d1+}-x_{ d2+})$. 

 In the first factor in  \p{b2.2} we  need the limits
\begin{align}\label{b.17}
&  (\bar n_2 x_{23})  = x_{2+} x_{3+}\,   \sigma_2\,, \nt
&  x^2_{23}-i\ep x_{23}^0 =-2x_{2+} x_{3+}  \left[(n_2 n_3) - x_{2-}\, \sigma_2 - x_{3-} \, \sigma_3  +i\ep\, \varepsilon_{23} \right],\nt
&\text{where} \ \  \sigma_2= 1/{x_{3+}} -{(\bar n_2 n_3)}/{x_{2+}}\,, \quad \sigma_3=1/{x_{2+}} -{(\bar n_3 n_2)}/{x_{3+}}\,.
\end{align}  
 We have pulled out the factor $ x_{2+}  x_{3+}$, which cancels that  from the definitions \p{apB4}. We get
\begin{align}\label{eq25}
&  \frac{ (\bar n_2 x_{23})}{(x^2_{23}-i\ep x_{23}^0)^2} = {- \sigma_2    \over 4[x_{2-}\, \sigma_2 + x_{3-} \, \sigma_3 -i\ep\, \varepsilon_{23}] } \ \times \   {- x_{2-}\, \sigma_2 - x_{3-} \, \sigma_3  +i\ep  \over  \left[(n_2 n_3) - x_{2-}\, \sigma_2 - x_{3-} \, \sigma_3  +i\ep\, \varepsilon_{23} \right]^2}\,.
\end{align}
 The second factor on the right-hand side is a delta sequence whose limit is $\delta(n_2 n_3)$.\footnote{Consider the sequence of functions $f_\eta(y)=\q(y)\eta/(y+\eta)^2$  and their primitives $F_\eta(y)= \q(y) y /(y+\eta)$, where $y\in \mathbb{R}$ and $\eta \in \mathbb{C}$. Clearly, $\lim_{\eta\to0}F_\eta(y) = \q(y)$, hence  $\lim_{\eta\to0}f_\eta(y) = \q'(y) = \delta(y)$. In \p{eq25}  $y=(n_2 n_3) \geq 0$  and $\eta = - x_{2-}\, \sigma_2 - x_{3-} \, \sigma_3  +i\ep \varepsilon_{23}  \in \mathbb{C}$. This delta function satisfies  the Lorentz covariant definition $\int \ep_{\mu\nu\lambda\rho} dn^\mu dn^\nu dn^\lambda n^\rho \, \delta(n^2)\, \delta(nn')\, \phi(n) = \phi(n')$ where $\phi(n)$ is a homogenous test function, $\phi(\om n)=\om^{-1} \phi(n)$. } On its support $(\bar n_2 n_3)=(\bar n_3 n_2)=1$ (up to  rescaling) and $(n_2 n_3)=(\bar n_2 \bar n_3)=0$. The first factor in \p{eq25} simplifies and the sign of $x_{d1+}-x_{ d2+}$ drops out,
\begin{align}\label{b210}
&{ \sigma_2    \over x_{2-}\, \sigma_2 + x_{3-} \, \sigma_3 -i\ep\, \varepsilon_{23} }  \ \stackrel{\delta(n_2 n_3)}{\longrightarrow}\ {1 \over x_{2-} - x_{3-}   - i\ep } \,.
\end{align}

In the limit  the  second factor  in \p{b2.2} becomes (we set $x_1 \equiv x$ and  $x_4=0$)
\begin{align}\label{b211}
  \frac1{ x^2 -i\ep x^0} \left[\frac1{ \left( x_{2-}  - (n_2 x)   +i\ep \right) (x_{3-}-i\ep )} + \frac1{ \left( x_{2-}  -i\ep \right)   \left( x_{3-} - (n_3 x)     +i\ep \right)}\right] \,.
\end{align}
The positions of the poles in \p{b210} and \p{b211} are such that only the first term  in \p{b211} contributes to the time integrals   and we get\footnote{Note that if  we replace the charge detector by another scalar detector, we do not find a delta sequence.  Moreover, the time integrals diverge. In  \cite{Belitsky:2013bja} this divergence  is avoided by selecting the $SU(4)$ channel {\bf 105} in the SSC, in which  there is no  cross-talk diagram.}
\begin{align}\label{b2.11}
{\int dx_{2-} dx_{3-}\, \lim_{x_{2+},x_{3+}\to\infty} x^2_{2+}  x^2_{3+}\ \text{Eq.~\p{b2.2}} \ \Rightarrow \quad \frac{\delta(n_2 n_3)}{(x^2 -i\ep x^0)((n_2 x)-i\ep )}}\,.
\end{align}
Finally, the  Fourier transform $x \to q$ of  \p{b2.11} yields the expected contact term in the QSC,
\begin{align}\label{}
\int \frac{d^4x\,    e^{iqx}\, \delta(n_2 n_3)}{ (-x^2 + i\ep  x^0)((n_2 x)-i\ep )} \sim \frac{\q(q^0) \q(q^2)}{(n_2 q)}\, \delta(n_2 n_3)  \sim  \frac{q^2}{ (n_2 q)^2 (n_3 q)}\, \delta(z)\,.
\end{align}

This simple example explains the origin of the contact term $\delta(z)$ in the event shapes at Born level. For a more general discussion see \cite{Korchemsky:2021htm}.

\section{Four-point correlation functions of half-BPS scalar operators}\label{AB}

In this appendix, we summarize the properties of the four-point correlation functions in $\mathcal N=4$ SYM that we use in computing the energy correlations. 

The four-point correlation functions of  interest are 
\begin{align}\label{gampl}
 G_{K_1 K_2 K_3 K_4} & = \langle O_{K_1}(x_1,Y_1) \,O_{K_2}(x_2,Y_2)\, O_{K_3}(x_3,Y_3)\, O_{K_4}(x_4,Y_4)\rangle \, ,
\end{align}
where $O_{K}(x,Y)$ are half-BPS operators defined in \p{e11}.  They are annihilated by half of the $\cN=4$ supersymmetry charges and their scaling dimension $\Delta=K$ is protected.  

The half-BPS scalar operator \p{e11} is the lowest component of a short $\cN=4$ supermultiplet. Among the operators \re{e11} with different weights $K$, the one with $K=2$ plays a special role. The corresponding $\cN=4$ supermultiplet becomes ultrashort \cite{Andrianopoli:1999vr}. It contains all the conserved currents of the  theory, including the $SU(4)$ R-symmetry current of spin one and the energy-momentum tensor of spin two, hence the name  ``stress-energy supermultiplet".
In virtue of $\cN=4$ superconformal symmetry, the four-point correlation functions of the operators belonging to the same supermultiplet are related to each other by Ward identities. 

In application to the energy correlations, we encounter the four-point correlation function 
$\vev{O_{K}(1)T_{\mu_1\nu_1}(2) T_{\mu_2\nu_2}(3) O_{K}(4)}$ where the half-BPS operators define the source/sink  and the stress-energy tensors describe the calorimeters. Applying the $\cN=4$ superconformal Ward identities, this correlator  can be expressed in terms of the four-point function of scalar operators $\vev{O_{K}(1)O_{2} (2) O_{2} (3) O_{K}(4)}$. Below we summarize  the properties of the latter.  
  
The correlator $G_{K22K}$ splits into the sum of two terms
\begin{align}\label{GK22K}
G_{K22K}=\vev{O_{K}(1)O_{2} (2) O_{2} (3) O_{K}(4)}=G^{0}_{K22K}+G^{\rm loop}_{K22K}\,.
\end{align}
The first, Born-level part is a rational function of the space-time coordinates and a polynomial in the auxiliary variables $Y$,  independent of the coupling. The second,  coupling dependent correction part  involves non-trivial functions originating from Feynman integrals. 

The expression for the Born term $G^0_{K22K}$ can be obtained by Wick contracting the scalar fields belonging to the four half-BPS operators. It is a polynomial of degree $(K+2)$ in the free scalar propagators
\begin{align}\label{25}
d_{ij} = d_{ji} \equiv \vev{\phi(x_i,Y_i) \phi(x_j,Y_j)} =  \frac{Y_i\cdot Y_j}{x^2_{ij}}\,,
\end{align}
where  $x^2_{ij}=(x_i-x_j)^2$. The coefficients of this polynomial depend on the rank of the gauge group $N_c$. In the planar limit, for $N_c\to\infty$, the connected part of $G^0_{K22K}$ is given by
\begin{align} 
G^0_{K22K}= \frac1{2} K^2(K-1)\, N_c^K  \,  d_{14}^{K-2} d_{12} d_{13} d_{23} d_{24}\,,
\end{align}
where the $K-$dependent combinatorial factor counts the number of Wick contractions.

The interacting (coupling dependent) part of \re{GK22K} takes the following form in the planar limit (see \cite{Chicherin:2015edu} and references therein)
\begin{align}\label{211'}
G^{\rm loop}_{K22K} = 2\, \left(\frac{N_c}{2}\right)^{K-2}\,  K^2\, R_{1234}\times d_{14}^{K-2}F_K(x,\lambda)\,.
\end{align}
Here $R$ is a universal rational prefactor carrying $SU(4)$  weight 2 and conformal weight 1 at each point,
\begin{align} \label{R4}
R_{1234}&= d_{13}^2 d_{24}^2 x_{13}^2x_{24}^2 + d_{12}d_{13}d_{24}d_{34}\left(x_{14}^2 x_{23}^2 -x_{13}^2x_{24}^2- x_{12}^2x_{34}^2\right) +(1\leftrightarrow 2) \, + \, (1\leftrightarrow 4)\,,
\end{align}  
where the last two terms are obtained by exchanging the pair of coordinates $(x_i,Y_i)$ of the points indicated in  the parentheses. 
According to its definition, $R_{1234}$ is completely symmetric under the exchange of any pair of points. For the special choice $Y_2 =  Y_3$  the relation \re{R4} simplifies as
\begin{align}\label{apC8}
R_{1234} \Big|_{Y_2 =  Y_3}= d_{12}d_{13}d_{24}d_{34} x_{14}^2x_{23}^2\,.
\end{align}
As explained in \cite{Belitsky:2013bja}, 
putting $Y_2 = Y_3$  is equivalent to projecting the product of the two operators $O_{2} (2) O_{2} (3)$ in \re{GK22K} onto the irreducible  representation $\bm{105}=[0,4,0]$ of $SO(6) \sim SU(4)$.  The importance of this choice is that  the scalar-scalar correlation $\langle \cO \cO \rangle_{K=2}$   in this channel coincides with the EEC. This property is however lost if the  sources have  $K>2$.

The coupling-dependent function $F_K(x,\lambda)$ is an $SU(4)$ singlet, i.e. it  does not depend on the $Y-$coordinates, and has conformal weight $1$ at all four points $x_i$ (for $i=1,\dots,4$). At weak coupling, it admits the perturbative expansion
\begin{align}\label{FK-exp}
F_K=\sum_{\ell=1}^\infty  \left(\frac{\lambda}{4\pi^2}\right)^\ell F_K^{(\ell)}\,,
\end{align}
where  the functions $F_K^{(\ell)}$ can be expanded over a basis of conformal $\ell-$loop four-point integrals. At $K=2$ the expansion \re{FK-exp} was derived in \cite{Eden:2011we,Eden:2012tu,Bourjaily:2015bpz,Bourjaily:2016evz} up to ten loops. For $K\ge 3$ the analogous expansion is known up to five loops \cite{Chicherin:2015edu,Chicherin:2018avq}.~\footnote{An interesting interpretation of these results, together with a possible extension to any loop order, was proposed in \cite{Caron-Huot:2021usw}.} The conformal integrals are independent of $K$ but they appear in the expression for  $F_K(x,\lambda)$ with  $K-$dependent coefficients. 

The explicit expressions of the function in \re{FK-exp} for $\ell=1,2$ are
\begin{align}\notag\label{Fs}
{}& F_K^{(1)}\  \, =g_{1234}\,,
\\\notag
{}& F_{K=2}^{(2)} =2  h_{1 2;3 4} + 2 h_{1 3;2 4}    +2 h_{1 4;2 3}  
  + 
 \frac12 \lr{x_{12}^2x_{34}^2+   x_{13}^2 x_{24}^2+  x_{14}^2x_{23}^2}  [g_{1 2 3 4}]^2\,,
\\
{}& F_{K>2}^{(2)} = 2 h_{1 2;3 4} + 2 h_{1 3;2 4}    +  h_{1 4;23}
  + 
 \frac12 \lr{   x_{13}^2 x_{24}^2+  x_{12}^2x_{43}^2}  [g_{1 2 3 4}]^2\,,
\end{align}
where the notation was introduced for the one- and two-loop conformal integrals
\begin{align} \notag\label{g-h}
& g_{1234}  = - 
\int \frac{d^4x_5}{x_{15}^2 x_{25}^2 x_{35}^2 x_{45}^2} \,,  \qquad 
\\
& h_{1 2;3 4}  =  x^2_{34} 
\int \frac{d^4x_5 \, d^4x_6}{(x_{15}^2 x_{35}^2 x_{45}^2) x_{56}^2
(x_{26}^2 x_{36}^2 x_{46}^2)}  \ .
\end{align}
Combining together \re{FK-exp} and \re{Fs},
one can verify that the function $F_{K=2}$ is invariant under the exchange of any pair of points in $(1,2,3,4)$, whereas $F_{K>2}$ is symmetric in the two pairs of points 
$(1,4)$ and $(2,3)$ separately, in agreement with the Bose properties of the correlation function \re{GK22K}. 

As follows from \re{Fs}, the one-loop correction in \re{FK-exp} is independent of $K$. At two loops, the functions
$F_{K=2}^{(2)}$ and $F_{K>2}^{(2)}$ are given by different linear combinations of the same conformal integrals. 
The three-loop results of \cite{Chicherin:2015edu} show that the functions $F_{K}^{(3)}$ are different for $K=2$ and $K=3$, and again become the same for all  $K\geq4$. 
The proposed generalization of \cite{Caron-Huot:2021usw} indicates that a similar pattern continues at higher loops. 
Namely, the function $F_{K}^{(\ell)}$ ceases to depend on $K$ for $K\ge \ell+1$.
This feature allows us to study the event shapes recursively, see Section~\ref{s42}. 

\subsection*{Mellin representation}     
 
The product $\Phi_K=x_{14}^2x_{23}^2 F_K(x,\lambda)$ is invariant under the conformal transformations and, therefore, it depends on the two cross-ratios \re{cr}. For our purposes, it is convenient to use the Mellin representation 
\begin{align}
F_K(x,\lambda) = {1\over x_{14}^2x_{23}^2} \int {dj_1 dj_2\over (2\pi i)^2}M_K(j_1,j_2)
\lr{x_{12}^2 x_{34}^2 \over x_{14}^2 x_{23}^2}^{j_1} \lr{x_{13}^2 x_{24}^2 \over x_{14}^2 x_{23}^2}^{j_2}\,, % \label{appB.3}
\end{align}
or equivalently $ \mathcal M[F_K]= M_K(j_1,j_2)$.
Here the integration contours are the same as in \re{Phi}.
The symmetry of $F_K(x,\lambda)$ under the exchange of $x_1$ and $x_4$ translates into $M_K(j_1,j_2)=M_K(j_2,j_1)$.
 
The Mellin amplitude $M_K(j_1,j_2)$ admits a weak-coupling expansion  analogous to \re{FK-exp}, see \p{eq:Mexpandla}. To find the corresponding functions $M^{(\ell)}_K(j_1,j_2)$, it is sufficient to know the Mellin transforms of the various conformal integrals in \re{Fs}. 
Let us denote the Mellin amplitudes for the integrals \re{g-h} as
\begin{align}\notag
& \mathcal M[g_{1234}] = M^{(1)}(j_1,j_2)\,,   
 \\[2mm]
& \mathcal M[h_{14;23}] = M^{(2)}(j_1,j_2) \,.
\end{align}
Their explicit expressions are given below, see \re{m1}. Then, the Mellin amplitudes of the remaining conformal integrals in \re{Fs} are
given by
\begin{align}\notag
{}& \mathcal M[h_{12;34}] = M^{(2)}(-1-j_1-j_2,j_2)\,,
 \\\notag
{}& \mathcal M[h_{13;24}] = M^{(2)}(-1-j_1-j_2,j_1)\,,
 \\\notag
{}& \mathcal M[x_{23}^2x_{14}^2 g_{1234}^2] = \widetilde{M}^{(2)}(j_1,j_2) \,, %=\int \frac{dj'_1 dj'_2}{(2\pi i)^2} M^{(1)}(j_1 - j'_1, j_2 - j'_2) M^{(1)}(j'_1,j'_2) 
\\\notag
{}& \mathcal M[x_{13}^2x_{24}^2 g_{1234}^2] = \widetilde{M}^{(2)}(j_1,j_2-1)\,,
\\
{}& \mathcal M[x_{12}^2x_{34}^2 g_{1234}^2]  = \widetilde{M}^{(2)}(j_1-1,j_2) \,.
\end{align}
The functions $M^{(1)}$, $M^{(2)}$ and $\widetilde M^{(2)}$ introduced above  are  
\begin{align}
M^{(1)}(j_1,j_2) {}& =-\frac{1}{4}\left[\Gamma(-j_1)\Gamma(-j_2) \Gamma(1+j_1+j_2)  \right]^2\,, \notag
\\\notag
 M^{(2)}(j_1,j_2) {}& =  -\frac{1}{4}\Gamma(-j_1)\Gamma(-j_2) \Gamma(1+j_1+j_2) 
\\\notag
{}& \times \int \frac{dj'_1 dj'_2}{(2\pi i)^2} \frac{\Gamma(j'_1-j_1)\Gamma(j'_2-j_2)\Gamma(1+j_1+j_2-j'_1-j'_2)}{\Gamma(1-j'_1)\Gamma(1-j'_2)\Gamma(1+j'_1+j'_2)} M^{(1)}(j'_1,j'_2)\,,
\\
 \widetilde{M}^{(2)}(j_1,j_2) {}&= \int \frac{dj'_1 dj'_2}{(2\pi i)^2} M^{(1)}(j_1 - j'_1, j_2 - j'_2) M^{(1)}(j'_1,j'_2)\,. \label{m1}
\end{align}
We apply the above relations to find from \re{Fs} the Mellin amplitudes of the one- and two-loop $F_K$,
\begin{align}\notag\label{M12}
M^{(1)}_K(j_1,j_2)\ \, {}&= M^{(1)}(j_1,j_2)\,,
\\[2mm]\notag
M^{(2)}_{K=2}(j_1,j_2) {}&= 2\left[ M^{(2)}(j_1,j_2)+ M^{(2)}(-1-j_1-j_2,j_2) +M^{(2)}(-1-j_1-j_2,j_1)\right]
\\\notag
{}&+\frac12 \left[ \widetilde{M}^{(2)}(j_1,j_2) + \widetilde{M}^{(2)}(j_1,j_2-1)+\widetilde{M}^{(2)}(j_1-1,j_2) \right]\,,
\\\notag
M^{(2)}_{K>2}(j_1,j_2) {}&= M^{(2)}(j_1,j_2)+ 2\left[ M^{(2)}(-1-j_1-j_2,j_2) +M^{(2)}(-1-j_1-j_2,j_1)\right] 
\\
{}&+\frac12 \left[\widetilde{M}^{(2)}(j_1,j_2-1)+\widetilde{M}^{(2)}(j_1-1,j_2) \right]\,.
\end{align}

\smallskip
 
\subsection*{Energy correlations in the Mellin representation}

According to \re{MK}, the EEC is given by the convolution of the Mellin amplitude $M_K(j_1,j_2)$ and the detector kernel $\mathcal K_K(j_1+j_2,z)$ from \re{K-2F1}. Taking into account the symmetry of the integrand of \re{MK} under the exchange  $j_1 \leftrightarrow j_2$, we get from \re{M12}
\begin{align}
{}& {\cal F}_{K=2}^{(2)}(z)  = \int \frac{dj_1 d j_2}{{(2\pi i)}^2} 
\left[{\Gamma(1-j_1-j_2)\over \Gamma(1-j_1)\Gamma(1-j_2)} \right]^2
{\cal K}_{K=2} (j_1+j_2,z)
 \notag\\
{}& \times
\Bigl[  2M^{(2)}(j_1,j_2) + 4 M^{(2)}(-1-j_1-j_2,j_1)
+ \frac{1}{2} \widetilde{M}^{(2)}(j_1,j_2) + \widetilde{M}^{(2)}(j_1,j_2-1)\Bigr]   \,.\label{MellinF2phys}
\end{align}
Here $M^{(2)}(j_1,j_2)$ and $\widetilde{M}^{(2)}(j_1,j_2)$ are the Mellin amplitudes \re{m1}  and the detector kernel is
\begin{align} 
& \cK_{K=2}(j,z) =\redq  \frac{2}{\pi} \sin (\pi  j) z^{-2-j} (1-z)^{j-1}  \,. \label{4.17}
\end{align}
The auxiliary function ${\cal F}_{\rm aux}^{(2)}(z)$ introduced in \p{F-aux} is given by the convolution of the Mellin amplitude $M^{(2)}_{K>2}(j_1,j_2)$ and the detector kernel ${\cal K}_{K=2} (j_1+j_2,z)$,
\begin{align}
 {\cal F}_{\rm aux}^{(2)}(z)   {}&= \int \frac{dj_1 d j_2}{{(2\pi i)}^2} 
\left[{\Gamma(1-j_1-j_2)\over \Gamma(1-j_1)\Gamma(1-j_2)} \right]^2
{\cal K}_{K=2} (j_1+j_2,z)
 \notag\\
{}& \times
\Bigl[  M^{(2)}(j_1,j_2) + 4 M^{(2)}(-1-j_1-j_2,j_1)
+ \widetilde{M}^{(2)}(j_1,j_2-1)\Bigr] \,. \label{MellinF2nonphys}
\end{align}
Evaluating the Mellin integrals \re{MellinF2phys} and \re{MellinF2nonphys}, one arrives at 
\re{EEC2loopK2phys} and \re{EEC2loopK2nonphys}, respectively.

\section{Detector kernel of the Mellin representation} \label{appA}

In this appendix, we present some details of the derivation of the Mellin representation \re{MK} and \re{K-2F1} of the energy correlation. We recall that the operator $\mathcal E(n)$ in \re{EE-corr} measures the flux of the energy in the direction specified by the null vector $n^\mu$ and it is built out of the energy-momentum tensor. 

If the underlying theory contains a conserved current of spin $s$,\footnote{If $s=0$ the term `conserved current', strictly speaking, does not apply. We justify this name by the fact that in $\cN=4$ SYM the operator $O_2$ of spin 0 belongs to the supermultiplet of the conserved energy-momentum tensor, supersymmetry  and R-symmetry currents.\label{ftnt:spins0}}  we can define the flow operator $\mathcal J_s(n)$ in an analogous manner.  For spin $s=2$  it coincides with $\mathcal E(n)$ whereas for arbitrary spin $s$ it measures the flow of the corresponding conserved charge.\footnote{For the definition of the flow operators with spin $s=0,1,2$ in  $\cN=4$ SYM  see \p{E-new} and \p{apB4}. Namely, ${\cal J}_{s=0} = {\cal O}$, ${\cal J}_{s=1} = {\cal Q}$, ${\cal J}_{s=2} = {\cal E}$.}  We generalize \re{EE-corr} and define the correlation of  charges with different spins
\begin{align}\label{vev-JJ} 
\vev{\mathcal J_{s_1} (n_1)\mathcal J_{s_2}(n_2) } _K  = \sigma^{-1}_{\rm tot}(q)  \int d^4 x\, \e^{iqx} \vev{O_K(x)\mathcal J_{s_1} (n_1)\mathcal J_{s_2}(n_2)\bar O_K(0)} \,,
\end{align}
where the total cross-section $\sigma_{\rm tot}(q)$ is given by \re{e14}.

In $\cN=4$ SYM the charges $\mathcal J_s$ are members  of the stress-energy multiplet. As a consequence, the correlation functions $\vev{O_K(x)\mathcal J_{s_1} (n_1)\mathcal J_{s_2}(n_2)\bar O_K(0)}$ with different spins $s_1$ and $s_2$ are related to each other by $\cN=4$ superconformal Ward identities. The general solution to these identities was derived in \cite{Belitsky:2014zha},
\begin{align}\label{JJ}
\vev{O_K(x)\mathcal J_{s_1} (n_1)\mathcal J_{s_2}(n_2)\bar O_K(0)}= {2^{s_1+1}i^{s_1+s_2}\over  (n_1n_2)^{s_1+1}} {(xn_2)^{s_1-s_2}\over (x^2)^{s_1+K-1}} {d\over d\gamma^{s_1}} 
 (1-\gamma)^{s_1} \gamma^{s_2}{d\over d\gamma^{s_2}} 
 \mathcal G_K(\gamma) \,,
\end{align}
where $\gamma$ is defined in \re{gamma} and the function $\mathcal G_K(\gamma)$ is independent of the spins  $s_1\ge s_2$. %$s_1$ and $s_2$. Note that the relation \re{JJ} is only valid for $s_1\ge s_2$.  

The relation \re{JJ} was derived in Ref.~\cite{Belitsky:2014zha} for sources of weight $K=2$. %, it is easily adapted to sources of arbitrary weight $K$. 
As discussed around \p{211'},  the generalization to arbitrary weight $K$ can be achieved by simply   attaching $(K-2)$   additional scalar propagators stretched between the source/sink  operators, thus introducing the dependence of $\mathcal G_K(\gamma)$ on the total weight. Going through  the analysis in \cite{Belitsky:2014zha} we see that the origin of the relation \p{JJ} has to do with the supermultiplet buildup at the detector points. At the source/sink points we stick to the lowest weight state of the supermultiplet, so the result is modified in an obvious way  by the additional weight $(K-2)$ at these points.

Because the function $\mathcal G_K(\gamma)$ is independent of the spins, we can obtain it by setting $s_1=s_2=0$ in \re{JJ}. In this case, the flow operators $\mathcal J_{s_i=0} (n_i)$ are related to the half-BPS scalar operators $O_{K=2}(x_i)$, see \p{apB4}, and the left-hand side of \re{JJ} is given by 
the correlation function $\vev{O_K(x)O_2(x_2) O_2(x_3)\bar O_K(0)}$ \footnote{For reasons explained  in \cite{Belitsky:2013bja,Belitsky:2014zha} (see also \p{apC8})  one restricts the four-point scalar correlator to the channel {\bf 105} in the R-symmetry decomposition. } integrated over $x_2$ and $x_3$. It is convenient to use the Mellin representation of this correlation function 
\begin{align}\notag
\vev{O_K(x_1)O_2(x_2) O_2(x_3)\bar O_K(x_4)} {}& = {1\over (x_{14}^2)^{K-2} x_{12}^2 x_{13}^2 x_{24}^2x_{34}^2} 
\\
{}& \times 
\int {dj_1 dj_2\over (2\pi i)^2}M_K(j_1,j_2)
\lr{x_{12}^2 x_{34}^2 \over x_{14}^2 x_{23}^2}^{j_1} \lr{x_{13}^2 x_{24}^2 \over x_{14}^2 x_{23}^2}^{j_2} \label{appB.3}
\end{align}
with $x_1=x$ and $x_4=0$. This relation has been obtained in Euclidean signature. Before we can proceed with the detector limit and time integration from the detector definitions \p{E-new} and \p{apB4}, we have to perform the analytic continuation to Minkowski space-time. As explained in \cite{Mack:2009mi, Belitsky:2013xxa, Belitsky:2013bja}, this amounts to replacing $x^2\to x^2-i0 x^0$ and $(xn_i)\to (xn_i)-i0$ in \re{JJ} and \re{appB.3}. The sign of the `$i0$' prescription is determined   by the order of the operators inside the correlation function \re{JJ}. Now we can repeat the calculation of  \cite{Belitsky:2014zha} to get
\begin{align}\label{calG}
\mathcal G_K(\gamma) = - {1\over 16\pi^3}\int {dj_1 dj_2\over (2\pi i)^2} \left[{\Gamma(1-j_1-j_2)\over \Gamma(1-j_1)\Gamma(1-j_2)} \right]^2 M_K(j_1,j_2) \gamma^{j_1+j_2-1} \,,
\end{align}
where $\gamma$ is given by \re{gamma}.

The last step is the Fourier transform in \re{vev-JJ}.   Expanding the derivatives in \re{JJ} we find that it is given by a linear combination of integrals of the form
\begin{align}\notag
I(a,b,c) &= \int d^4x \e^{iqx} {(xn_2)^{a}\over (x^2-i0 x^0)^{b}} \gamma^c
\\
&= \lr{(n_1n_2)/ 2}^{-c} \int d^4x \e^{iqx} {(xn_1-i0)^{c}(xn_2-i0)^{a+c}\over (x^2-i0 x^0)^{b+c}} \,.
\end{align}
Using the Schwinger parameterization they can be evaluated as 
\begin{align}\notag\label{I-int}
& I(a,b,c) = \lr{(n_1n_2)/ 2}^{-c} {(-1)^b i^{-a}\over \Gamma(-c)\Gamma(-a-c)}2\pi^3 {(q^2/4)^{b+c-2}\over \Gamma(b+c)\Gamma(b+c-1)}
\\
& \times\lr{q^2\over 2(qn_1)}^{-c}\lr{q^2\over 2(qn_2)}^{-a-c}
\int_0^1 d\tau_1  d\tau_2 \, \tau_1^{-c-1}\tau_2^{-a-c-1}
(1-\tau_1-\tau_2+z \tau_1 \tau_2)^{b+c-2}\,,
\end{align}
where the integration goes over the region $1-\tau_1-\tau_2+z \tau_1 \tau_2\ge 0$.

Combining all factors together we find from \re{vev-JJ} 
\begin{align}\notag
\vev{\mathcal J_{s_1}(n_1)\mathcal J_{s_2}(n_2)}_K  
{}& = {(q^2)^{s_1+s_2}\over 2(4\pi)^2 (qn_1)^{s_1+1}(qn_2)^{s_2+1}}  
\\
{}& \times \int {dj_1 dj_2\over (2\pi i)^2} \left[{\Gamma(1-j_1-j_2)\over \Gamma(1-j_1)\Gamma(1-j_2)} \right]^2 M_K (j_1,j_2) {\mathcal K}_K^{(s_1,s_2)}(j_1+j_2,z)  \,.\label{e412}
\end{align}
Here the dependence on the coupling constant resides in the Mellin amplitude $M_K (j_1,j_2)$. The dependence on the  variable $z$ and the spins $s_1, s_2$ comes from the detector kernel 
\begin{align}\notag\label{K-2F1'}
& {\mathcal K}_K^{(s_1,s_2)}(j,z) =  \sum_{k=0}^{s_1}(-1)^k \lr{s_1\atop k} \frac{   \Gamma (j) \Gamma (j+k)}{ 
   \Gamma (j-s_2) \Gamma (j+k-s_1)  }
 \\
&\times  {{ z^{1-j-k}}\,\Gamma(K+1)\Gamma(K-1) \over \Gamma (K-2+j+k) 
   \Gamma (K+s_1+s_2-1-j-k)}
   \,
   _2F_1\left({s_1+1-j-k, s_2+1-j-k\atop K+s_1+s_2-1-j-k}\Big|z\right),
\end{align}
where the hypergeometric function arises from the integration in \re{I-int}.

For $s_1=s_2=2$ the relation \re{e412} coincides with \re{e1.6} and \re{MK}, and the kernel \re{K-2F1'} reduces to \re{K-2F1}. 
At large $K$ the relation \re{K-2F1'} simplifies as
\begin{align}  \label{K-large}
\mathcal K_{K }^{(s_1,s_2)}(j,z) &= K^{3-s_1-s_2} 
\sum_{k=0}^{s_1} \lr{s_1\atop k}  \frac{(-1)^k  z^{1-j-k}  \Gamma (j) \Gamma (j+k)}{
   \Gamma (j-s_2) \Gamma (j+k-s_1)  } \,. % +\dots
\end{align}
 Notice that it scales as $1/K$ for $s_1=s_2=2$.

\section{Plus-distributions}\label{Gel}

 In this appendix, we recall the definition 
of the plus-distributions that appear in the expressions for the contact terms discussed in Section~\ref{s6}. 

 Following \cite{GelShil58Gen}, we define  
\begin{align} \label{D2}
   {}&  \int_0^1 dz\, \left[{1\over z}\right]_+ 
 \phi(z)= \int_0^1 {dz\over z}[\phi(z)-\phi(0)]\,, 
 \\\label{D3}
  {}&  \int_0^1 dz\, \left[{\log^k (z) \over z}\right]_+ 
 \phi(z)= \int_0^1 {dz\over z}\, \log^k (z) [\phi(z)-\phi(0)]\,,
\end{align}
where $\phi(z)$ is a  test function. 
The distributions $[1/(1-z)]_+$ and $[\log^k(1-z)/(1-z)]_+$ are defined in the same manner. 

In general, for ${\rm Re} \,\alpha >-n-1$ and $\alpha\neq -1, -2, \ldots, -n$ the distribution $z_+^\alpha$ is defined as \cite{GelShil58Gen}
\begin{align}\label{D1}
\int_0^1 dz\, z_+^\alpha \,
 \phi(z)  &= \int_0^1 dz\, z^\alpha \left[ \phi(z) -
 \sum_{k=0}^{n-1}   \frac{z^{k} }{k!  }\phi^{(k)}(0)
  \right] + \sum_{k=1}^n \frac{\phi^{(k-1)}(0)}{(k-1)! (\alpha+k)}\,,  
\end{align}
where $\phi^{(k)}(0)$ denotes the $k$-th derivative,
and similarly for the distribution $(1-z)_+^\alpha$. Note that $z^{-1}_+$ is {not} the value of  $z^\alpha_+$ at $\alpha=-1$. The distribution $z^\alpha_+$ admits the following Laurent series expansion in the vicinity of $\alpha=-1$: 
\begin{align}\label{a16}
 z^\alpha_+ = \frac{\delta(z)}{\alpha+1} + z^{-1}_+ +(\alpha+1) [z^{-1} \log z]_+ + \ldots + \frac{(\alpha+1)^k}{k!} [z^{-1} \log^k z]_+ + \ldots  
\end{align}

%%%%%%%%%
\bibliography{EEC}

\providecommand{\href}[2]{#2}\begingroup\raggedright\begin{thebibliography}{10}

\bibitem{Neill:2022lqx}
D.~Neill, G.~Vita, I.~Vitev and H.~X. Zhu, \emph{{Energy-Energy Correlators for
  Precision QCD}},  in \emph{{Snowmass 2021}}, 3, 2022.
\newblock \href{https://arxiv.org/abs/2203.07113}{{\tt 2203.07113}}.

\bibitem{Basham:1978bw}
C.~L. Basham, L.~S. Brown, S.~D. Ellis and S.~T. Love, \emph{{Energy
  Correlations in electron - Positron Annihilation: Testing QCD}},
  \href{http://dx.doi.org/10.1103/PhysRevLett.41.1585}{\emph{Phys. Rev. Lett.}
  {\bf 41} (1978) 1585}.

\bibitem{Basham:1978zq}
C.~L. Basham, L.~S. Brown, S.~D. Ellis and S.~T. Love, \emph{{Energy
  Correlations in electron-Positron Annihilation in Quantum Chromodynamics:
  Asymptotically Free Perturbation Theory}},
  \href{http://dx.doi.org/10.1103/PhysRevD.19.2018}{\emph{Phys. Rev. D} {\bf
  19} (1979) 2018}.

\bibitem{Komiske:2022enw}
P.~T. Komiske, I.~Moult, J.~Thaler and H.~X. Zhu, \emph{{Analyzing N-Point
  Energy Correlators inside Jets with CMS Open Data}},
  \href{http://dx.doi.org/10.1103/PhysRevLett.130.051901}{\emph{Phys. Rev.
  Lett.} {\bf 130} (2023) 051901},
  [\href{https://arxiv.org/abs/2201.07800}{{\tt 2201.07800}}].

\bibitem{Hofman:2008ar}
D.~M. Hofman and J.~Maldacena, \emph{{Conformal collider physics: Energy and
  charge correlations}},
  \href{http://dx.doi.org/10.1088/1126-6708/2008/05/012}{\emph{JHEP} {\bf 05}
  (2008) 012}, [\href{https://arxiv.org/abs/0803.1467}{{\tt 0803.1467}}].

\bibitem{Dixon:2019uzg}
L.~J. Dixon, I.~Moult and H.~X. Zhu, \emph{{Collinear limit of the
  energy-energy correlator}},
  \href{http://dx.doi.org/10.1103/PhysRevD.100.014009}{\emph{Phys. Rev. D} {\bf
  100} (2019) 014009}, [\href{https://arxiv.org/abs/1905.01310}{{\tt
  1905.01310}}].

\bibitem{Korchemsky:2019nzm}
G.~P. Korchemsky, \emph{{Energy correlations in the end-point region}},
  \href{http://dx.doi.org/10.1007/JHEP01(2020)008}{\emph{JHEP} {\bf 01} (2020)
  008}, [\href{https://arxiv.org/abs/1905.01444}{{\tt 1905.01444}}].

\bibitem{Kologlu:2019mfz}
M.~Kologlu, P.~Kravchuk, D.~Simmons-Duffin and A.~Zhiboedov, \emph{{The
  light-ray OPE and conformal colliders}},
  \href{http://dx.doi.org/10.1007/JHEP01(2021)128}{\emph{JHEP} {\bf 01} (2021)
  128}, [\href{https://arxiv.org/abs/1905.01311}{{\tt 1905.01311}}].

\bibitem{Chang:2020qpj}
C.-H. Chang, M.~Kologlu, P.~Kravchuk, D.~Simmons-Duffin and A.~Zhiboedov,
  \emph{{Transverse spin in the light-ray OPE}},
  \href{http://dx.doi.org/10.1007/JHEP05(2022)059}{\emph{JHEP} {\bf 05} (2022)
  059}, [\href{https://arxiv.org/abs/2010.04726}{{\tt 2010.04726}}].

\bibitem{Caron-Huot:2017vep}
S.~Caron-Huot, \emph{{Analyticity in Spin in Conformal Theories}},
  \href{http://dx.doi.org/10.1007/JHEP09(2017)078}{\emph{JHEP} {\bf 09} (2017)
  078}, [\href{https://arxiv.org/abs/1703.00278}{{\tt 1703.00278}}].

\bibitem{Kravchuk:2018htv}
P.~Kravchuk and D.~Simmons-Duffin, \emph{{Light-ray operators in conformal
  field theory}}, \href{http://dx.doi.org/10.1007/JHEP11(2018)102}{\emph{JHEP}
  {\bf 11} (2018) 102}, [\href{https://arxiv.org/abs/1805.00098}{{\tt
  1805.00098}}].

\bibitem{Sveshnikov:1995vi}
N.~A. Sveshnikov and F.~V. Tkachov, \emph{{Jets and quantum field theory}},
  \href{http://dx.doi.org/10.1016/0370-2693(96)00558-8}{\emph{Phys. Lett. B}
  {\bf 382} (1996) 403--408}, [\href{https://arxiv.org/abs/hep-ph/9512370}{{\tt
  hep-ph/9512370}}].

\bibitem{Korchemsky:1997sy}
G.~P. Korchemsky, G.~Oderda and G.~F. Sterman, \emph{{Power corrections and
  nonlocal operators}}, \href{http://dx.doi.org/10.1063/1.53732}{\emph{AIP
  Conf. Proc.} {\bf 407} (1997) 988},
  [\href{https://arxiv.org/abs/hep-ph/9708346}{{\tt hep-ph/9708346}}].

\bibitem{Korchemsky:1999kt}
G.~P. Korchemsky and G.~F. Sterman, \emph{{Power corrections to event shapes
  and factorization}},
  \href{http://dx.doi.org/10.1016/S0550-3213(99)00308-9}{\emph{Nucl. Phys. B}
  {\bf 555} (1999) 335--351}, [\href{https://arxiv.org/abs/hep-ph/9902341}{{\tt
  hep-ph/9902341}}].

\bibitem{LeBellac:1991cq}
M.~Le~Bellac, \emph{{Quantum and statistical field theory}}.
\newblock 1991.

\bibitem{Belitsky:2013xxa}
A.~V. Belitsky, S.~Hohenegger, G.~P. Korchemsky, E.~Sokatchev and A.~Zhiboedov,
  \emph{{From correlation functions to event shapes}},
  \href{http://dx.doi.org/10.1016/j.nuclphysb.2014.04.020}{\emph{Nucl. Phys. B}
  {\bf 884} (2014) 305--343}, [\href{https://arxiv.org/abs/1309.0769}{{\tt
  1309.0769}}].

\bibitem{Belitsky:2013bja}
A.~V. Belitsky, S.~Hohenegger, G.~P. Korchemsky, E.~Sokatchev and A.~Zhiboedov,
  \emph{{Event shapes in $\mathcal{N} = 4$ super-Yang-Mills theory}},
  \href{http://dx.doi.org/10.1016/j.nuclphysb.2014.04.019}{\emph{Nucl. Phys. B}
  {\bf 884} (2014) 206--256}, [\href{https://arxiv.org/abs/1309.1424}{{\tt
  1309.1424}}].

\bibitem{Belitsky:2013ofa}
A.~V. Belitsky, S.~Hohenegger, G.~P. Korchemsky, E.~Sokatchev and A.~Zhiboedov,
  \emph{{Energy-Energy Correlations in N=4 Supersymmetric Yang-Mills Theory}},
  \href{http://dx.doi.org/10.1103/PhysRevLett.112.071601}{\emph{Phys. Rev.
  Lett.} {\bf 112} (2014) 071601}, [\href{https://arxiv.org/abs/1311.6800}{{\tt
  1311.6800}}].

\bibitem{Belitsky:2014zha}
A.~V. Belitsky, S.~Hohenegger, G.~P. Korchemsky and E.~Sokatchev, \emph{{N=4
  superconformal Ward identities for correlation functions}},
  \href{http://dx.doi.org/10.1016/j.nuclphysb.2016.01.008}{\emph{Nucl. Phys. B}
  {\bf 904} (2016) 176--215}, [\href{https://arxiv.org/abs/1409.2502}{{\tt
  1409.2502}}].

\bibitem{Henn:2019gkr}
J.~M. Henn, E.~Sokatchev, K.~Yan and A.~Zhiboedov, \emph{{Energy-energy
  correlation in $N$=4 super Yang-Mills theory at next-to-next-to-leading
  order}}, \href{http://dx.doi.org/10.1103/PhysRevD.100.036010}{\emph{Phys.
  Rev. D} {\bf 100} (2019) 036010},
  [\href{https://arxiv.org/abs/1903.05314}{{\tt 1903.05314}}].

\bibitem{RattazziToAppear}
F.~Eren, A.~Monin, R.~Rattazzi and M.~Walters, \emph{{to appear}}, .

\bibitem{Lashkari:2016vgj}
N.~Lashkari, A.~Dymarsky and H.~Liu, \emph{{Eigenstate Thermalization
  Hypothesis in Conformal Field Theory}},
  \href{http://dx.doi.org/10.1088/1742-5468/aab020}{\emph{J. Stat. Mech.} {\bf
  1803} (2018) 033101}, [\href{https://arxiv.org/abs/1610.00302}{{\tt
  1610.00302}}].

\bibitem{El-Showk:2011yvt}
S.~El-Showk and K.~Papadodimas, \emph{{Emergent Spacetime and Holographic
  CFTs}}, \href{http://dx.doi.org/10.1007/JHEP10(2012)106}{\emph{JHEP} {\bf 10}
  (2012) 106}, [\href{https://arxiv.org/abs/1101.4163}{{\tt 1101.4163}}].

\bibitem{Goncalves:2014ffa}
V.~Gon\c{c}alves, \emph{{Four point function of $\mathcal{N}=4$ stress-tensor
  multiplet at strong coupling}},
  \href{http://dx.doi.org/10.1007/JHEP04(2015)150}{\emph{JHEP} {\bf 04} (2015)
  150}, [\href{https://arxiv.org/abs/1411.1675}{{\tt 1411.1675}}].

\bibitem{Korchemsky:2015ssa}
G.~P. Korchemsky and E.~Sokatchev, \emph{{Four-point correlation function of
  stress-energy tensors in $ \mathcal{N}=4 $ superconformal theories}},
  \href{http://dx.doi.org/10.1007/JHEP12(2015)133}{\emph{JHEP} {\bf 12} (2015)
  133}, [\href{https://arxiv.org/abs/1504.07904}{{\tt 1504.07904}}].

\bibitem{Korchemsky:2021okt}
G.~P. Korchemsky, E.~Sokatchev and A.~Zhiboedov, \emph{{Generalizing event
  shapes: in search of lost collider time}},
  \href{http://dx.doi.org/10.1007/JHEP08(2022)188}{\emph{JHEP} {\bf 08} (2022)
  188}, [\href{https://arxiv.org/abs/2106.14899}{{\tt 2106.14899}}].

\bibitem{Korchemsky:2021htm}
G.~P. Korchemsky and A.~Zhiboedov, \emph{{On the light-ray algebra in conformal
  field theories}},
  \href{http://dx.doi.org/10.1007/JHEP02(2022)140}{\emph{JHEP} {\bf 02} (2022)
  140}, [\href{https://arxiv.org/abs/2109.13269}{{\tt 2109.13269}}].

\bibitem{Gehrmann:2000zt}
T.~Gehrmann and E.~Remiddi, \emph{{Two loop master integrals for gamma*
  ---\ensuremath{>} 3 jets: The Planar topologies}},
  \href{http://dx.doi.org/10.1016/S0550-3213(01)00057-8}{\emph{Nucl. Phys. B}
  {\bf 601} (2001) 248--286}, [\href{https://arxiv.org/abs/hep-ph/0008287}{{\tt
  hep-ph/0008287}}].

\bibitem{Remiddi:1999ew}
E.~Remiddi and J.~A.~M. Vermaseren, \emph{{Harmonic polylogarithms}},
  \href{http://dx.doi.org/10.1142/S0217751X00000367}{\emph{Int. J. Mod. Phys.
  A} {\bf 15} (2000) 725--754},
  [\href{https://arxiv.org/abs/hep-ph/9905237}{{\tt hep-ph/9905237}}].

\bibitem{Panzer:2014caa}
E.~Panzer, \emph{{Algorithms for the symbolic integration of hyperlogarithms
  with applications to Feynman integrals}},
  \href{http://dx.doi.org/10.1016/j.cpc.2014.10.019}{\emph{Comput. Phys.
  Commun.} {\bf 188} (2015) 148--166},
  [\href{https://arxiv.org/abs/1403.3385}{{\tt 1403.3385}}].

\bibitem{Dixon:2018qgp}
L.~J. Dixon, M.-X. Luo, V.~Shtabovenko, T.-Z. Yang and H.~X. Zhu,
  \emph{{Analytical Computation of Energy-Energy Correlation at Next-to-Leading
  Order in QCD}},
  \href{http://dx.doi.org/10.1103/PhysRevLett.120.102001}{\emph{Phys. Rev.
  Lett.} {\bf 120} (2018) 102001},
  [\href{https://arxiv.org/abs/1801.03219}{{\tt 1801.03219}}].

\bibitem{Aprile:2020luw}
F.~Aprile and P.~Vieira, \emph{{Large $p$ explorations. From SUGRA to big
  STRINGS in Mellin space}},
  \href{http://dx.doi.org/10.1007/JHEP12(2020)206}{\emph{JHEP} {\bf 12} (2020)
  206}, [\href{https://arxiv.org/abs/2007.09176}{{\tt 2007.09176}}].

\bibitem{Buchbinder:2010ek}
E.~I. Buchbinder and A.~A. Tseytlin, \emph{{Semiclassical four-point functions
  in $AdS_{5}$ x $S^{5}$}},
  \href{http://dx.doi.org/10.1007/JHEP02(2011)072}{\emph{JHEP} {\bf 02} (2011)
  072}, [\href{https://arxiv.org/abs/1012.3740}{{\tt 1012.3740}}].

\bibitem{Caetano:2011eb}
J.~Caetano and J.~Escobedo, \emph{{On four-point functions and integrability in
  N=4 SYM: from weak to strong coupling}},
  \href{http://dx.doi.org/10.1007/JHEP09(2011)080}{\emph{JHEP} {\bf 09} (2011)
  080}, [\href{https://arxiv.org/abs/1107.5580}{{\tt 1107.5580}}].

\bibitem{Rastelli:2016nze}
L.~Rastelli and X.~Zhou, \emph{{Mellin amplitudes for $AdS_5\times S^5$}},
  \href{http://dx.doi.org/10.1103/PhysRevLett.118.091602}{\emph{Phys. Rev.
  Lett.} {\bf 118} (2017) 091602},
  [\href{https://arxiv.org/abs/1608.06624}{{\tt 1608.06624}}].

\bibitem{Rastelli:2017udc}
L.~Rastelli and X.~Zhou, \emph{{How to Succeed at Holographic Correlators
  Without Really Trying}},
  \href{http://dx.doi.org/10.1007/JHEP04(2018)014}{\emph{JHEP} {\bf 04} (2018)
  014}, [\href{https://arxiv.org/abs/1710.05923}{{\tt 1710.05923}}].

\bibitem{Binder:2019jwn}
D.~J. Binder, S.~M. Chester, S.~S. Pufu and Y.~Wang, \emph{{$ \mathcal{N} $ = 4
  Super-Yang-Mills correlators at strong coupling from string theory and
  localization}}, \href{http://dx.doi.org/10.1007/JHEP12(2019)119}{\emph{JHEP}
  {\bf 12} (2019) 119}, [\href{https://arxiv.org/abs/1902.06263}{{\tt
  1902.06263}}].

\bibitem{Penedones:2010ue}
J.~Penedones, \emph{{Writing CFT correlation functions as AdS scattering
  amplitudes}}, \href{http://dx.doi.org/10.1007/JHEP03(2011)025}{\emph{JHEP}
  {\bf 03} (2011) 025}, [\href{https://arxiv.org/abs/1011.1485}{{\tt
  1011.1485}}].

\bibitem{Penedones:2016voo}
J.~Penedones, \emph{{TASI lectures on AdS/CFT}},  in \emph{{Theoretical
  Advanced Study Institute in Elementary Particle Physics}: {New Frontiers in
  Fields and Strings}}, pp.~75--136, 2017.
\newblock \href{https://arxiv.org/abs/1608.04948}{{\tt 1608.04948}}.
\newblock \href{http://dx.doi.org/10.1142/9789813149441_0002}{DOI}.

\bibitem{Rychkov:2020rcd}
S.~Rychkov, \emph{{3D Ising Model: a view from the Conformal Bootstrap
  Island}}, \href{http://dx.doi.org/10.5802/crphys.23}{\emph{Comptes Rendus
  Physique} {\bf 21} (2020) 185--198},
  [\href{https://arxiv.org/abs/2007.14315}{{\tt 2007.14315}}].

\bibitem{Streater:1989vi}
R.~F. Streater and A.~S. Wightman, \emph{{PCT, spin and statistics, and all
  that}}.
\newblock 1989.

\bibitem{Jafferis:2017zna}
D.~Jafferis, B.~Mukhametzhanov and A.~Zhiboedov, \emph{{Conformal Bootstrap At
  Large Charge}}, \href{http://dx.doi.org/10.1007/JHEP05(2018)043}{\emph{JHEP}
  {\bf 05} (2018) 043}, [\href{https://arxiv.org/abs/1710.11161}{{\tt
  1710.11161}}].

\bibitem{Iliesiu:2018fao}
L.~Iliesiu, M.~Kolo\u{g}lu, R.~Mahajan, E.~Perlmutter and D.~Simmons-Duffin,
  \emph{{The Conformal Bootstrap at Finite Temperature}},
  \href{http://dx.doi.org/10.1007/JHEP10(2018)070}{\emph{JHEP} {\bf 10} (2018)
  070}, [\href{https://arxiv.org/abs/1802.10266}{{\tt 1802.10266}}].

\bibitem{Hellerman:2015nra}
S.~Hellerman, D.~Orlando, S.~Reffert and M.~Watanabe, \emph{{On the CFT
  Operator Spectrum at Large Global Charge}},
  \href{http://dx.doi.org/10.1007/JHEP12(2015)071}{\emph{JHEP} {\bf 12} (2015)
  071}, [\href{https://arxiv.org/abs/1505.01537}{{\tt 1505.01537}}].

\bibitem{Monin:2016jmo}
A.~Monin, D.~Pirtskhalava, R.~Rattazzi and F.~K. Seibold, \emph{{Semiclassics,
  Goldstone Bosons and CFT data}},
  \href{http://dx.doi.org/10.1007/JHEP06(2017)011}{\emph{JHEP} {\bf 06} (2017)
  011}, [\href{https://arxiv.org/abs/1611.02912}{{\tt 1611.02912}}].

\bibitem{Kologlu:2019bco}
M.~Kologlu, P.~Kravchuk, D.~Simmons-Duffin and A.~Zhiboedov, \emph{{Shocks,
  Superconvergence, and a Stringy Equivalence Principle}},
  \href{http://dx.doi.org/10.1007/JHEP11(2020)096}{\emph{JHEP} {\bf 11} (2020)
  096}, [\href{https://arxiv.org/abs/1904.05905}{{\tt 1904.05905}}].

\bibitem{Liu:2020tpf}
J.~Liu, D.~Meltzer, D.~Poland and D.~Simmons-Duffin, \emph{{The Lorentzian
  inversion formula and the spectrum of the 3d O(2) CFT}},
  \href{http://dx.doi.org/10.1007/JHEP09(2020)115}{\emph{JHEP} {\bf 09} (2020)
  115}, [\href{https://arxiv.org/abs/2007.07914}{{\tt 2007.07914}}].

\bibitem{Caron-Huot:2022eqs}
S.~Caron-Huot, M.~Kologlu, P.~Kravchuk, D.~Meltzer and D.~Simmons-Duffin,
  \emph{{Detectors in weakly-coupled field theories}},
  \href{http://dx.doi.org/10.1007/JHEP04(2023)014}{\emph{JHEP} {\bf 04} (2023)
  014}, [\href{https://arxiv.org/abs/2209.00008}{{\tt 2209.00008}}].

\bibitem{Chen:2020vvp}
H.~Chen, I.~Moult, X.~Zhang and H.~X. Zhu, \emph{{Rethinking jets with energy
  correlators: Tracks, resummation, and analytic continuation}},
  \href{http://dx.doi.org/10.1103/PhysRevD.102.054012}{\emph{Phys. Rev. D} {\bf
  102} (2020) 054012}, [\href{https://arxiv.org/abs/2004.11381}{{\tt
  2004.11381}}].

\bibitem{Hatta:2012kn}
Y.~Hatta, E.~Iancu, A.~H. Mueller and D.~N. Triantafyllopoulos, \emph{{Jet
  evolution from weak to strong coupling}},
  \href{http://dx.doi.org/10.1007/JHEP12(2012)114}{\emph{JHEP} {\bf 12} (2012)
  114}, [\href{https://arxiv.org/abs/1210.1534}{{\tt 1210.1534}}].

\bibitem{Caetano:2023zwe}
J.~a. Caetano, S.~Komatsu and Y.~Wang, \emph{{Large Charge 't Hooft Limit of
  $\mathcal{N}=4$ Super-Yang-Mills}},
  \href{https://arxiv.org/abs/2306.00929}{{\tt 2306.00929}}.

\bibitem{Andrianopoli:1999vr}
L.~Andrianopoli, S.~Ferrara, E.~Sokatchev and B.~Zupnik, \emph{{Shortening of
  primary operators in N extended SCFT(4) and harmonic superspace
  analyticity}}, \href{http://dx.doi.org/10.4310/ATMP.1999.v3.n4.a8}{\emph{Adv.
  Theor. Math. Phys.} {\bf 4} (2000) 1149--1197},
  [\href{https://arxiv.org/abs/hep-th/9912007}{{\tt hep-th/9912007}}].

\bibitem{Chicherin:2015edu}
D.~Chicherin, J.~Drummond, P.~Heslop and E.~Sokatchev, \emph{{All three-loop
  four-point correlators of half-BPS operators in planar $ \mathcal{N} $ = 4
  SYM}}, \href{http://dx.doi.org/10.1007/JHEP08(2016)053}{\emph{JHEP} {\bf 08}
  (2016) 053}, [\href{https://arxiv.org/abs/1512.02926}{{\tt 1512.02926}}].

\bibitem{Eden:2011we}
B.~Eden, P.~Heslop, G.~P. Korchemsky and E.~Sokatchev, \emph{{Hidden symmetry
  of four-point correlation functions and amplitudes in N=4 SYM}},
  \href{http://dx.doi.org/10.1016/j.nuclphysb.2012.04.007}{\emph{Nucl. Phys. B}
  {\bf 862} (2012) 193--231}, [\href{https://arxiv.org/abs/1108.3557}{{\tt
  1108.3557}}].

\bibitem{Eden:2012tu}
B.~Eden, P.~Heslop, G.~P. Korchemsky and E.~Sokatchev, \emph{{Constructing the
  correlation function of four stress-tensor multiplets and the four-particle
  amplitude in N=4 SYM}},
  \href{http://dx.doi.org/10.1016/j.nuclphysb.2012.04.013}{\emph{Nucl. Phys. B}
  {\bf 862} (2012) 450--503}, [\href{https://arxiv.org/abs/1201.5329}{{\tt
  1201.5329}}].

\bibitem{Bourjaily:2015bpz}
J.~L. Bourjaily, P.~Heslop and V.-V. Tran, \emph{{Perturbation Theory at Eight
  Loops: Novel Structures and the Breakdown of Manifest Conformality in N=4
  Supersymmetric Yang-Mills Theory}},
  \href{http://dx.doi.org/10.1103/PhysRevLett.116.191602}{\emph{Phys. Rev.
  Lett.} {\bf 116} (2016) 191602},
  [\href{https://arxiv.org/abs/1512.07912}{{\tt 1512.07912}}].

\bibitem{Bourjaily:2016evz}
J.~L. Bourjaily, P.~Heslop and V.-V. Tran, \emph{{Amplitudes and Correlators to
  Ten Loops Using Simple, Graphical Bootstraps}},
  \href{http://dx.doi.org/10.1007/JHEP11(2016)125}{\emph{JHEP} {\bf 11} (2016)
  125}, [\href{https://arxiv.org/abs/1609.00007}{{\tt 1609.00007}}].

\bibitem{Chicherin:2018avq}
D.~Chicherin, A.~Georgoudis, V.~Gon\c{c}alves and R.~Pereira, \emph{{All
  five-loop planar four-point functions of half-BPS operators in $\mathcal N=4$
  SYM}}, \href{http://dx.doi.org/10.1007/JHEP11(2018)069}{\emph{JHEP} {\bf 11}
  (2018) 069}, [\href{https://arxiv.org/abs/1809.00551}{{\tt 1809.00551}}].

\bibitem{Caron-Huot:2021usw}
S.~Caron-Huot and F.~Coronado, \emph{{Ten dimensional symmetry of $ \mathcal{N}
  $ = 4 SYM correlators}},
  \href{http://dx.doi.org/10.1007/JHEP03(2022)151}{\emph{JHEP} {\bf 03} (2022)
  151}, [\href{https://arxiv.org/abs/2106.03892}{{\tt 2106.03892}}].

\bibitem{Mack:2009mi}
G.~Mack, \emph{{D-independent representation of Conformal Field Theories in D
  dimensions via transformation to auxiliary Dual Resonance Models. Scalar
  amplitudes}},  \href{https://arxiv.org/abs/0907.2407}{{\tt 0907.2407}}.

\bibitem{GelShil58Gen}
I.~Gelfand and G.~Shilov, \emph{Generalized functions, vol. 1 academic press},
  {\emph{New York} {\bf 1967} (1964) }.

\end{thebibliography}\endgroup
\bibliographystyle{JHEP}

\end{document}